\documentclass[aps,pra,twocolumn,showpacs,letterpaper,tighten,float,superscriptaddress,longbibliography,footinbib]{revtex4-1}
\usepackage{amssymb}
\usepackage{amsbsy}
\usepackage{amsmath}
\usepackage{mathrsfs}
\usepackage{epsfig}
\usepackage{graphicx}
\usepackage{array}
\usepackage{textcomp}
\usepackage{color}
\usepackage{braket}
\usepackage{bm,hyperref}
\usepackage[justification=raggedright,font=small]{caption}
\usepackage[titletoc,toc,title]{appendix}
\usepackage[normalem]{ulem}

\usepackage[labelformat=simple]{subcaption}

\definecolor{myblue}{rgb}{.93, .93, 1}

\definecolor{darkgreen}{rgb}{0,0.7,0}

\newcommand{\beq}{\begin{equation}}
\newcommand{\eeq}{\end{equation}}

\newcommand{\bpm}{\begin{pmatrix}}
\newcommand{\epm}{\end{pmatrix}}
\newcommand{\bmm}{\begin{matrix}}
\newcommand{\emm}{\end{matrix}}

\renewcommand{\>}{\rangle}
\newcommand{\mD}{\mathcal{D}}
\newcommand{\mC}{\mathcal{C}}
\newcommand{\mZ}{\mathcal{Z}}
\newcommand{\mP}{\mathcal{P}}
\newcommand{\mbZ}{\mathbb{Z}}

\graphicspath{{./Figures/}}

\begin{document}

\title{Cage-Net Fracton Models}
\author{Abhinav Prem}\thanks{These authors contributed equally to this work.}
\affiliation{
Department of Physics and Center for Theory of Quantum Matter,
University of Colorado, Boulder, Colorado 80309, USA
}
\author{Sheng-Jie Huang}\thanks{These authors contributed equally to this work.}
\affiliation{
Department of Physics and Center for Theory of Quantum Matter,
University of Colorado, Boulder, Colorado 80309, USA
}
\author{Hao Song}
\affiliation{Departamento de F\'isica Te\'orica, Universidad Complutense, 28040 Madrid, Spain}
\author{Michael Hermele}
\affiliation{
Department of Physics and Center for Theory of Quantum Matter,
University of Colorado, Boulder, Colorado 80309, USA
}
\date{\today}

\begin{abstract}

We introduce a class of gapped three-dimensional models, dubbed ``cage-net fracton models," which host immobile fracton excitations in addition to non-Abelian particles with restricted mobility. Starting from layers of two-dimensional string-net models, whose spectrum includes non-Abelian anyons, we condense extended one-dimensional ``flux-strings" built out of point-like excitations. Flux-string condensation generalizes the concept of anyon condensation familiar from conventional topological order and allows us to establish properties of the fracton ordered (equivalently, flux-string condensed) phase, such as its ground state wave function and spectrum of excitations. Through the examples of doubled Ising and SU(2)$_k$ cage-net models, we demonstrate the existence of  strictly immobile Abelian fractons and of non-Abelian particles restricted to move only along one dimension.  In the doubled Ising cage-net model, we show that these restricted-mobility non-Abelian excitations are a fundamentally three-dimensional phenomenon, as they cannot be understood as bound states amongst two-dimensional non-Abelian anyons and Abelian particles.  We further show that the ground state wave function of such phases can be understood as a fluctuating network of extended objects -- cages -- and strings, which we dub a cage-net condensate. Besides having implications for topological quantum computation in three dimensions, our work may also point the way towards more general insights into quantum phases of matter with fracton order.

\end{abstract}

\maketitle



\section{Introduction and Motivation}
\label{intro}

A topologically ordered quantum phase of matter in arbitrary spatial dimensions is defined as one which exhibits a finite gap to all excitations in the thermodynamic limit, has a finite but non-trivial ground state degeneracy on a topologically non-trivial manifold---such that no local operators can distinguish between the degenerate ground states---and supports fractionalized quasi-particles which cannot be locally created. Starting with the discovery of the fractional quantum Hall effect~\cite{Tsui, Laughlin}, significant theoretical attention has been paid to the characterization and classification of such quantum phases of matter, which are intrinsically tied to a pattern of long-range entanglement in their many-body ground states~\cite{wen1989,wen1990,wen1990groundstate,preskill,levin2006,chen2010,wen2013}. The possibility of realizing quasi-particles exhibiting non-Abelian braiding statistics~\cite{mooreread,kitaev,gurarie} and the potential application of topological states for fault-tolerant quantum computation~\cite{kitaev2,ColorCode2D,ColorCode3D,nayakreview,readreview} has further spurred progress in understanding topological order. In $d=2$ spatial dimensions in particular, a clear picture of topological order has emerged, from its realization in quantum Hall states~\cite{Laughlin,wenniu,mooreread}, bosonic spin liquids~\cite{wen2002,balents2002,moessner,levinwen}, and superconductors (with dynamical electromagnetism)~\cite{wen1991,sondhi,moroz} to its description in terms of topological quantum field theories~\cite{witten1989} or within the formalism of tensor category theory, for both bosonic~\cite{kitaev,Rowell,wenbos} and fermionic~\cite{lan2016} topological order. In addition, much has now been understood about the interplay of symmetry and topology through the study of symmetry enriched topological (SET) phases in 2d~\cite{wen2002,levin,essin,ran,lu,barkeshli,xu,hung,chengSET,Chen2015AnomalousSET,hsongThesis,hsongSGSET,hsongPGSPT,wenrev}.

Recently, a frisson of excitement and a layer of intrigue has been added to the study of topological order in three spatial dimensions owing to the theoretical discovery of a novel class of quantum phases, which have been the focus of intense theoretical research~\cite{chamon,castelnovo,bravyi,haah,haah2,yoshida,fracton1,fracton2,sub,genem,williamson,prem,han,sagar,hsieh,slagle1,nonabelian,decipher,balents,slagle2,screening,chiral,prem2,regnault,valbert,devakul,regnault2,han2,albert,leomichael,gromov,shirley,slagle3,han3,bulmash,twisted,yizhi1,devakulfractal,yizhi2,pinch}.
Discovered initially in certain exactly solvable lattice spin models, these exotic phases host point-like excitations which are fundamentally immobile or which are confined to move only along sub-dimensional manifolds. The immobility of certain excitations, dubbed ``fractons," stems from the lack of any one-dimensional string-like operator at the ends of which they may be created. Instead, depending on the specific model, fractons are created either at the corners of two-dimensional membrane~\cite{chamon,fracton1,fracton2} or fractal~\cite{haah,yoshida} operators, with the corresponding models referred to as ``type-I" or ``type-II" fracton models respectively in the taxonomy of Vijay, Haah, and Fu~\cite{fracton2}. While type-I models host additional topologically charged excitations which can move only along $c<d$ sub-manifolds and are hence termed ``dim-$c$" excitations, type-II models have \textit{no} mobile excitations which carry topological charge.

Despite the striking appearance of such exotic quasi-particles, gapped fracton models~\footnote{These exclude tensor gauge theories discussed in Refs.~\cite{sub,genem,screening,chiral,prem2,leomichael,gromov} which also display much of the fracton phenomenology but are gapless.} display many features familiar from topological order---they have a gap to all excitations, display long-range entanglement in their ground state, and support topologically charged excitations which cannot be created locally. However, unlike topologically ordered phases which have a finite ground state degeneracy on non-trivial manifolds, fracton phases have a ground state degeneracy on the 3-torus that grows sub-extensively with volume. Since their ground state degeneracy explicitly depends on the geometry of the manifold, fracton phases of matter are strictly speaking \textit{not} topologically ordered. Indeed, it has been recently demonstrated that certain type-I fracton models may acquire a robust ground state degeneracy even on topologically trivial manifolds, albeit in the presence of spatial curvature~\cite{slagle2}.

Given that fracton phases have much in common with topologically ordered phases, it is natural to ask whether fracton models can be understood in terms of conventional topological phases and their degress of freedom. Indeed, Refs.~\cite{han,sagar} answered this question in the affirmative, showing explicitly that the paradigmatic X-Cube model, which displays type-I fracton physics, can be constructed by suitably coupling layers of $d=2$ $\mathbb{Z}_2$ topological order. In particular, the coupling is understood as the condensation of one-dimensional extended objects built out of the excitations of the $d=2$ topologically ordered layers. Based on this, the authors of Refs.~\cite{han,sagar} proposed novel Abelian $d=3$ fracton models built from coupled layers of $d=2$ topological orders.

In this work, we generalize the layer construction of Refs.~\cite{han,sagar} and construct novel $d=3$ type-I fracton phases by coupling together layers of $d=2$ non-Abelian topological orders. More specifically, we consider layers of Levin-Wen string-net models~\cite{levinwen}, which describe a large class of bosonic topological orders in two spatial dimensions, and condense extended one-dimensional strings which are composed out of anyons in the $d=2$ layers. We note that Ref.~\cite{han} already used a coupled-layer construction based on the doubled semion string-net model to construct a semionic version of the X-Cube model. Our work goes beyond this construction by considering string-net models whose excitations are non-Abelian anyons. Based on general principles of anyon condensation in tensor categories~\cite{bais,eliens,kong,neupert,burnell}, we then establish the existence of deconfined dim-1 excitations with non-Abelian braiding statistics in the condensed $d=3$ fracton phases.

Some other works have discussed fracton models with non-Abelian excitations. Ref.~\cite{nonabelian} introduced a model that intertwines the Majorana checkerboard fracton model~\cite{fracton1} with layers of $p + ip$ superconductors.  In this model, the Majorana checkerboard fractons become non-Abelian excitations. Ref.~\cite{nonabelian} also introduced a class of models based on coupled layers of 2d quantum double models and claimed that these models support immobile non-Abelian fracton excitations. As we discuss in more detail in Sec.~\ref{concls}, in our opinion, Ref.~\cite{nonabelian} did not take the necessary steps to establish the existence of non-Abelian fractons in this class of models, and therefore we believe this claim should be viewed as a proposal, yet to be established. In contrast, in our work, we discuss what it means for excitations to be non-Abelian in gapped fracton phases, and establish the presence of non-Abelian sub-dimensional excitations in our models.  Finally we note that, very recently, some of us, together with Martin-Delgado, have introduced fracton models based on twisted gauge theories that support immobile non-Abelian fractons~\cite{twisted}.

Crucially, the non-Abelian dim-1 excitations in our models are a fundamentally three-dimensional phenomenon, a result we establish for the simplest of our models, which is based on layers of doubled Ising string-net models.  We show that these excitations of this model cannot be understood as bound states of dim-2 excitations, or as bound states of non-Abelian dim-2 excitations with Abelian sub-dimensional particles. The presence of such excitations, which we dub as being \textit{intrinsically} sub-dimensional and \textit{inextricably} non-Abelian, demonstrates that this model displays a novel non-Abelian fracton order.

Along these lines, recent work has introduced the notion of a \emph{foliated fracton phase} \cite{shirley}.  Two fracton phases $A$ and $B$ are equivalent as foliated fracton phases if $A$ stacked with decoupled layers of 2d topologically ordered states is adiabatically connected to $B$, stacked with a possibly different set of 2d topologically ordered layers.  The intrinsically sub-dimensional and inextricably non-Abelian nature of the dim-1 excitations implies that our fracton model supporting these excitations is not equivalent -- in the sense above -- to \emph{any}  Abelian foliated fracton phase. Our results thus establish the existence of non-Abelian foliated fracton phases.

Investigating these models allows us to establish the structure of the ground state wave function in the fracton phase, which we propose can be understood as a condensate of fluctuating ``cage-nets." This extends the notion of ``string-net" wave functions to the case of fracton phases. Potentially, this may enable progress in understanding the mathematical structure underlying gapped fracton phases, and constructions of non-Abelian fracton phases that are beyond a coupled-layer approach.

The rest of the paper is organized as follows: in Sec.~\ref{strnet}, we review the basic ideas underlying string-net models. Similarly, in Sec.~\ref{review} we review the current understanding of fracton models, focusing in particular on the X-Cube model introduced by Vijay, Haah, and Fu~\cite{fracton2}. Using this model as an example, we will introduce the layer-construction approach for studying certain fracton phases and will also elucidate the nature of the ground state wave function. In Sec.~\ref{ising}, we will then introduce a new fracton model, which we construct from coupled layers of $d=2$ doubled Ising string-net models. Borrowing ideas from anyon condensation in $d=2$ topological orders, we then establish the excitation spectrum of the fracton phase, obtained from condensing extended one-dimensional string-like excitations, and explicitly demonstrate the existence of non-Abelian dim-1 particles. Sec.~\ref{sec:intrinsic} contains one of the central results of this paper, where we establish the intrinsic and inextricable nature of the non-Abelian dim-1 excitations, thereby highlighting that these excitations are a fundamentally three dimensional phenomenon. Our construction is then generalized to the case of SU(2)$_k$ string-net models, after which we introduce the concept of the ``cage-net" wave function as the ground state wave function for the class of type-I fracton phases studied here. We end by discussing implications of our work for the field of fractons and by exploring open questions and future directions. 


\section{A Review of String-Net Models}
\label{strnet}

In this section, we review Levin-Wen string-net models~\cite{levinwen}. We start by reviewing the models and their essential features generally, while concurrently establishing the notations used throughout this paper. We then discuss excitations in the restricted class of string-net models whose input is a unitary modular tensor category; specifically, we describe the general procedure for constructing string operators for flux excitations in these models. Since we will rely heavily on this framework for our construction of cage-net fracton models, readers familiar with string-nets may wish to skim this section, while those interested in further details are referred to the original paper by Levin and Wen~\cite{levinwen}.

Levin and Wen's construction is a general procedure for identifying fixed-point ground state wave functions, and corresponding Hamiltonians, for a large class of topologically ordered phases in $2+1$ dimensions. The basic idea is to define the ground state wave function implicitly via certain local constraints, which are designed to enforce topological invariance. Within this approach, the ground states are understood as infrared fixed points under renormalization-group (RG) flows. Such an approach has the advantage of capturing the universal physical properties of topologically ordered phases, without being mired in ultraviolet complexities. These wave functions are ground states of local Hamiltonians which are given by the sum of mutually commuting terms, and are thus exactly soluble. The Levin-Wen models constructed thusly are equivalent to Hamiltonian constructions of the Turaev-Viro model~\cite{turaev,turaev2010,wang2010} and certain doubled Chern-Simons topological quantum field theories~\footnote{Generalisations of the original string-net construction have been discussed in the literature and are generally believed to describe all ``doubled'' phases~\cite{kitaevkong,kong2014,linlevin2014,lan2014,lin2017}. However, these models cannot describe topological phases whose edges are necessarily gapless, as occurs \emph{e.g.} when the thermal Hall conductance is non-zero.}.

\begin{figure}[t]
	\centering
	\includegraphics[width=4cm]{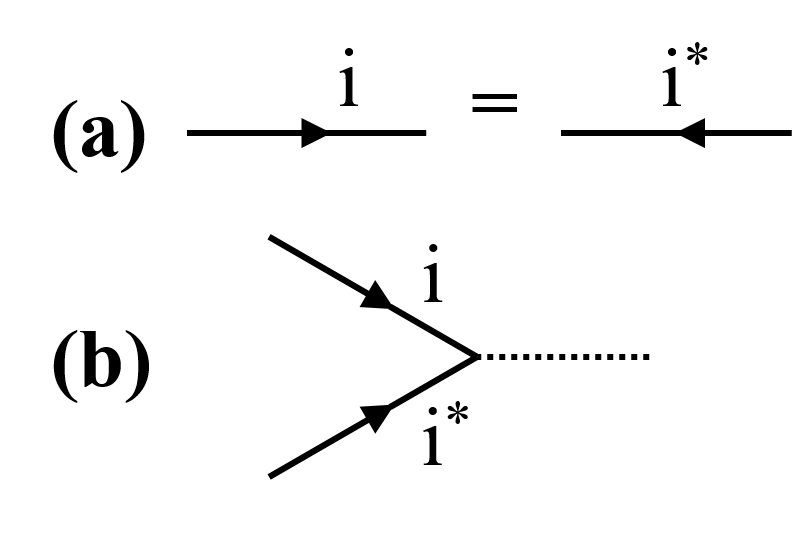}
	\caption{(a) The dual string $i^*$ has an orientation opposite to that of $i$. (b) The branching rules associated with null strings are defined such that $\delta_{ij0} = 1$ iff $i=j^*$ and vanishes otherwise.}
	\label{fig:01}
\end{figure}

More specifically, a string-net is a fixed trivalent graph embedded in two-dimensional space, where each edge carries an orientation and is labelled by a string-type $j$. There are a finite number of string types $j=0,1,\dots,N$, where each label may be thought of as a particle species propagating along the edge. Further, each string type $j$ has an associated unique ``conjugate" or ``dual" string $j^*$, such that reversing the orientation of an edge corresponds to the mapping $j\mapsto j^*$ (see Fig.~\ref{fig:01}(a)). This mapping $j\mapsto j^*$ satisfies $(j)^{**} = j$ and we require that the null string type $0$ is self dual, $0^* = 0$, since it is equivalent to having no string at all. In the language of category theory, the string labels are the simple objects of a unitary fusion category $\mC$, an algebraic structure that generalizes the properties of irreducible group representations under the tensor product. For example, the strings may be labelled by group elements of a finite group or the irreducible representations of a finite group or quantum group, with the null string $0$ labelling the identity element of the group in the former case, and the trivial representation in the latter.

A string-net is required to satisfy branching rules encoded in a three-index object $\delta_{ijk}$, associated with each triple of strings $\{i,j,k\}$ oriented inward toward a vertex.  We follow Ref.~\cite{levinwen} and assume $\delta_{i j k}$ is invariant under arbitrary permutations of the indices; more generally, this assumption can be relaxed~\cite{kitaevkong,kong2014,lan2014}.  If $\delta_{ijk} = 1$, the triple is allowed to meet at a node of the graph and $\delta_{ijk} = 0$ if such a configuration is forbidden.  In string-nets based on the irreducible representations of a group, $\delta_{ijk} = 1$ if and only if the tensor product $i\otimes j \otimes k$ contains the trivial representation. Triples containing the null string $(i,j,0)$ are allowed only when $i = j^*$, \emph{i.e.} $\delta_{ij0} = 0$ if $i\neq j^*$ and $\delta_{jj^* 0} = 1$, as depicted in Fig.~\ref{fig:01}(b). Each string label is associated with a real number $d_j \geq 1$ called the quantum dimension, and the branching rules satisfy
\begin{align}
\sum_{k} \frac{d_k}{d_i d_j} \delta_{ijk^*} &= 1, \nonumber \\
\sum_{i,j} \frac{d_i d_j}{d_k} \delta_{ijk^*} & = \mD^2, 
\end{align}
where $\mD = \sum_k \sqrt{d_k^2}$ is the total quantum dimension.

String-net models are quantum models defined on a trivalent lattice. The Hilbert space of each link is $N+1$-dimensional, with orthonormal basis states $| j \rangle$ labelled by the string types $j = 0, \dots, N$. The full Hilbert space is simply a tensor product of the link Hilbert spaces, with string configurations labeling orthonormal basis states. String-net models energetically favor string configurations that satisfy the branching rules at each vertex; such configurations are string-nets, and are denoted $X$. The corresponding states $|X \rangle$ span a low-energy subspace of the full Hilbert space. Ground states of sting-net models are of the form $\ket{\Phi} = \sum_{X} \Phi(X)^* \ket{X}$, where $\Phi(X)\equiv \braket{\Phi | X }$ is the probability amplitude of the state $\Phi$ of being in the string-net configuration $X$. We also refer to $\Phi(X)$ as a wave function. We note that two string configurations that are identical up to reversing string orientations and relabelling $j\to j^*$ are considered equivalent up to a phase factor~\cite{levinwen}.

In order to construct fixed point string-net wave functions, which describe ground states of certain exactly soluble Hamiltonians, Levin and Wen imposed local constraints -- in addition to the branching rules -- designed to enforce topological invariance of the wave function. Given a state $\ket{\Phi} = \sum_X \Phi(X)^* \ket{X}$, the local constraints on $\Phi(X)$ are graphically depicted as
\begin{align}
\Phi \bpm \includegraphics[height=0.3in]{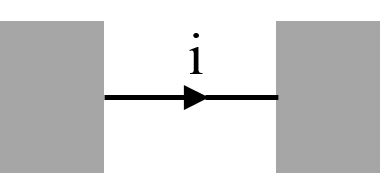} \epm  =&
\Phi \bpm \includegraphics[height=0.3in]{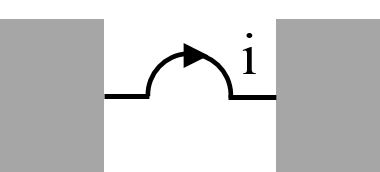} \epm,
\label{topinv} \\
\Phi \bpm \includegraphics[height=0.3in]{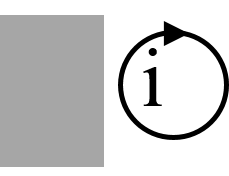} \epm  =&
\vartheta_i\Phi \bpm \includegraphics[height=0.3in]{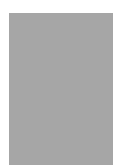} \epm,
\label{scaleinv} \\
 \Phi \bpm \includegraphics[height=0.3in]{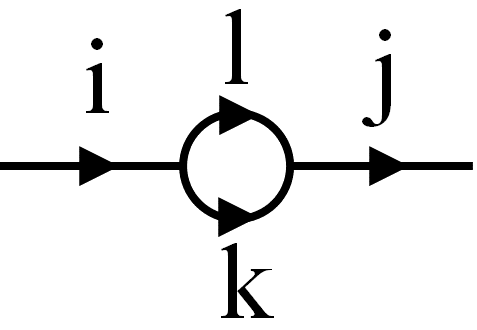} \epm  =&
\delta_{ij} \Phi \bpm \includegraphics[height=0.3in]{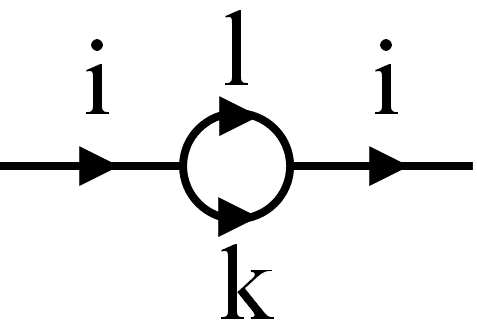} \epm,
\label{bubble} \\
 \Phi \bpm \includegraphics[height=0.3in]{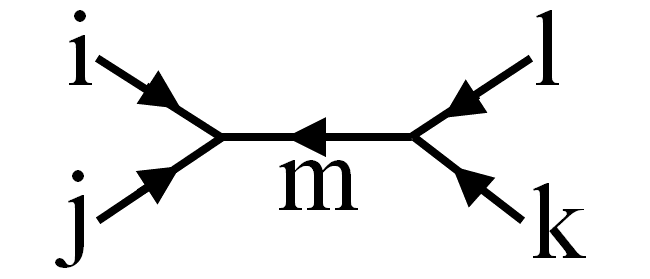} \epm  =&
\sum_{n} F^{ijm}_{kln} \Phi \bpm \includegraphics[height=0.3in]{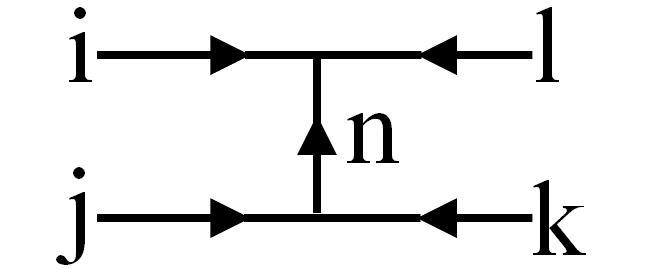} \epm.
\label{fusion} 
\end{align}

These local constraints establish equivalence classes between string-net configurations, since two configurations which can be transformed into each other through the above local relations are defined to be equivalent. Intuitively, Eq.~\eqref{topinv} comes from demanding that the wave function be topologically invariant \emph{i.e.} the amplitude should be the same for configurations which can be continuously deformed into one another. Eq.~\eqref{scaleinv} implies that a disconnected loop contributes only a scaling constant to the amplitude. Here $\vartheta_i = \kappa_i d_i$, where $\kappa_i$ is the Frobenius-Schur indicator when $i = i^*$ and can be gauge transformed to $\kappa_i = 1$ otherwise~\cite{lan2014}. For the rest of this paper, we will only consider cases where $\vartheta_i = d_i$. Eq.~\eqref{bubble} reflects scale invariance---since a closed string disappears at length scales large compared to the string size, the bubble becomes irrelevant at long length scales. Eq.~\eqref{fusion} is required to uniquely specify the ground state wave function and can be motivated by crossing symmetry in conformal field theories. Here, the recoupling tensor $F^{ijm}_{kln}\in \mathbb{C}$ is the quantum $6j$-symbol. The quantum dimension $d_j$ and the tensor $F$ generalize the ordinary vector space dimensions and $6j$-symbols associated with irreducible group representations. By definition, $F_{kln}^{ijm} = 0$ if any of the branchings $(i,j,m),(l,k,m^*),(i,l,n),(j,k,n^*)$ are forbidden by the branching rules. 

The fundamental idea underlying the local constraints Eqs.~\eqref{topinv}-\eqref{fusion} is that the amplitude for any string-net configuration can be related to the amplitude of the vacuum configuration by multiple applications of these local rules. Adopting the convention $\Phi\text{(vacuum)} = 1$, the ground state wave function is then uniquely determined (on a manifold with trivial topology) by the set of rules~\eqref{topinv}-\eqref{fusion}. Equivalently, the universal properties of $\Phi$ are captured by the fusion data $(d_i, F_{kln}^{ijk})$ satisfying the consistency conditions~\cite{levinwen}
\begin{align}
\label{conds}
F^{ijk}_{j^* i^* 0} &= \frac{v_k}{v_i v_j} \delta_{ijk} ,\nonumber \\
F^{ijm}_{kln} = F^{klm^*}_{jin} &= F^{jim}_{lkn^*} = F^{imj}_{k^*nl} \frac{v_m v_n}{v_j v_l} ,\nonumber \\
\sum_{r,s=0}^N F_{mlr}^{kp^*q} F_{mr^*s}^{jip} &= \sum_{n,r,s=0}^N F_{qkn}^{jip} F_{mls}^{n^*iq} F_{slr}^{kjn},
\end{align}
where $v_i = v_i^* = \sqrt{d_i}$ (with $v_0 = 1$) and
\beq
\delta_{ijk} =
\begin{cases}
1,& (i,j,k) \text{ allowed,} \\
0,& (i,j,k) \text{ forbidden.}
\end{cases}
\eeq

The first condition in Eq.~\eqref{conds} is a normalization condition, the second is the tetrahedral symmetry, and the third is the pentagon identity. There is a one-to-one correspondence between solutions of Eq.~\eqref{conds} and $2+1D$ string-net condensed phases. For instance, it is known that if the string labels $j$ run over all irreducible group representations of a finite group $G$, $d_j$ are the corresponding dimensions of the group representations, and $F_{kln}^{ijm}$ are the $6j$ symbols for the group, then the Levin-Wen state can be mapped to Kitaev's quantum double model describing a deconfined gauge theory with gauge group $G$~\cite{Aguado2009}. 

\begin{figure}[t]
	\includegraphics[width=0.5\textwidth]{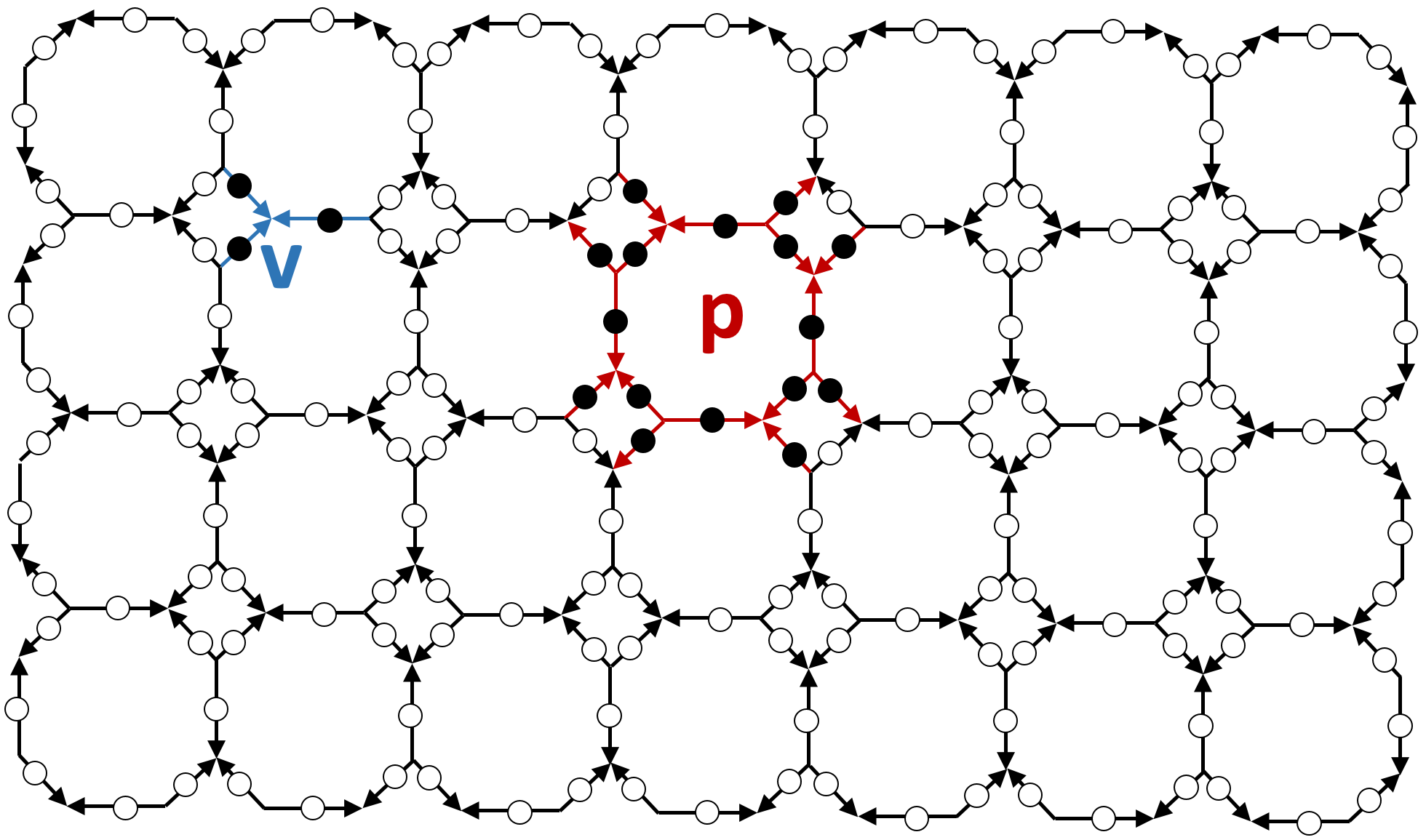}
	\caption{Two-dimensional truncated square lattice on which we define the string-net models. The vertex projector $A_v$ acts on the three spins adjacent to the vertex $v$ and enforces the branching rules. The plaquette term $B_p$ acts on the 16 spins adjacent to the plaquette $p$ and provides dynamics to the string-net configurations.}
	\label{fig:02}	
	\end{figure}

Thus, we have seen that the input data required to specify a string-net model constitutes the set of string-types $j=0,1,\dots,N$, fusion rules, and $F$-tensors satisfying the consistency conditions~\eqref{conds}. Mathematically, this input data corresponds to a unitary fusion category $\mC$, where the distinct string-types correspond to distinct simple objects in $\mC$. For each ground state wave function $\Phi$ associated with the fusion category $\mC$, or equivalently to the fusion data $(d_i, F_{kln}^{ijk})$ satisfying the constraints~\eqref{conds}, there also exists an exactly solvable Hamiltonian for which $\Phi$ is the ground state. While string-net models are usually defined on a honeycomb lattice, following the construction of a semionic X-Cube model based on doubled semion string-net layers \cite{han}, we will instead define the string-net model on the truncated square lattice shown in Fig.~\ref{fig:02}. The reasons for this choice of lattice, which is also sometimes referred to as the square-octagon lattice, will become apparent in later sections.

The Levin-Wen Hamiltonian is defined as
\beq
\label{lwh}
H = -\sum_v A_v - \sum_{p} B_p,
\eeq
where the first sum runs over vertices $v$, and the second over plaquettes $p$, including both the truncated-square and diamond plaquettes. The vertex term $A_v$ acts on the three strings adjacent to a vertex $v$ and is a projector enforcing the branching rules,
\beq
A_v \left| \bmm \includegraphics[height=0.3in]{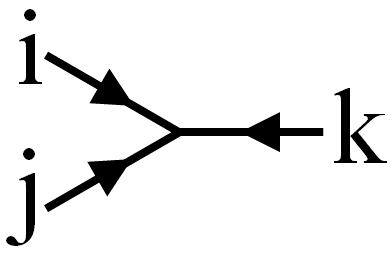} \emm \right\rangle = 
\delta_{ijk} \left| \bmm \includegraphics[height=0.3in]{vertex} \emm \right\rangle.
\label{eqn:av}
\eeq
This term measures the ``electric charge" at each vertex $v$ and favors states with no charge. Thus, the low energy Hilbert space in the presence of this constraint constitutes the set of all allowed string-net configurations.

The plaquette projector $B_p$ represents the kinetic part of the Hamiltonian, which provides dynamics to the string-net configurations and makes them condense. This is understood as a magnetic flux projector, which measures the magnetic flux through each plaquette and prefers states with no flux. We focus on truncated square plaquettes; the description of $B_p$ for the diamond plaquettes is identical, upon making the obvious modifications. The operator $B_p$ is defined as
\beq
\label{plaq}
B_p = \sum_{s=1}^N \frac{d_s}{\mD^2} B_p^s,
\eeq
where $B_p^s$ acts on the strings forming the plaquette $p$ as well as on the outer legs of $p$. Graphically, $B_p^s$ has a simple interpretation as an operator which adds an isolated loop of string-type $s$ inside the plaquette $p$,
\begin{align}
B_p^s \Bigg|
\bmm\includegraphics[height=0.8in]{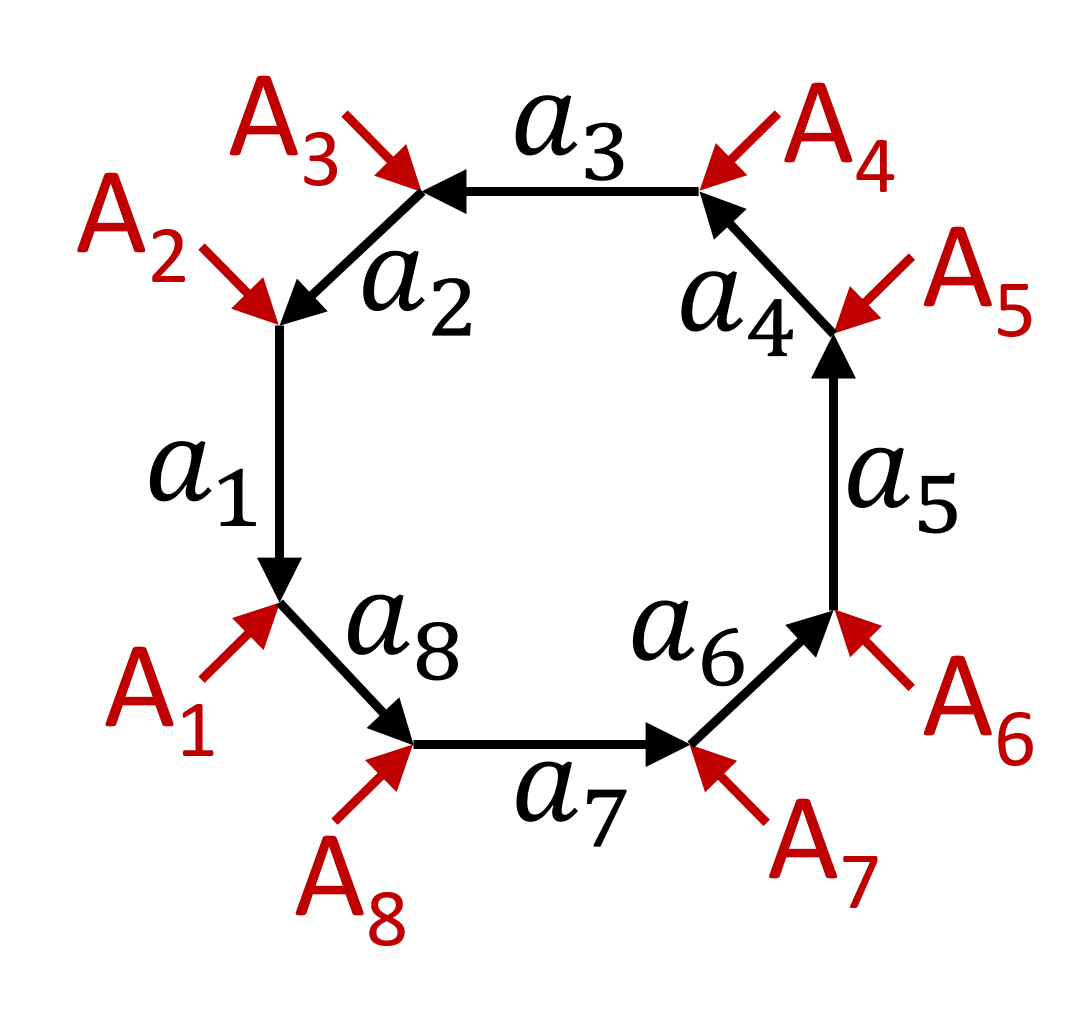}\emm \Bigg\> &=  \Bigg|
\bmm\includegraphics[height=0.8in]{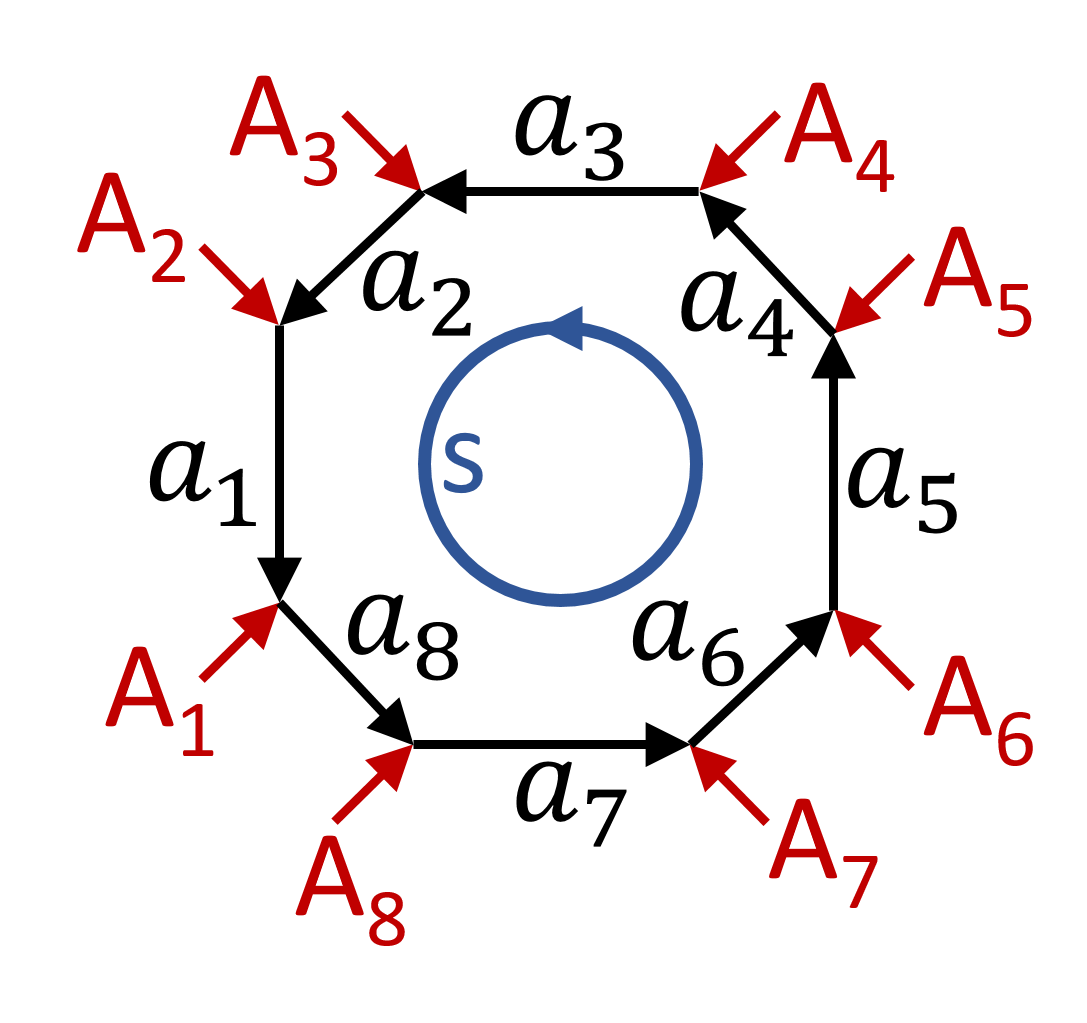}\emm \Bigg\>.
\end{align}
In order to make sense of and work with states like that on the right-hand side, one introduces a ``fattened'' lattice where the degrees of freedom are continuum string-nets (see Ref.~\cite{levinwen} for details). Labelling the links so that the 8 internal legs (within the truncated square) are labelled by $a_i$ and the 8 external legs are labelled by $A_i$, where $i = 1,\dots,8$, we can follow the analysis of Ref.~\cite{levinwen} in order to show that 
\begin{align}
B_p^s \Bigg| \bmm\includegraphics[height=0.8in]{Bc1}\emm \Bigg\> = &
\sum_{\{a'_i\}_{i=1}^8} \left[\prod_{i=1}^8 F_{s^* a'_{i+1} a_i^{\prime *}}^{A_{i+1} a_i^* a_{i+1}} \right] \nonumber \\
& \times \Bigg| \bmm\includegraphics[height=0.8in]{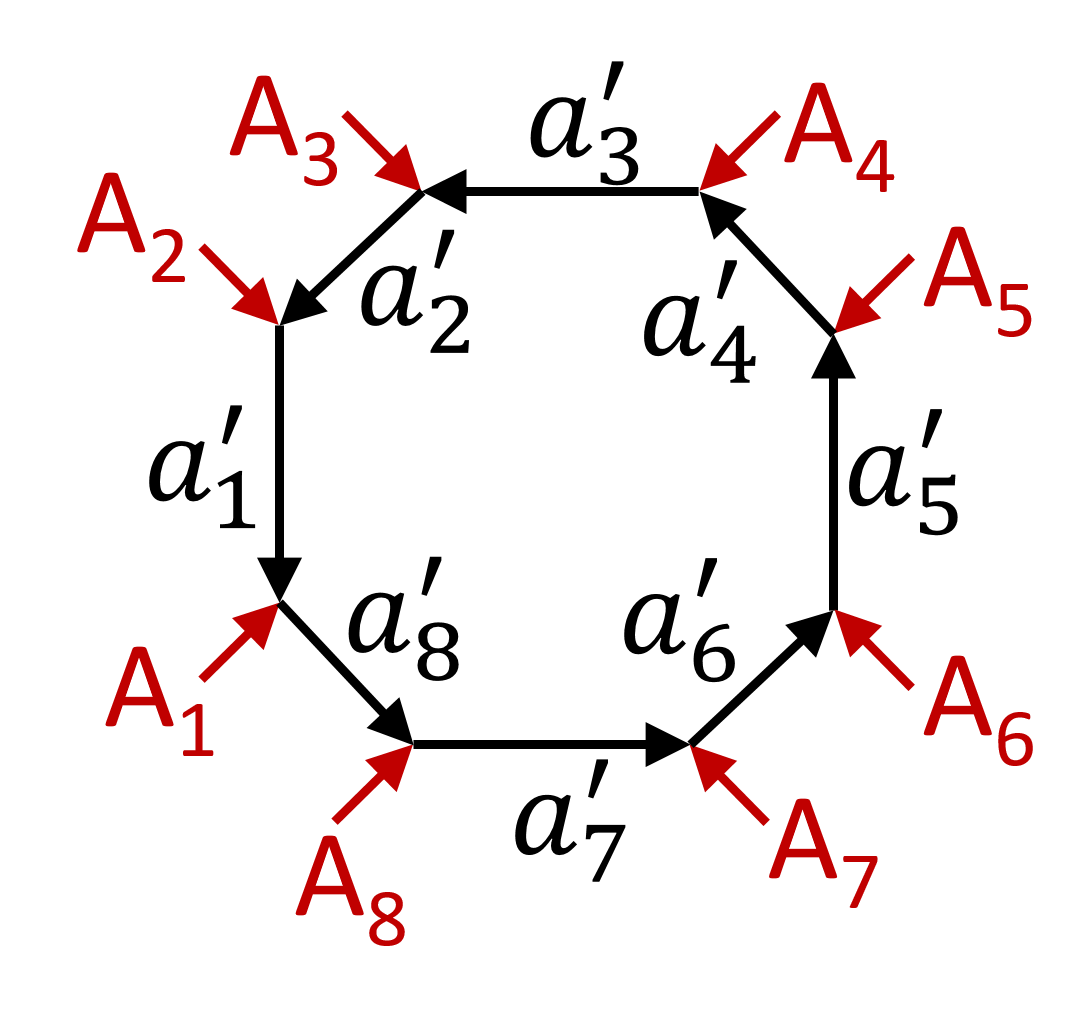}\emm \Bigg\>, 
\end{align}
where addition is defined mod 8 \emph{i.e.}, $A_9 = A_1$ and $a_9 = a_1$. Finally, we require that the Hamiltonian~\eqref{lwh} be Hermitian, which corresponds to the following constraint on the $F$ tensor:
\beq
F_{k^* l^* n^*}^{i^* j^* m^*} = \left(F_{klm}^{ijm}\right)^*.
\eeq

The string-net Hamiltonian is exactly solvable, as it can be explicitly shown that it is composed of mutually commuting terms,
\beq
[A_v,A_{v'}] = 0,\,[B_p,B_{p'}] = 0,\,[A_v,B_p] = 0,\quad \forall v,v',p,p'.
\eeq
Additionally, since the terms $A_v$ and $B_p$ are projection operators (they have eigenvalues $0,1$), the ground state is defined by the following conditions:
\beq
A_v \ket{\Phi} = \ket{\Phi},\,B_p \ket{\Phi} = \ket{\Phi},\quad \forall v,p.
\eeq
The ground state is unique on a topologically trivial manifold but on topologically non-trivial manifolds, there exists a ground state degeneracy~\cite{levinwen,Hu2012}. Any state with eigenvalue $0$ for at least one vertex $v$ or plaquette $p$ is an excited state, hence also establishing the presence of a finite energy gap to all excitations.

\subsection{Restriction to Unitary \textit{Modular} Tensor Categories}
\label{unimod}

Elementary excitations in $2+1D$ topologically ordered phases of matter are anyons, such as Laughlin quasiparticles and quasiholes~\cite{Laughlin}, and the $e,m$ excitations in the Kitaev toric code model~\cite{kitaev2}. In general, anyonic excitations are algebraically described by unitary modular tensor categories (UMTCs), which are unitary fusion categories with additional structure describing braiding of excitations.  This additional structure is characterized in part by a unitary $S$-matrix~\cite{Rowell,bonderson}. In a UMTC $\mathcal{A}$, the only excitation which braids trivially with itself and all other anyonic excitations corresponds to $0 \in \mathcal{A}$.

As discussed earlier, string net models take as their input a unitary fusion category $\mC$, which consists of a set of string-types, fusion rules, and an $F$-tensor satisfying the consistency conditions~\eqref{conds}. The output of string-net models are the anyon excitations, which are objects of the UMTC $\mZ(\mC)$, the Drinfeld center of $\mC$~\cite{kitaevkong,lan2014}. For instance, the UMTC describing excitations in the toric code is the Drinfeld center of $\mbZ_2$. More generally, in Kitaev's quantum double models, anyons are described by the irreducible group representations of the quantum double of a finite group $G$, $\mZ(G)$.

Here, we forgo a general discussion of excitations in string-net models and instead focus only on the sub-class of these models from which we will construct fracton models. Specifically, we take as the input of the string-net construction a unitary \textit{modular} tensor category $\mC$ \emph{i.e.} our starting point is a unitary fusion category that admits braiding, and is thus equipped with a unitary $S$-matrix. The resulting anyons for this class of string-nets are objects in the Drinfeld center of $\mC$, $\mZ(\mC) = \mC \times \bar{\mC}$~\cite{drinfeld2010}. In this case the anyons in $\mathcal{Z}(\mC)$ can be labelled by an ordered pair $(a,b)$ where $a\in \mC$ and $b\in \bar{\mC}$. It is often convenient to use the notation $a\bar{b}$ instead of $(a,b)$ and we will henceforth use these interchangeably. The $F$ and $R$ tensors for the output category $\mZ(\mC)$ describing excitations in this class of models are provided in Appendix~\ref{app:FandRdetails}.

In less abstract terms, the class of models we consider includes lattice versions of discretized versions of doubled Chern-Simons theories, each of which is a chiral Chern-Simons theory together with its mirror image. Well-known examples from this class of string-net models are based on the semion, Fibonacci, Ising, and SU(2)$_n$ UMTCs. The input category for the toric code model is $not$ modular, and in general, neither are the input categories for string-net models which are equivalent to discrete non-Abelian gauge theories.

We  now describe the excitations of the string-net Hamiltonian~\eqref{lwh} defined on the truncated square lattice (see Fig.~\ref{fig:02}). Excited states of this model have at least one vertex projector $A_v$ or plaquette projector $B_p$ with eigenvalue 0. The low-lying excitations of this model appear either as pairs of vertex defects---electric charges---where $A_v$ has eigenvalue 0 for two vertices $v,v'$ or as pairs of plaquette defects---magnetic fluxes---where $B_p$ has eigenvalue 0 for two plaquettes $p,p'$. The operators creating pairs of defects in these models are Wilson string-like operators, which act on all edges along some path $P$ connecting the two defects. For the purposes of our coupled-layer construction, we will be primarily interested in flux excitations, \emph{i.e.} those where only $B_p$ projectors are violated.  Furthermore, we will  focus on Abelian fluxes. Here we will describe the construction of string operators for Abelian fluxes, which violate precisely two plaquettes at their end-points, leaving all other terms in the Hamiltonian untouched. 

In order to do this, we first introduce the $S$-matrix. Given an input UMTC $\mC$ with $n+1$ simple objects which correspond to the string-labels in the string-net model, its $S$-matrix is defined graphically as
\beq
S_{ab} = \frac{1}{\mD} \bmm \includegraphics[height=0.3in]{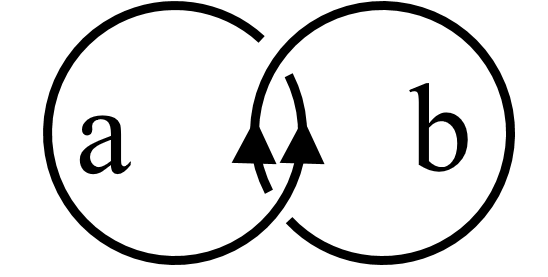} \emm,
\eeq
or in terms of the topological spin
\beq
\theta_a = \frac{1}{d_a} \bmm \includegraphics[height=0.3in]{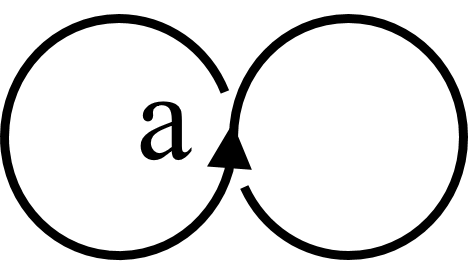} \emm,  \label{eqn:topspin}
\eeq
as
\beq
S_{ab} = \frac{1}{\theta_a \theta_b} \sum_c \frac{d_c}{\mD} \theta_c \delta_{abc^*}. 
\eeq
We note that these anyon world-line diagrams are distinct from, though related to, diagrams for string-nets. The unitary $S$-matrix has the properties
\beq
S_{ab} = S_{ba} = S_{a^* b}^*,
\eeq
\beq
S_{0a} = \frac{d_a}{\mD},\quad d_a = \frac{S_{0a}}{S_{00}}.
\eeq
We can also remove closed loops from strings as follows:
\beq
\bmm \includegraphics[height=0.3in]{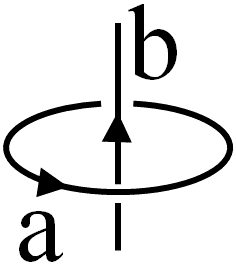} \emm = \frac{S_{ab}}{S_{0b}}\bmm \includegraphics[height=0.3in]{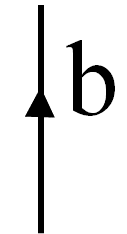} \emm
\eeq

In string-net models that take as their input a UMTC $\mC$, with string types $j=0,1,\dots,n$, for each string-type $a$ there exists a corresponding flux excitation $\phi_a$, which is expected to correspond to the anyon $(a,a)$~\cite{Hu2015,burnell2,burnell1}. The set of such fluxes $\{\phi_a\}$ can be identified by first observing that the operators $B_p^s$ (see Eq.~\eqref{plaq}) form a commutative algebra~\cite{levinwen}
\beq
B_p^i B_p^j = \sum_{k=0}^n \delta_{ijk} B_p^k.
\eeq
From the generators of this algebra, we can now define the flux projectors~\cite{Hu2015}
\beq
\mP_p^{i} = \sum_{j=0}^n S_{i0}S_{ij}B_p^j.
\eeq
These projectors satisfy the properties
\beq
\mP_p^{i} \mP_p^{j} = \delta_{ij} \mP_p^{j}, \quad \sum_{i=0}^n \mP_p^{i} = 1,
\eeq
which can be established using elementary properties of the $S$-matrix and the Verlinde formula~\cite{verlinde}
\beq
\delta_{i j k^*} = \sum_l \frac{S_{i l} S_{j l} S^*_{k l}}{S_{0 l}} \text{.}
\eeq
Moreover, $\mP_p^{0} = B_p$ is a projector onto states with trivial flux at plaquette $p$, with $\phi_0$ the trivial flux excitation. In general, a non-trivial flux $\phi_a$ is present at plaquette $p$ in some state $\ket{\psi}$ iff $\mP_p^{a} \ket{\psi} = \ket{\psi}$ for $a\neq0$.  Each flux $\phi_a$ has a unique anti-particle $\bar{\phi}_{a}$, with excitations always created from the ground state in particle-antiparticle pairs. 

\begin{figure}[t]
	\centering
	\includegraphics[width=0.5\textwidth]{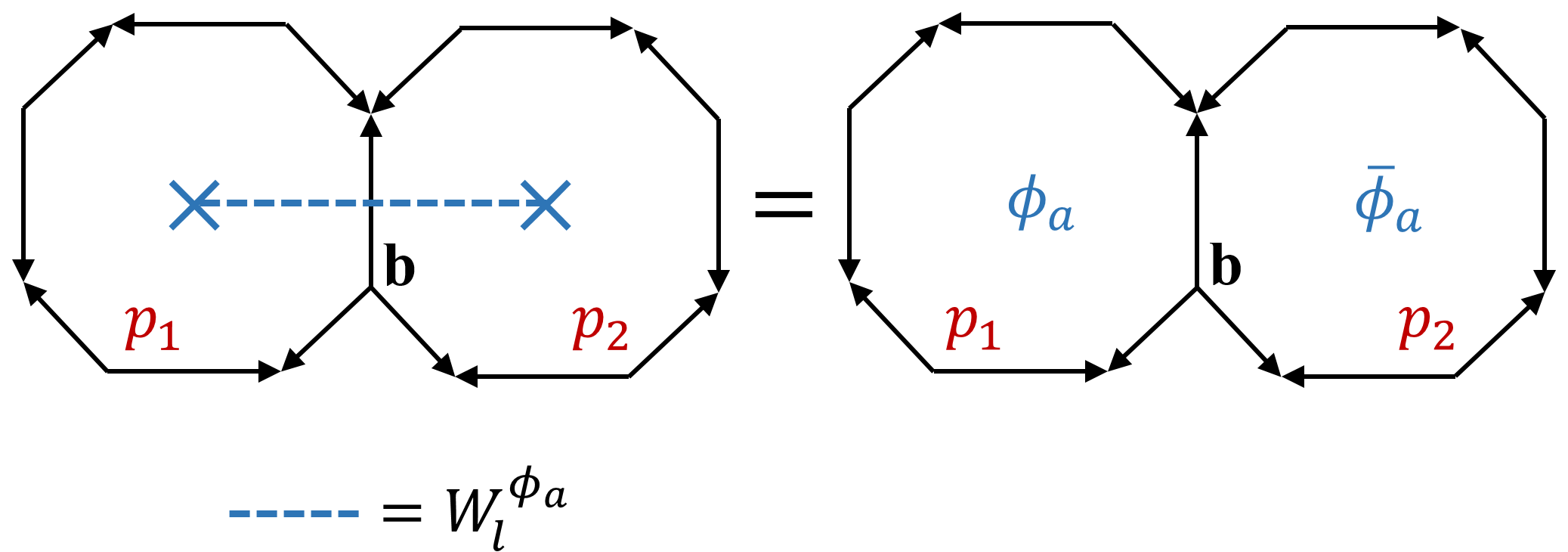}
	\caption{Action of the fundamental string $W_l^{\phi_a}$ on the link $l$ separating plaquettes $p_1$ and $p_2$, where $l$ carries the string-label $b$. Assuming that we start from a state in which there are no fluxes at $p_1$ and $p_2$, the operator $W_l^{\phi_a}$ creates a pair of fluxes $\phi_a$ and $\bar{\phi}_a$ on these plaquettes.}
	\label{fig:03}
\end{figure} 
\begin{figure*}
  \includegraphics[width=\textwidth]{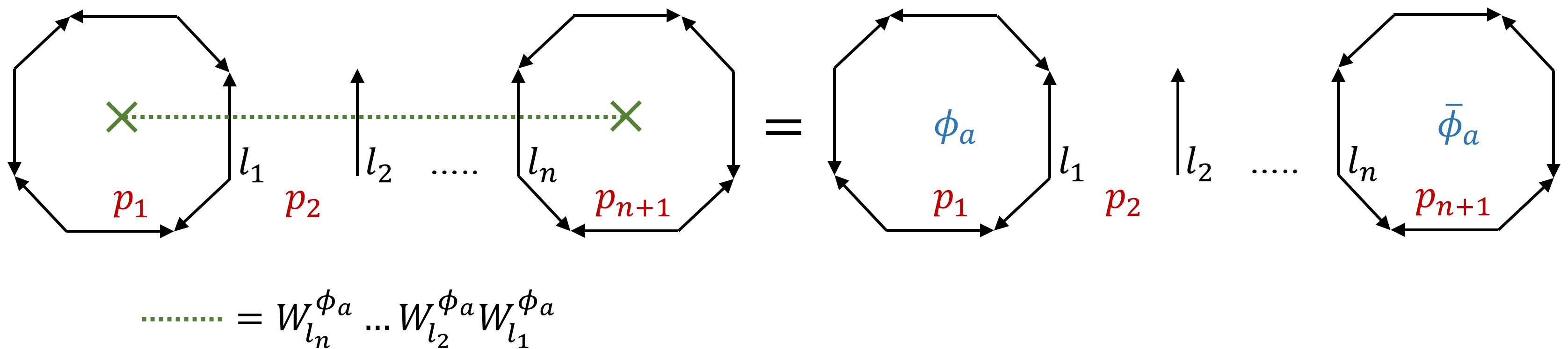}
  \caption{Path-independent string operator which creates a pair of fluxes, with a flux isolated at each end of the string.}
  \label{fig:04}
\end{figure*}

Now we assume that $a$ (and hence $\phi_a \simeq (a,a)$) is Abelian, and construct a Wilson string operator for $\phi_a$. A segment of Wilson string operator acting on a link $l$ is defined as
\beq
\label{fun-string}
W_l^{\phi_a} \left|\,\, \bmm \includegraphics[height=0.3in]{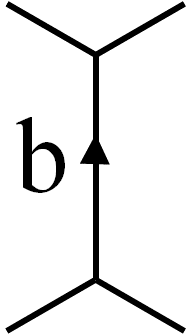} \emm \,\,\right\rangle
= \frac{S_{a0}S_{ab}}{S_{00}S_{b0}} \left| \,\,\bmm \includegraphics[height=0.3in]{string1} \emm \,\,\right\rangle
= e^{i \Theta_{a b}} \left| \,\,\bmm \includegraphics[height=0.3in]{string1} \emm \,\,\right\rangle \text{,}
\eeq
where $b$ is the string-label on link $l$, and where $\Theta_{a b}$ is the Abelian statistics angle for a full braid of $a$ around $b$. Equivalently,
\beq
\label{eqn:flux_string_loop}
W_l^{\phi_a} \left|\,\, \bmm \includegraphics[height=0.3in]{string1} \emm \,\,\right\rangle
=  \left| \,\,\bmm \includegraphics[height=0.3in]{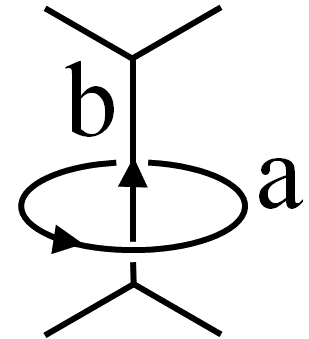} \emm \,\,\right\rangle.
\eeq
Since the operator $W_l^{\phi_a}$ only affects the two neighbouring plaquettes $p_1$ and $p_2$ sharing the link $l$, and since this definition holds for any plaquette shape, we omit the rest of the lattice in the graphical representation of the string operator~\eqref{fun-string}. 

To understand the effect of acting with $W_l^{\phi_a}$ on the ground state, the following relations are useful
\begin{eqnarray}
W_l^{\phi_a} B^s_{p_1} &=& e^{i \Theta_{a s}} B^s_{p_1} W^{\phi_a}_l \nonumber \\
W_l^{\phi_a} B^s_{p_2} &=& e^{-i \Theta_{a s}} B^s_{p_2} W^{\phi_a}_l \label{eqn:Wl_Bsp_relns} \\
W_l^{\phi_a} B^s_{p'} &=&  B^s_{p'} W^{\phi_a}_l \text{,} \nonumber
\end{eqnarray}
where $p_1$ and $p_2$ are as shown in Fig.~\ref{fig:03}, and $p'$ is any plaquette $p' \neq p_1, p_2$.  These equations are straightforward to establish using the graphical representation of $B^s_p$. Using these relations and elementary properties of the $S$-matrix, it follows that~\cite{Hu2015}
\begin{align}
\mP_{p_1}^{b} W_l^{\phi_a} \ket{\Phi} &= \delta_{b, a} W_l^{\phi_a} \ket{\Phi}, \nonumber \\
\mP_{p_2}^{b} W_l^{\phi_a} \ket{\Phi} &= \delta_{b, a^*} W_l^{\phi_a} \ket{\Phi}, \nonumber \\
\mP_{p'}^{b} W_l^{\phi_a} \ket{\Phi} &= \delta_{b, 0} W_l^{\phi_a} \ket{\Phi}, \label{eqn:Wl_P_relns}
\end{align}
where $\Phi$ is a ground state of the string-net Hamiltonian. It follows that $W_l^{\phi_a}$ creates a pair of fluxes, $\phi_a$ and $\bar{\phi}_a$, respectively, on the plaquettes $p_1$ and $p_2$. As an aside, we note that the relations Eq.~\eqref{eqn:Wl_P_relns} hold in general, \emph{i.e.} without relying on our assumption that $a$ is Abelian~\cite{Hu2015}. However, Eq.~\eqref{eqn:Wl_Bsp_relns} only holds, and indeed only makes sense, when $a$ is Abelian.

Now, we consider a product of $W^{\phi_a}_l$ along some path, as shown in Fig.~\ref{fig:04}. It follows immediately from Eq.~\eqref{eqn:Wl_Bsp_relns} that this product does not create any excitations away from its endpoints. Therefore, this product is a Wilson string operator that creates fluxes $\phi_a$ and $\bar{\phi}_a$ at the endpoints. Given the identification $\phi_a \simeq (a,a)$, we expect that the topological spin $\theta_{\phi_a} = 1$; that is, $\phi_a$ is an Abelian boson.  Indeed, this is easily verified by using Eq.~\eqref{eqn:topspin} to compute $\theta_{\phi_a}$ from the string operator.


\section{A Review of Fracton Phases}
\label{review}

In this section, we review recent progress in the field of fracton phases of matter. We focus primarily on the example of the X-Cube model~\cite{fracton2}, which is closely related to the cage-net fracton models that we introduce. Readers familiar with the recent field of fractons are encouraged to peruse this section, as we introduce the concept of a ``cage-net" wave function here. For a recent review of fracton physics taking a broader perspective, we refer the reader to~\cite{fractonreview}.

Fracton phases of matter represent a new class of quantum phases of matter which extend and challenge existing notions regarding topological order in three spatial dimensions. Originally discovered in exactly solvable $3d$ lattice models~\cite{chamon,bravyi,haah,haah2,yoshida,fracton1,fracton2}, these gapped systems are distinguished by the presence of point-like fractionalized excitations---fractons---which cannot move without creating additional topological excitations and are hence fundamentally immobile. In contrast with anyons in two-dimensional topologically ordered systems, where anyons are created at the ends of a Wilson string operator and can thus move by repeated applications of a local line-like operator, there exists no local line-like operator that creates a pair of fractons. Instead, fractons are created at the corners of membrane or fractal operators, endowing isolated fractons with their characteristic immobility. In addition to fractons, these systems often host additional excitations which may only move along sub-dimensional manifolds and are hence referred to as ``sub-dimensional" excitations.

An important distinction between conventional $3d$ topologically ordered phases and fracton phases is that the ground state degeneracy of the former on a non-trivial spatial manifold is a finite constant determined only by the topology of the manifold, while the same is not true for the latter. Indeed, all known $3d$ gapped fracton phases exhibit a ground state degeneracy on the 3-torus that grows sub-extensively with system size.  As for conventional topologically ordered phases, the degenerate ground states are locally indistinguishable.  However, the sub-extensive ground state degeneracy demonstrates that gapped fracton phases lie beyond a description in terms of topological quantum field theory (TQFT). This is remarkable, as it overturns the conventional wisdom that all gapped quantum phases of matter admit a TQFT description. 

Fracton phases have been broadly divided into type-I and type-II varieties~\cite{fracton2}. In type-I phases, fractons are separated by the application of a membrane-like operator and there exist additional topological excitations with sub-dimensional mobility. In type-II phases, fractons are created at the ends of a fractal operator and all topological excitations are strictly immobile. Within this taxonomy, all models considered in this paper belong to type-I fracton phases. 

\subsection{An Example: the X-Cube model}
\label{type1}

\begin{figure}[t]
\includegraphics[width=8cm]{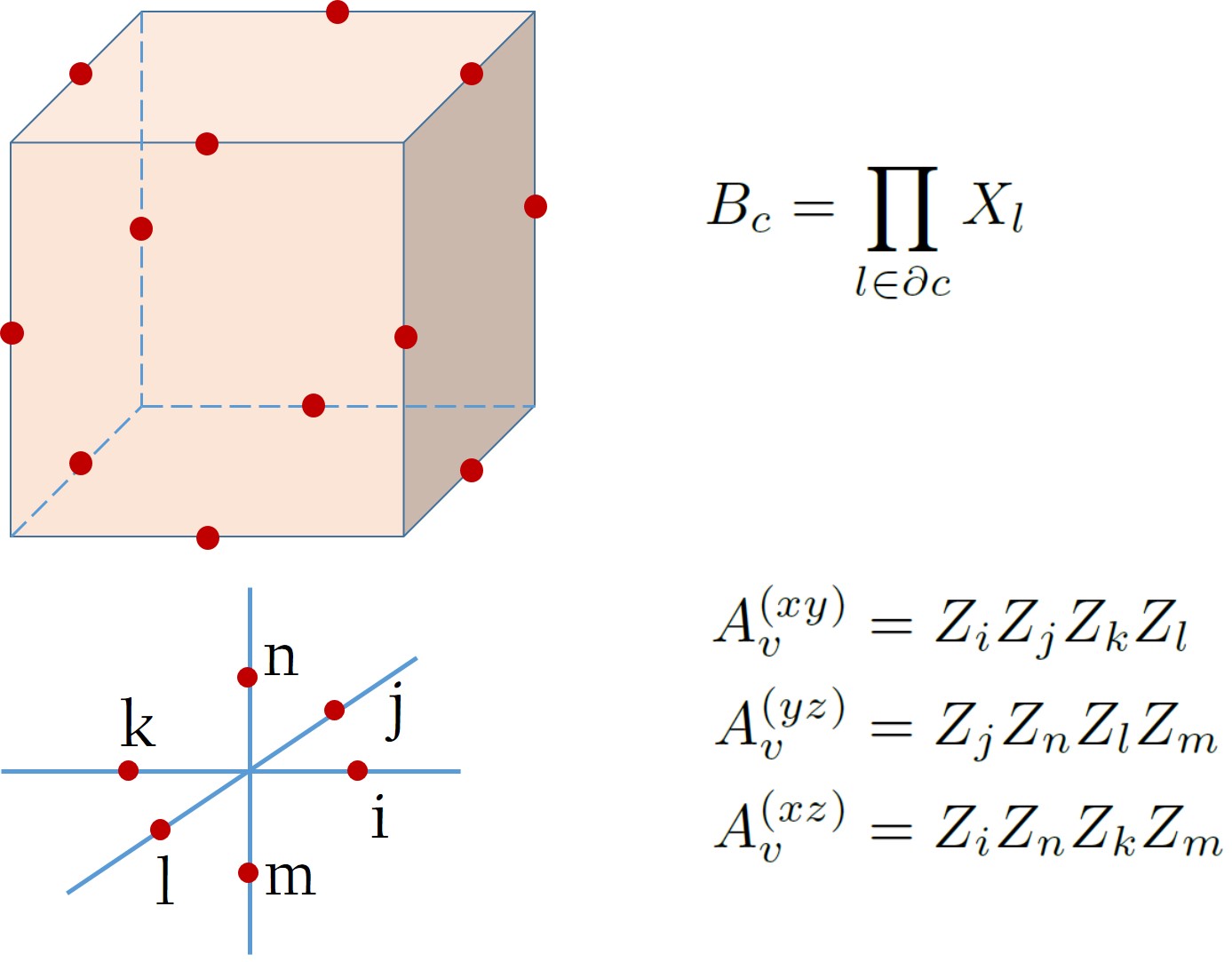}
\centering
\caption{The X-Cube model is represented by spins $\sigma$ placed on the links of a cubic lattice and is given by the sum of a twelve-spin Pauli-$x$ operator at each cube $c$ and planar four-spin Pauli-$z$ operators at each vertex $v$.}
\label{XCube}
\end{figure}

We will henceforth concentrate only on type-I gapped fracton models in $d=3$ spatial dimensions, focusing in particular on the paradigmatic $X$-Cube model~\cite{fracton2}. The X-Cube model is an exactly solvable lattice spin model defined on the simple cubic lattice. A single Ising spin (\emph{i.e.} qubit) lives on each link. The Hamiltonian is 
\begin{align}
H_{XC} = - \sum_{v,k} A_v^{(k)} - \sum_c B_c ,
\label{HXC}
\end{align}
where the terms are described in Fig.~\ref{XCube}, with $X_l$ ($Z_l$) the Pauli-$x$ (Pauli-$z$) operator acting on the spin on link $l$. In the first term, the sum is over all vertices $v$ and over the three orientations $k = xy, yz, xz$, while the second term involves a sum over all cubes $c$.

\begin{figure}[t]
\includegraphics[width=0.3\textwidth]{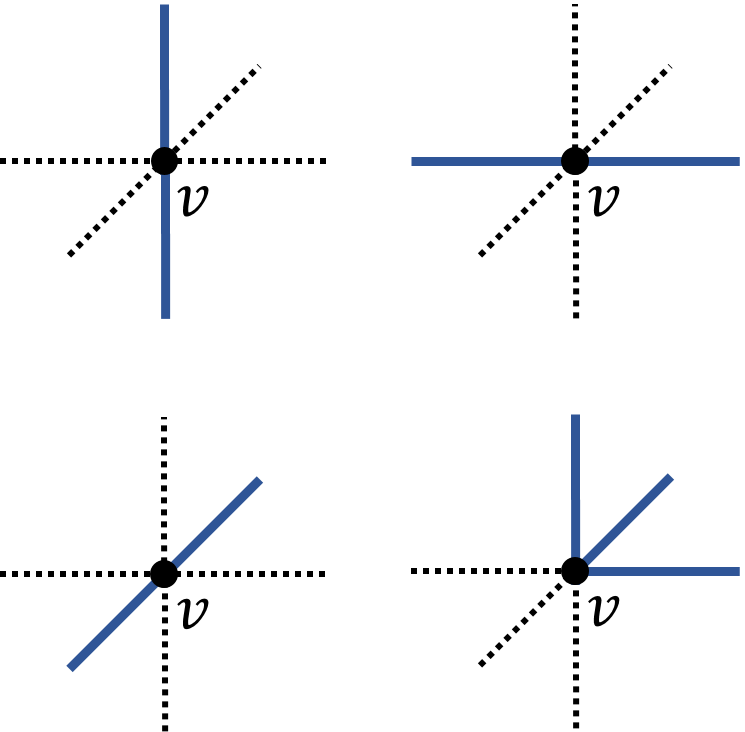}
\centering
\caption{Generating set for all allowed string configurations in the X-Cube model, with all other configurations obtained from combinations of these. Dashed black lines indicate the absence of a string on that link while thick blue lines indicate the presence of one.}
\label{vertexallowed}
\end{figure}

The Hamiltonian~\eqref{HXC} is exactly solvable since it is the sum of mutually commuting operators, \emph{i.e.} $[B_c,B_{c'}] = [A_v^{(k)},A_{v'}^{(k')}] = [A_v^{(k)}, B_c] = 0$, and since each of these operators is a product of Pauli operators, they each have eigenvalues $\pm 1$. A ground state $\Phi$ satisfies the stabilizer constraints
\beq
B_c \ket{\Phi} = \ket{\Phi}, \quad A_v^{(k)}\ket{\Phi} = \ket{\Phi},\quad \forall c,v,k.
\eeq
Let us work in the $Z$ eigenbasis, where we will represent a link with spin $\downarrow$ as a string, and a link with spin $\uparrow$ as no string. The constraint $A_v^{(k)} = +1$ implies than an even number of the four links adjacent to $v$, and lying in the plane $k$, are occupied with a string.  Note that each link $l$ adjacent to $v$ appears in two different $A^{(k)}_v$ vertex operators. A generating set of all allowed string configurations at a vertex $v$ is shown in Fig.~\ref{vertexallowed}; all the allowed configurations can be constructed by superposing these and noting that the strings have a $\mbZ_2$ character, so that having two strings on the same link is equivalent to having none.

\begin{figure}[b]
\centering
\begin{subfigure}[b]{0.26\textwidth}
\includegraphics[width=\textwidth]{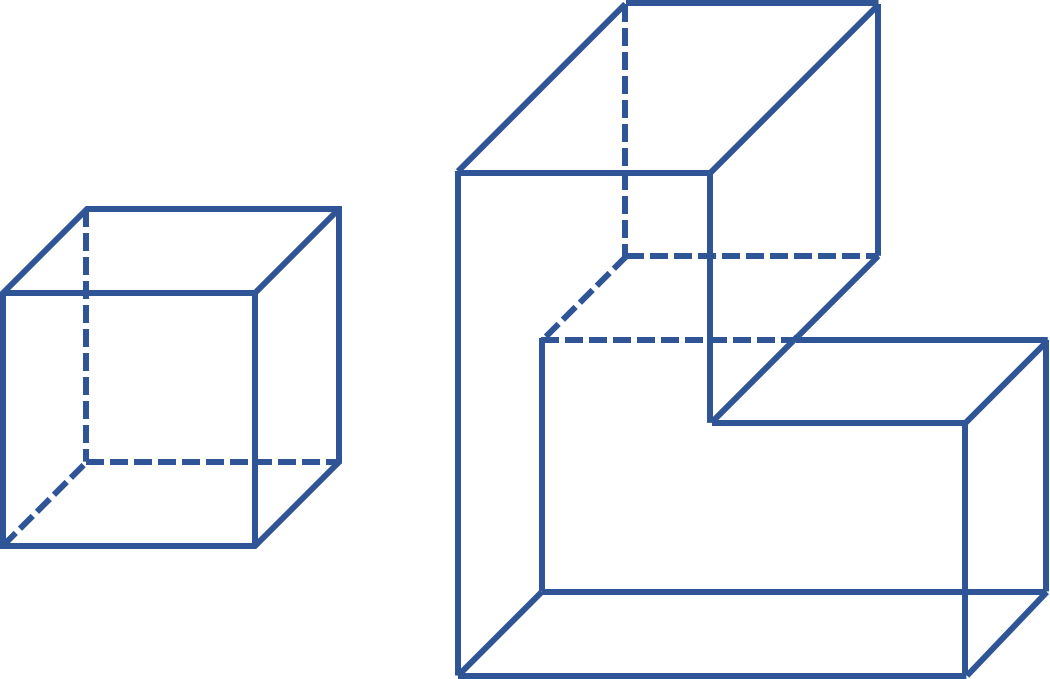}
\caption{}
\label{skeleton1}
\end{subfigure}\quad\quad
\begin{subfigure}[b]{0.18\textwidth}
\includegraphics[width=\textwidth]{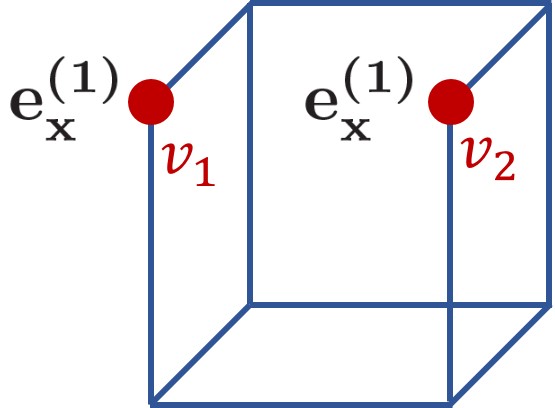}
\caption{}
\label{skeleton2}
\end{subfigure}
\caption{(a) A representative example of a ``cage-net," which is a state which minimizes the vertex terms $A_v^{(k)}$. The ground state wave function of the X-Cube model can be thought of as a cage-net condensate. (b) An incomplete cage, which is missing a string parallel to the x-axis. Such a configuration violates the terms $A_{v_i}^{(xy)}$ and $A_{v_i}^{(xz)}$ for $i=1,2$ and thus possesses a pair of charge excitations at the vertices $v_1$ and $v_2$. The underlying lattice has been omitted for clarity.}
\end{figure}

From Fig.~\ref{vertexallowed}, we see that a simple allowed configuration is one where a string passes ``straight through'' $v$, \emph{i.e.} where there are exactly two collinear strings adjacent to $v$.  These are the only allowed configurations with two strings adjacent to $v$. In particular, a configuration where a string ``turns a corner'' at $v$, \emph{i.e.} where there are two perpendicular strings adjacent to $v$, is not allowed. Instead, a configuration with three mutually orthogonal strings adjacent to $v$ is allowed. From these observations, it is clear that the states that minimize the vertex term are not simply closed loop configurations. Instead, as can be inferred from Fig.~\ref{vertexallowed}, the states that minimize the vertex term are ``cages," or configurations of strings forming the edges of a rectangular prism.  More precisely, ignoring global issues that depend on boundary condtions, the string configurations minimizing the vertex term are superpositions of such cages. Such a configuration is exemplified in Fig.~\ref{skeleton1}. Configurations where a string along the boundary is missing are associated with ``electric charge'' excitations, which violate the vertex term and, as we will shortly see, are restricted to move only along lines. This is depicted in Fig.~\ref{skeleton2}.

The cube operator $B_c$ flips all the spins at the edges of a cube, and graphically can be understood as creating or destroying the cage bounding the cube $c$. Hence, acting with the cube operators mixes states with different allowed cage configurations, with the ground state of the X-Cube model given by the equal weight superposition of all such allowed cage configurations. Mirroring our discussion of $d=2$ topologically ordered phases, where the ground state is understood as a ``string-net" condensate, we see that the ground state of a phase with fracton order is described as a ``cage-net" condensate. This condensate wave function takes the explicit form
\beq
\ket{\Psi} = \prod_c \frac{1 + B_c}{\sqrt{2}}\ket{\uparrow \uparrow\dots \uparrow}, \label{eqn:cage-net-gs}
\eeq
where $\ket{\uparrow \uparrow\dots \uparrow}$ denotes the vacuum state with all spins pointing up. With periodic boundary conditions, there are degenerate ground states, which can be obtained from Eq.~\eqref{eqn:cage-net-gs} by flipping spins along straight lines which wind around the system. On a 3-torus with linear dimension $L$, the ground state degeneracy ${\rm GSD}$ satisfies
$\log_2 {\rm GSD} = 6L - 3$~\cite{fracton2}.

\begin{figure}[t]
\centering
\begin{subfigure}[b]{0.23\textwidth}
\includegraphics[width=\textwidth]{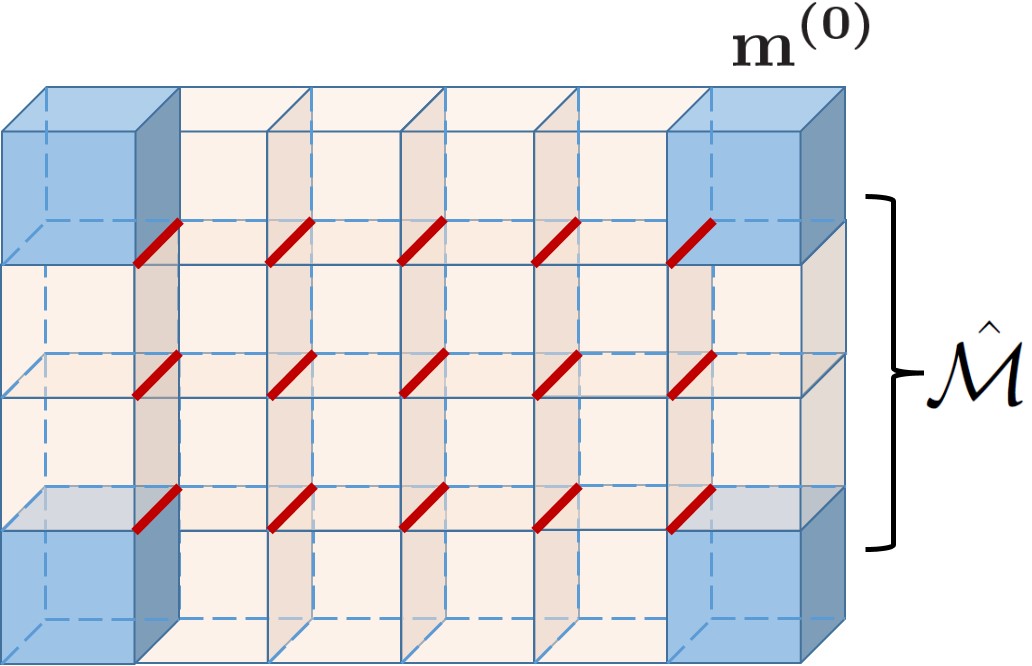}
\caption{}
\label{membrane-1}
\end{subfigure}\quad\quad
\begin{subfigure}[b]{0.21\textwidth}
\includegraphics[width=\textwidth]{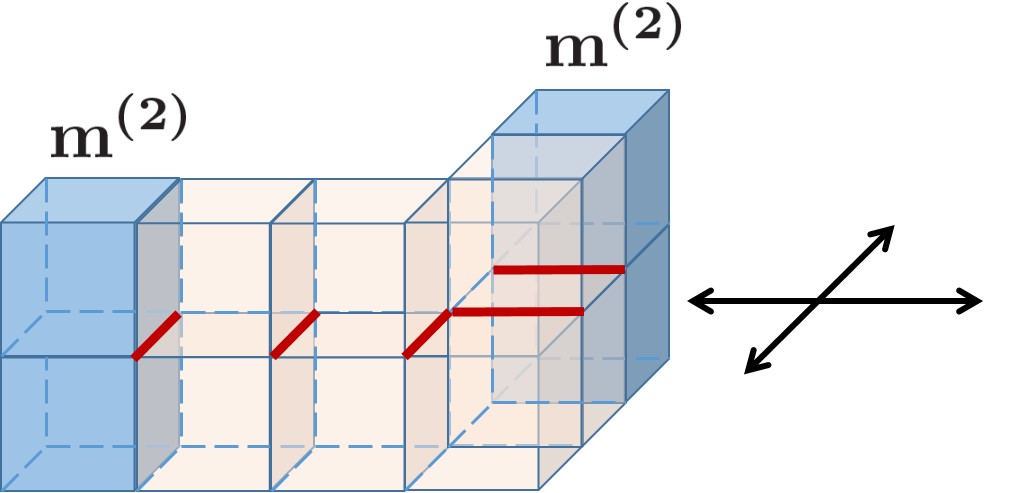}
\caption{}
\label{membrane-2}
\end{subfigure}
\caption{Topological excitations of the X-Cube model are depicted in (a) and (b). 
Fractons $m^{(0)}$ are created at corners by acting on the ground state 
by a membrane operator $\mathcal{M}$ that is the product of $Z$ operators along red links. 
Composite topological excitations $m^{(2)}$ are created by a Wilson line operator and are thus mobile along two-dimensional sub-manifolds.}
\end{figure}

Fractons live at cubes $c$ where the $B_c$ eigenvalue is $-1$, as opposed to $+1$ in a ground state. However, there exists no local operator which can create a single pair of fractons. 
Indeed, applying a $Z$ operator to a link 
flips the $B_c$ eigenvalues of the four cubes sharing that link. Acting on the ground state with an operator $\hat{\mathcal{M}}$ formed by taking a product of $Z$s over a rectangular membrane creates four fractons at the corners of the membrane, as shown in Fig.~\ref{membrane-1}. A single fracton, denoted $m^{(0)}$ 
(where the superscript denotes that it is a dimension-0 excitation), 
is thus fundamentally immobile, 
as moving it would create additional topological excitations. However, pairs of fractons separated along a principal axis of the lattice are free to move, since the repeated application of local membrane operators acts as a hopping operator for such pairs. In particular, a product of $Z$ operators over a path contained in a $\{1 0 0 \}$ plane creates a pair of fractons at each end, as illustrated in Fig.~\ref{membrane-2}. Each pair is a composite excitation 
that can move in this two-dimensional plane, and which we refer to as a dimension-2 (dim-2) excitation $m^{(2)}$.

The electric charges in the X-Cube model correspond to violations of the vertex term $A_v^{(k)}$ and are denoted $e_\mu^{(1)}$, where $\mu=x,y,z$ labels the principal axes along which the charge can move. A charge $e_\mu^{(1)}$ at vertex $v$ corresponds to a state which violates $A_v^{(\mu \nu)}$ and $A_v^{(\mu \lambda)}$, where $\mu$, $\nu$ and $\lambda$ are all distinct. Such charge excitations are created at the ends of a Wilson line of $X$ operators which are confined to one-dimension. Thus, the charge excitations in the X-Cube model $e_\mu^{(1)}$ are dimension-1 particles which may only move along the $\mu$ direction without creating additional topological excitations. While a single charge excitation cannot turn a corner without creating other excitations, three charges with mutually perpendicular orientations can annihilate at a vertex.

With the ground state and excitation spectrum of the X-Cube model established, we now discuss the relation of this model to the $d=2$ toric code model~\cite{kitaev2}. Refs.~\cite{han,sagar} provided an explicit construction for the X-Cube model starting from layers of $d=2$ $\mbZ_2$ toric codes. Based on this construction, the excitations of the X-Cube model were understood in terms of those of decoupled toric code layers, via a particle-string, or $p$-string, condensation mechanism. Given that our aim in this paper is to construct new fracton phases from coupled layers of string-net models, we briefly review the ideas of Refs.~\cite{han,sagar} here.

Let us consider three independent stacks of $d=2$ toric code models, each defined on the square lattice, along the three principal axes of the simple cubic lattice. As a result of the stacking, two spins reside on each link of the cubic lattice formed by the stacked, inter-penetrating layers. For instance, a link parallel to the $z$-axis carries a spin which participates an $xz$-plane toric code and another spin which participates in a $yz$-plane toric code. Now, for a plane $P$, the toric code is defined as
\beq
H_P^{TC} = -\sum_{v\in P} A_v^{o(P)} - \sum_{p\in P}B_p,
\eeq
where $o(P)$ is the orientation of the plane $P$, given by the direction normal to $P$. Here, $B_p$ is the usual plaquette operator which is given by
\beq
B_p = \prod_{l\in p} X_l^{o(p)},
\eeq
where $X_l^{o(p)}$ is the Pauli-$x$ operator acting on the spin living on link $l$ in the plane with orientation $o(p)$, specified by the normal to $p$. Similarly, the vertex term $A_v$ is defined as
\beq
A_v^\mu = \prod_{l\in v} Z_l^\mu,
\eeq
where $Z_l$ is the Pauli-$z$ operator and where the product goes over the four links adjacent to the vertex $v$ and lying in the plane whose normal is in the direction $\mu$.

These layers of $\mbZ_2$ topological orders are then coupled together by a $ZZ$ coupling on each link $l$ as follows:
\beq
H = \sum_P H_P^{TC} - J \sum_{l} Z_l^{\mu_1} Z_l^{\mu_2}
\eeq
where $\mu_1$ and $\mu_2$ are lattice directions orthogonal to the direction of $l$. In the limit where $J\to\infty$, and treating $\sum_P H_P^{TC}$ as a perturbation, one recovers the X-Cube model~\eqref{HXC} at sixth order in degenerate perturbation theory. Rather than delve into details of this calculation, for which the reader is referred to~\cite{han}, it is useful to understand the physical effect of the coupling on the decoupled layers.

\begin{figure}[t]
\includegraphics[width=5cm]{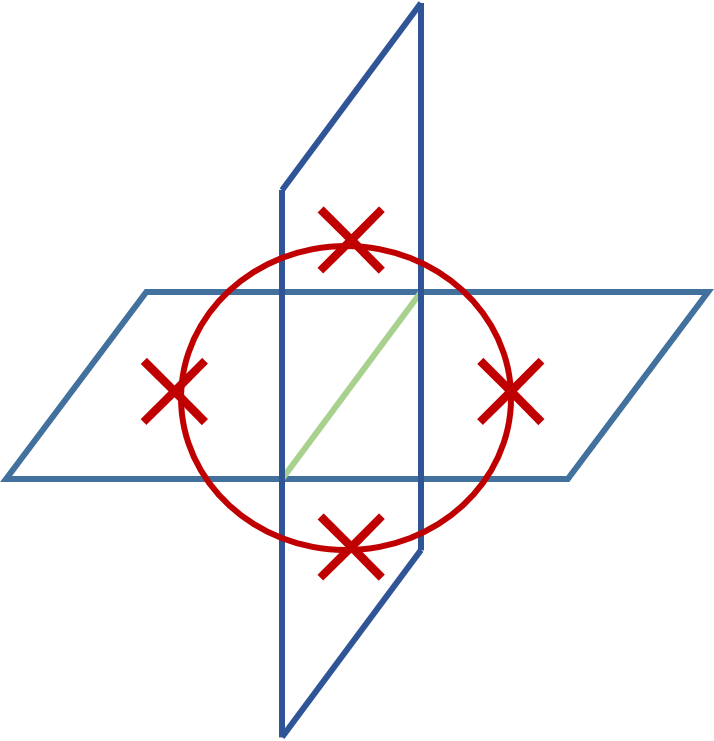}
\centering
\caption{The elementary $p$-string from which larger $p$-strings can be built. The X-Cube fracton phase can be understood as a $p$-string condensate of toric code $m$ excitations. The green line represents the action of the coupling term on the link $l$, which excites four $m$ particles (red crosses). These can be connected by a $p$-string, shown here by the closed red string.}
\label{pstring}
\end{figure}

Since the coupling term does not commute with the plaquette operator $B_p$, acting with the coupling term on a decoupled toric code ground state excites $m$ particles. Specifically, when acting on a link $l$ shared by two intersecting planes, the coupling term excites four $m$ particles, as shown schematically in Fig.~\ref{pstring}. Representing each $m$ particle by a line normal to the plane in which the $m$ particle moves, the lines from the four $m$ particles can be joined together to form a closed particle- or $p$-string. Figure~\ref{pstring} depicts the smallest $p$-string which results from the action of $Z_l^{\mu_1} Z_l^{\mu_2}$ on a single link. The $p$-string can grow and deform by acting with the coupling on an extended set of links. In this language, the X-Cube phase is thus understood as a condensate of extended one-dimensional string-like objects, the $p$-strings. Since we are condensing fluxes, the elementary charge excitations, which braid non-trivially with the $m$'s, will be confined in the $p$-string condensed phase. However, one can show that bound states of $e$'s coming from two perpendicular layers of toric codes will remain deconfined even in the condensed fracton phase. These bound states correspond precisely to the dimension-1 charge excitations of the X-Cube model. The fracton excitations can also be described within the $p$-string condensation picture as the ends of open $p$-strings. For decoupled toric codes, an open $p$-string corresponds to a ``stack'' of $m$ particles. In the condensed phase, the ``interior'' of the stack disappears into the condensate, leaving behind two fracton defects at the ends (see Ref.~\cite{han} for details). Thus, much insight can be gained into the excitations of the fracton phase based on the properties of the $d=2$ topological order from whence it came. In particular, the $p$-string picture allows us to use tools from the theory of anyon condensation in $d=2$ topologically ordered phases in order to infer the properties of excitations in the fracton phase.

This ends our review of fracton phases. We will now construct novel non-Abelian fracton phases of matter from layers of inter-penetrating string-net models, and through general principles of anyon condensation, will establish the existence of deconfined excitations with reduced mobility and non-Abelian fusion.

\section{Cage-net Fracton models}
\label{ising}

Based on string-net models and general principles of anyon condensation~\cite{bais,eliens,kong,neupert,burnell}, we now discuss a construction of non-Abelian fracton models. Following the discussion in Sec.~\ref{type1}, we consider layers of inter-penetrating string-net models, stacked along the $x$, $y$, and $z$-directions. From this, a non-Abelian fracton model can be obtained by condensing the $p$-strings made out of the flux excitation $\phi_{a}$, referred to as a $\phi_{a}$-string in the following sections.

We will focus on cases where the flux excitation $\phi_{a} \simeq (a,a)$ is an Abelian boson.  The corresponding Wilson string operator was described in Sec.~\ref{unimod}, and should be distinguished from the $\phi_a$-string, which is a one-dimensional object built from point-like $\phi_a$ excitations. Since the braiding for any other anyon in each 2d layer with the $\phi_{a}$-string is the same as its braiding with the $(a,a)$ anyon, the 2d particles in each layer that remain deconfined upon condensing $\phi_a$-strings are precisely those whose mutual statistics with $(a,a)$ is trivial. This is the same as in ordinary anyon condensation in 2d; however, unlike in that case, here point particles \emph{cannot} disappear into the condensate, because the condensate is one of extended $\phi_a$-strings and not of point anyons. In particular, the $(a,a)$ anyon itself remains as a gapped, deconfined excitation in each 2d layer.

While some of the 2d particles in each layer get confined, condensing $\phi_a$ strings leads to deconfined dim-1 particles, just as in the coupled layer construction of the X-Cube model. These excitations arise as bound states of two confined anyons in perpendicular layers, such that the composite has trivial statistics with the $\phi_a$ string. Moreover, we will see that that deconfined fractons are descendants of the $\phi_{a}$ flux.

In this section, we begin with a simple example based on the doubled Ising string-net model, where all the essential properties of excitations can be seen clearly. We  first give a short review of the model itself and discuss the flux condensation procedure. We then construct a doubled Ising non-Abelian fracton model via flux-string condensation. A straightforward generalization of this construction to SU(2)$_k$ is then presented, followed by a discussion of the ground state wave functions of these fracton models, which are naturally interpreted as condensates of fluctuating cage-nets.

\subsection{Doubled Ising string-net model}
\label{doubleising}

Strings in the doubled Ising string-net model are labelled by the anyons of the Ising anyon theory, $\{0, \sigma, \psi\}$. For the purposes of constructing fracton models, it will be useful to define the string-net model on a truncated square lattice, as depicted in Fig.~\ref{fig:02}. The branching rules  are given by
\begin{equation}
\delta_{000}=\delta_{0 \psi \psi}=\delta_{0 \sigma \sigma}=\delta_{\psi \sigma \sigma} = 1,
\end{equation}
together with cyclic permutations, and $\delta_{ijk}=0$ otherwise. The quantum dimensions are $d_0 = d_\psi = 1$ and $d_\sigma = \sqrt{2}$, and the $F$ tensor is given in Appendix~\ref{app:Ising-details}. The Hamiltonian is of the Levin-Wen form given in Eq.~\eqref{lwh}.

This gives a concrete realization of the doubled Ising topological order, whose excitation spectrum contains $9$ different types of anyons: $ \{ 0, \sigma, \bar{\sigma}, \psi, \bar{\psi}, \sigma\bar{\sigma}, \psi\bar{\sigma}, \sigma\bar{\psi}, \psi \bar{\psi} \}$. The fusion rules follow from those of a single chiral theory of Ising anyons, which are
\begin{eqnarray}
\psi \times \psi &=& 0, \\
\sigma \times \psi &=& \sigma, \\
\sigma \times \sigma &=& 0 + \psi \text{.}
\end{eqnarray}

One of the flux excitations  is $\phi_{\psi} = \psi \bar{\psi}$ with quantum dimension $d_{\psi \bar{\psi}} = 1$. To write down the $\psi \bar{\psi}$ flux Wilson string operator, we need the $S$-matrix of the Ising anyon theory:
\begin{eqnarray}
S = \frac{1}{2} \left( \begin{array}{ccc}
1 & \sqrt{2} & 1  \\
\sqrt{2} & 0 & -\sqrt{2} \\
1 & -\sqrt{2} & 1  \\
\end{array} \right).
\end{eqnarray}
From Eq.~\eqref{fun-string}, we see that a segment of the $\psi \bar{\psi}$ Wilson string operator takes the form 
\begin{equation}
W^{\psi \bar{\psi}}_{l} = (-1)^{n_{\sigma}(l)},
\label{eqn:flux_string_ising}
\end{equation}
where $n_{\sigma}(l)=1$ if the link $l$ is occupied by $\sigma$ string, and  $n_{\sigma}(l)=0$ otherwise.

Following Levin and Wen~\cite{levinwen}, we found explicit forms for the Wilson string operators of the $\sigma$, $\bar{\sigma}$, $\sigma \bar{\psi}$, and $\psi \bar{\sigma}$ excitations. The form of these string operators is described in Appendix~\ref{app:Ising-details}. All four string operators have the property that  $n_{\sigma}(l)$ is toggled between $0$ and $1$ on the path of links on which the string operator acts. It follows that these strings anti-commute with the $\psi \bar{\psi}$ Wilson strings at crossings, which is expected based on the $\theta = \pi$ mutual statistics between $\psi \bar{\psi}$ and each of $\sigma$, $\bar{\sigma}$, $\sigma \bar{\psi}$, and $\psi \bar{\sigma}$. Moreover, a short loop of any of the four string operators around a single plaquette $p$ reduces to $B_p^\sigma$.

\subsection{Condensing $\psi \bar{\psi}$ flux in doubled Ising string-net model}
\label{condense}

The $\psi \bar{\psi}$ flux condensation in the doubled Ising string-net model can be implemented following Refs.~\cite{burnell1,burnell2}. First, we decrease the gap for creating the $\psi \bar{\psi}$ flux by modifying the plaquette term into the following form,
\begin{equation}
B_{p}(J) = \frac{1}{2} (\mP_{p}^{0}+\mP_{p}^{\psi}) + \frac{1}{2} J (\mP_{p}^{0}-\mP_{p}^{\psi}),
\label{eqn:bj_ising}
\end{equation}
where $\mP_{p}^{\psi} = \frac{1}{4} (1-\sqrt{2} B_{p}^{\sigma} + B_{p}^{\psi})$ is the $\psi \bar{\psi}$ flux projector, which gives $1$ if the plaquette $p$ contains a $\psi \bar{\psi}$ flux, and $0$ otherwise.  Similarly, $\mP^0_p = B_p  = \frac{1}{4} (1+\sqrt{2} B_{p}^{\sigma} + B_{p}^{\psi})$ projects onto trivial flux at $p$. The coefficient $J$ sets the gap for the $\psi \bar{\psi}$ flux, which is tuned to a small positive number. Then, we condense the $\psi \bar{\psi}$ flux by adding the term
\begin{equation}
V=-V_{0} \sum_{l} (-1)^{n_{\sigma}(l)},
\label{eqn:flux_cond_ising}
\end{equation}
which creates a pair of $\psi \bar{\psi}$ fluxes on the plaquettes adjacent to $l$.
Upon making $V_0$ sufficiently large, the ground state prefers the existence of $\psi \bar{\psi}$ fluxes, which condense. One can see that Eq.~\eqref{eqn:flux_cond_ising} also gives an energy penalty to all links occupied by a $\sigma$ string. Therefore, in the $\psi \bar{\psi}$ condensed phase, the $\sigma$ string is removed from the low energy spectrum. In other words, condensing the $\psi \bar{\psi}$ flux confines the $\sigma$ string.

The full Hamiltonian, including the coupling~\eqref{eqn:flux_cond_ising}, becomes
\begin{equation}
H = - \sum_{v} A_{v} - \sum_{p} B_{p} (J) -V_{0} \sum_{l} (-1)^{n_{\sigma}(l)},
\label{eqn:h_cond_ising}
\end{equation}
where we can express $B_{p} (J) = \frac{1}{4} (1+B_{p}^{\psi}) +  \frac{\sqrt{2}}{4} J B_{p}^{\sigma}$. The term $\frac{\sqrt{2}}{4} J B_{p}^{\sigma}$ anti-commutes with Eq.~\eqref{eqn:flux_cond_ising} since it introduces $\sigma$ strings back into the ground state. As all the other terms commute with each other, the model is exactly solvable when $J=0$. Taking the limit $V_0 \to \infty$ and $J = 0$, only $0$ and $\psi$ strings remain in the low-energy Hilbert space, and we obtain an exactly solvable Hamiltonian for the condensed phase
\begin{equation}
H_{TC} = - \sum_{v} A_{v} - \frac{1}{4} \sum_{p} ( 1 + B^{\psi}_p ) \text{,}
\label{eqn:h_tc_ising}
\end{equation}
which is precisely the string-net Hamiltonian for the toric code model. Therefore, the result of $\psi \bar{\psi}$ flux condensation is the toric code topological order, which matches the result from the anyon condensation theory of Ref.~\cite{bais}.

Following Ref.~\cite{burnell}, we now give an alternate derivation of the excitation content in the $\psi \bar{\psi}$-condensed phase, from the properties of anyons in the doubled Ising theory, rather than from the microscopic string-net model. The $\sigma$ and $\bar{\sigma}$ excitations are confined due to their non-trivial braiding phase $\theta=\pi$ with the $\psi\bar{\psi}$ flux.  On the other hand, $\psi$ and $\bar{\psi}$ remain deconfined because they have trivial mutual statistics with $\psi \bar{\psi}$. Moreover, because $\psi$ can fuse with a condensed $\psi \bar{\psi}$ particle to become $\bar{\psi}$, $\psi$ and $\bar{\psi}$ are identified in the condensed phase; these particles become the fermionic anyon in the toric code. The bound states $\psi \bar{\sigma}$ and $\sigma \bar{\psi}$ are also confined since binding a confined particle $\sigma$ or $\bar{\sigma}$ to a deconfined particle $\psi$ or $\bar{\psi}$ results in a confined excitation.

$\sigma \bar{\sigma}$ remains deconfined since it braids trivially with $\psi\bar{\psi}$. However, it instead splits into two particles each with quantum dimension $1$. Here is one quick way to see why the splitting happens: Considering the fusion of two $\sigma \bar{\sigma}$ particles, we obtain $\sigma \bar{\sigma} \times \sigma \bar{\sigma} = 1 + \psi + \bar{\psi} + \psi \bar{\psi}$. Since $\psi \bar{\psi}$ is condensed, it is identified with the vacuum and so there are two copies of the vacuum in the vacuum fusion channel, implying that $\sigma \bar{\sigma}$ can not be an anyon of the simple type in the condensed phase. Instead, it turns out that $\sigma \bar{\sigma}$ splits into the $e$ and $m$ particles in the toric code~\cite{bais, burnell1, burnell2}. This can also be seen directly by examining the $\sigma \bar{\sigma}$ string operator in the string-net model~\cite{burnell1, burnell2}. 

\subsection{Doubled Ising cage-net fracton model}
\label{sec:su2_2-cage-net}

\begin{figure}
\includegraphics[width=0.4\textwidth]{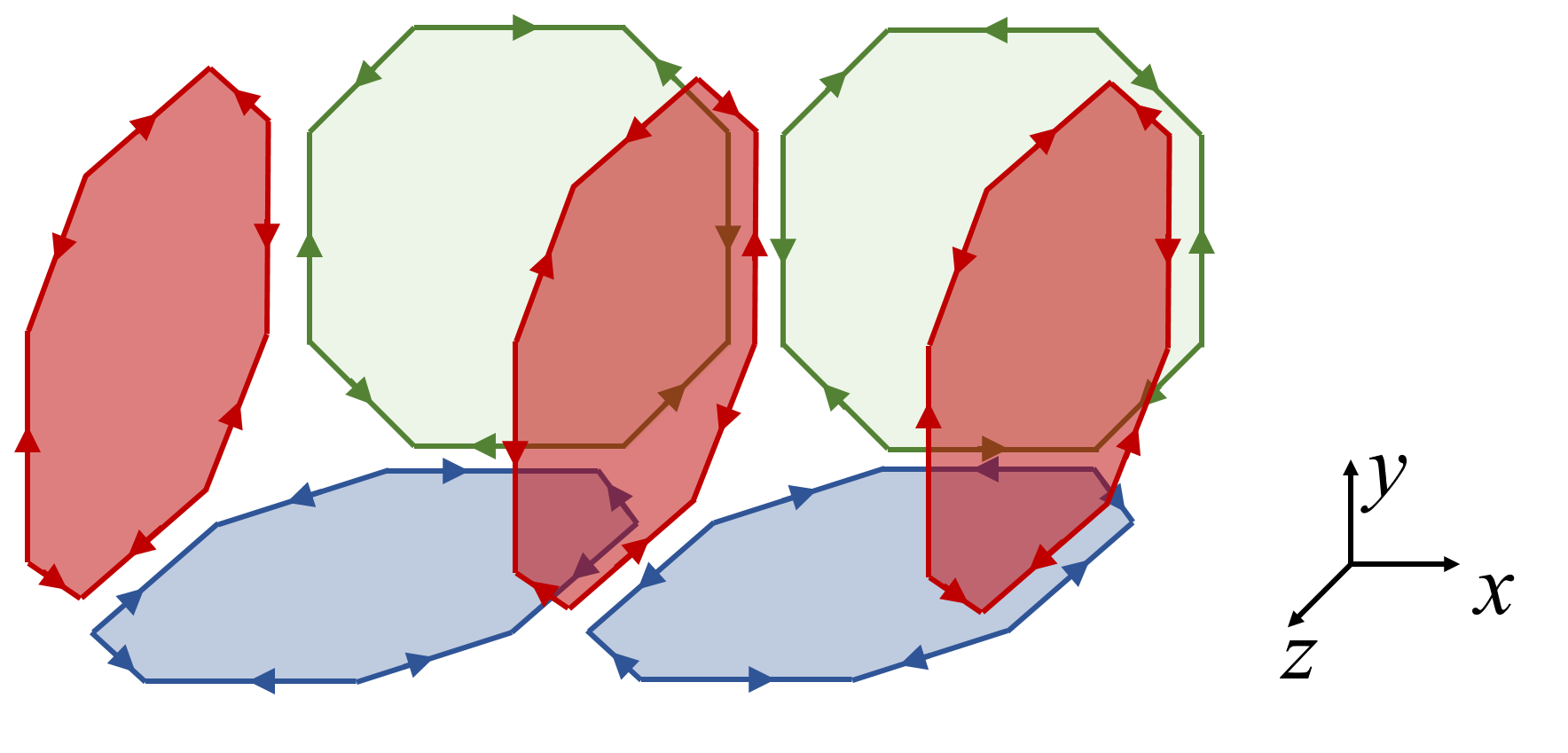}
\caption{Stacking $d=2$ layers of string-net models along the $x$, $y$, and $z$ directions. Links parallel to the principal axes $\mu = x,y,z$ carry two string degrees of freedom while the others carry a single string degree of freedom. For the doubled Ising string-net model (and for SU(2)$_k$ string-net models $\forall k$) the string orientations may be eliminated.}
\label{fig:stacking}
\end{figure}

In order to obtain the $d=3$ doubled Ising cage-net fracton model, we stack layers of $d=2$ doubled Ising string-net models along the $x$, $y$, and $z$ directions as shown in Fig.~\ref{fig:stacking}. The string-net Hamiltonian on each plane $P$ is given by
\begin{equation}
H_{P} = - \sum_{v \in P} A_{v}^{o(P)} - \sum_{p \in p_o(P)} B_{p} - \sum_{p \in p_d(P)} B_{p} \text{,}
\label{eqn:hp_su2_2donelayer}
\end{equation}
where $o(P)$ is the orientation of the plane $P$, specified by the direction normal to $P$. We will treat octagonal and diamond plaquettes differently, so the corresponding terms are written separately in the Hamiltonian, with $p_o(P)$ ($p_d(P)$) the set of octagonal (diamond) plaquettes in plane $P$. Strings belonging to a string-net model in plane $P$ are denoted $i^\nu$ where $\nu = o(P)$. For example, the string $\sigma^z$ belongs to the string-net model in the $xy$-plane. Links $l$ which are parallel to one of the principal axes $\mu = x,y,z$ carry two string degrees of freedom $i^\nu$ and $i^\lambda$, where $\mu$, $\nu$ and $\lambda$ are all different. The Hamiltonian obtained by stacking string-net layers along principal axes is simply the sum of Eq.~\eqref{eqn:hp_su2_2donelayer} over all planes,
\begin{equation}
H_{S} = \sum_{P} H_{P}.
\label{eqn:eqn:hp_su2_2ddecoupled}
\end{equation}

We now enter the $d=3$ fracton phase by condensing $p$-strings made out of $\psi \bar{\psi}$ fluxes. Similar to the flux condensation in the doubled Ising string-net model, we decrease the gap of the $\psi \bar{\psi}$ flux by modifying the plaquette terms and by adding the $\psi \bar{\psi}$ flux condensation term. Here, we only modify the octagonal plaquette terms, and only add the coupling between the layers on links parallel to the principal axes, which share strings from two different layers.
The resulting Hamiltonian is
\begin{equation}
H_{P} = \sum_{P} H'_{P} + V,
\label{eqn:hp}
\end{equation}
where $H'_{P}$ is the modified string-net Hamiltonian in which the gap of $\psi \bar{\psi}$ has been decreased on octagonal plaquettes,
\begin{eqnarray}
H'_{P} &=& - \sum_{v \in P} A_{v}^{o(P)} - \sum_{p \in p_{o}(P)} B_{p}(J) - \sum_{p \in p_d(P)} B_{p}
\\
&=& - \sum_{v \in P} A_{v}^{o(P)} - \frac{1}{2}  \sum_{p \in p_{o}(P)} \frac{1}{4} (1 + B_{p}^{\psi^{o(P)}}) 
\nonumber \\
&\,& - \frac{J}{2} \sum_{p \in p_o(P)} \frac{\sqrt{2}}{4} B_{p}^{\sigma^{o(P)}} - \sum_{p \in p_d(P)} B_{p} \text{.}
\label{eqn:hp_j}
\end{eqnarray}

\begin{figure}[t]
\includegraphics[width=0.3\textwidth]{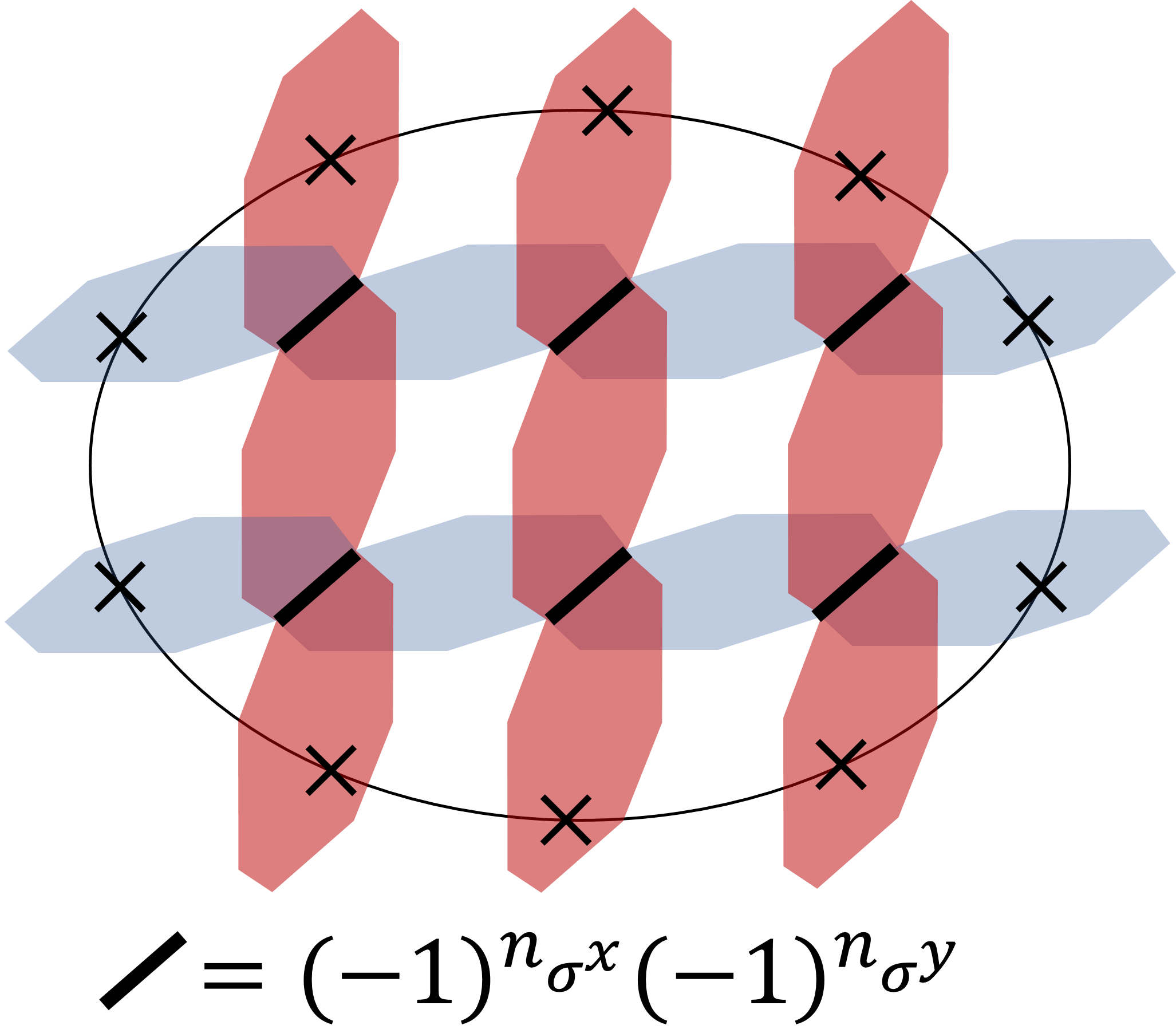}
\caption{A long $\psi \bar{\psi}$-string, created by acting with $(-1)^{n_{\sigma^x}(l)} (-1)^{n_{\sigma^y}(l)}$ operators along the links marked in black.}
\label{fig:bigstring}
\end{figure}

The term $V$ in Eq.~\eqref{eqn:hp} is a coupling term which implements the $\psi \bar{\psi}$-string condensation,
\begin{equation}
V=- V_{0} \sum_{l \in l_{o}} (-1)^{n_{\sigma^{\mu}}(l)} (-1)^{n_{\sigma^{\nu}}(l)},
\end{equation}
where $\mu$, $\nu$ denote the orientation of the planes intersecting at link $l$. Here, $l_{o}$ only includes links parallel to the principal axes, which are also the links shared by octagonal plaquettes. We define $n_{\sigma^\mu}(l) = 1$ when the link $l$ is occupied by a $\sigma^{\mu}$ string, and $n_{\sigma^\mu}(l) = 0$ otherwise. The coupling term $V$ creates a fundamental $\psi \bar{\psi}$-string consisting of four $\psi \bar{\psi}$ fluxes on the four octagonal plaquettes adjacent to the link $l$. A longer $\psi \bar{\psi}$ string can be created by repeated application of the operator $(-1)^{n_{\sigma^{\mu}}(l)} (-1)^{n_{\sigma^{\nu}}(l)}$ along links $l$ which form a rectangular membrane, as depicted in Fig.~\ref{fig:bigstring}.

We note that, for a single string-net layer, we can also modify only the octagonal plaquette terms, and only add a term creating pairs of $\psi \bar{\psi}$ particles on links parallel to the principal axes.  The effect of these terms is still to condense $\psi \bar{\psi}$ anyons, so while one does not obtain precisely the $\mathbb{Z}_2$ toric code string-net model from such a construction, the topological order in the condensed phase is the same as that of the toric code.

Returning to the cage-net fracton model, we take the limit $V_{0} \rightarrow \infty$ limit to be deep inside the fracton phase. For links $l \in l_o$ running along the $\mu$-direction, the coupling term favours configurations where the two string degrees of freedom $i^\nu$ and $i^\lambda$ on link $l$ are either both labelled by $\sigma$ or where neither string is labelled by $\sigma$. That is, the string configurations in the low-energy Hilbert space are $0, \psi^{\nu}, \psi^{\lambda}, \psi^{\nu} \psi^{\lambda}$ and $\sigma^{\nu} \sigma^{\lambda}$.  Links $l \notin l_o$ along the edges of the diamond plaquettes are contained only in a single plane, and are not affected by the coupling term.

\begin{figure}[t]
\includegraphics[width=0.4\textwidth]{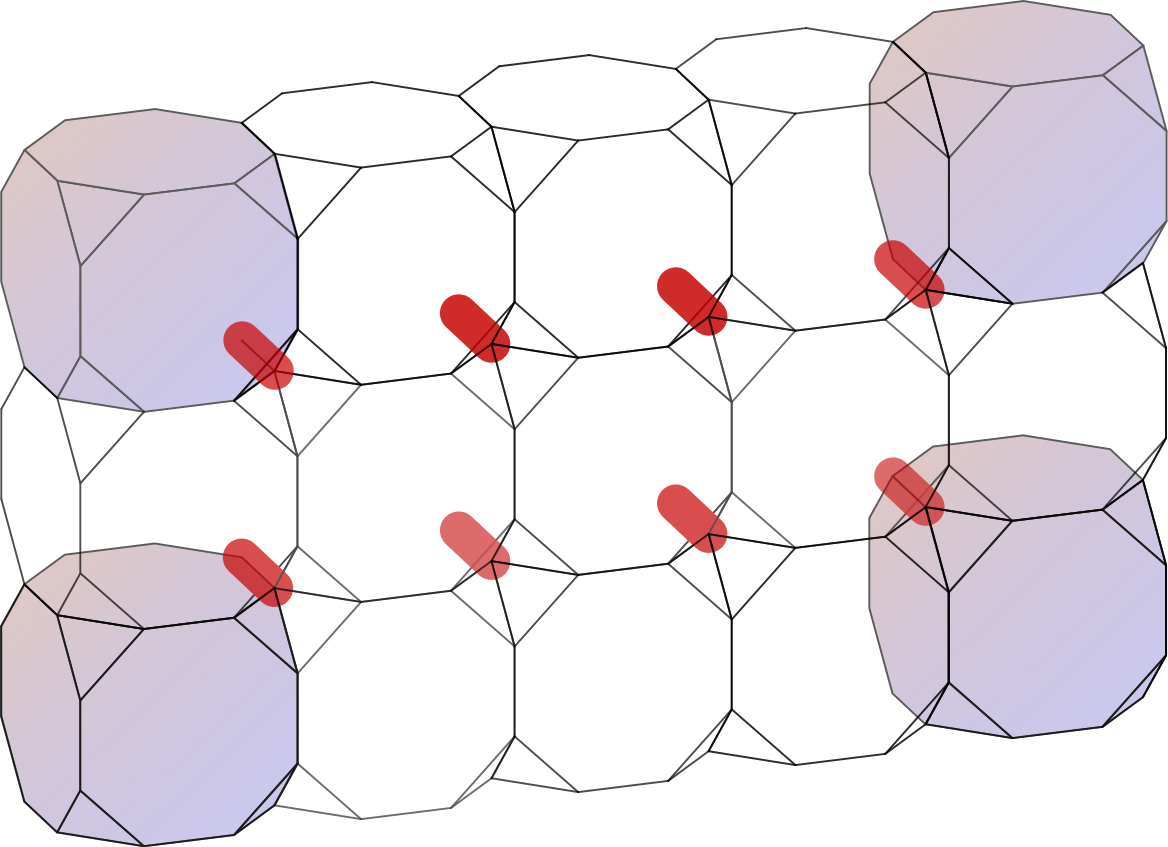}
\caption{Deconfined fracton excitations, represented by the purple cages, are created at the corners of a rectangular membrane operator composed of flux creation operators $V_l$ acting along the red links.}
\label{fig:fluxmembrane}
\end{figure}

We see that the term proportional to $J$ in the Hamiltonian Eq.~\eqref{eqn:hp_j}, which introduces $\sigma$ strings on the octagonal plaquettes, anti-commutes with the coupling term $V$. Treating the terms other than $V$ as perturbations, and carrying out degenerate perturbation theory, the leading non-trivial contribution of the $J$-term is a ``cage" term at sixth order in degenerate perturbation theory,
\begin{equation}
H_{cage} = - J_{c} \sum_{c} B_{c},
\label{eqn:H_ising_cage}
\end{equation}
where
\beq
B_{c} =  \prod_{p_{o} \in c} (\mP_{p}^{0} - \mP_{p}^{\psi}) = \prod_{p_{o} \in c} \frac{\sqrt{2}}{4} B_{p}^{\sigma^{o(p_o)}}.
\eeq
Here, each cage is a truncated cube (see Fig.~\ref{fig:fluxmembrane}), the product is over the six octagonal plaquettes on the boundary of the cage, and $o(p_o)$ is the normal direction to the plaquette $p_o$.

The cage operator $B_{c}$ has eigenvalue $+1$ if there are an even number of $\psi \bar{\psi}$ fluxes through the octagonal faces of the cage, and eigenvalue $-1$ if the number of fluxes is odd. $B_{c}$ commutes with the coupling  $V$, since each term in  $V$  overlaps with an even number of octagonal plaquettes belonging to the cage $c$. Hence, the effective Hamiltonian describing the doubled-Ising fracton phase is 
\begin{eqnarray}
H_{f} &=& - \sum_P \sum_{v \in P} A_{v}^{o(P)} -  \frac{1}{2} \sum_P  \sum_{p \in p_{o}(P)} \frac{1}{4} (1 + B_{p}^{\psi^{o(P)}}) \nonumber \\ &-& \sum_P \sum_{p \in p_d(P)} B_{p}  -  J_{c} \sum_{c} B_{c}.
\label{eqn:h_fraction_ising}
\end{eqnarray}

Let us now discuss the excitations in the fracton phase. We note that, strictly speaking, all excitations in this phase carry a position index, labelling their position on the lattice; this should be clear from the example of the X-cube model, where each fracton constitutes a distinct superselection sector, which is accounted for by a position index.  With this caveat, we omit the  position indices in what follows to avoid cumbersome notation, making them explicit only when needed for clarity.

Following the discussion at the beginning of Sec.~\ref{ising}, the particles $\psi$, $\bar{\psi}$, $\psi \bar{\psi}$, and $\sigma \bar{\sigma}$ remain deconfined in each $d = 2$ layer, as they have trivial braiding with the condensed $\psi \bar{\psi}$ string. The other dim-2 particles of each Ising string net layer ($\sigma$, $\bar{\sigma}$, $\psi \bar{\sigma}$ and $\sigma \bar{\psi}$) have non-trivial $\theta = \pi$ Abelian statistics with the $\psi \bar{\psi}$ string, and are hence confined. We emphasize that the excitations from each layer that survive condensation are different from the case of $\psi \bar{\psi}$ particle condensation in a single $d=2$ layer. There, because $\psi \bar{\psi}$ disappears into the condensate, the $\psi$ and $\bar{\psi}$ excitations are identified, and $\sigma \bar{\sigma}$ splits into Abelian anyons. Here, $\psi$ and $\bar{\psi}$ remain distinct excitations, and $\sigma \bar{\sigma}$ remains non-Abelian, as we see from the fusion rule $\sigma \bar{\sigma} \times \sigma \bar{\sigma} = 0 + \psi + \bar{\psi} + \psi \bar{\psi}$.

On a link $l \in l_o$, we define the flux creation operator
\beq
V_l = (-1)^{n_{\sigma} (l)},
\eeq
where $n_{\sigma}(l) = 1$ when the string type on link $l$ is $\sigma^{\mu} \sigma^{\nu}$, and $n_\sigma(l) = 0$ otherwise.  This term anti-commutes with the four cage operators $B_c$ sharing the link $l$, and thus creates four cage excitations.  Moreover, this term is the projection to the low-energy Hilbert space of the Wilson string segment $W_l$ creating $\psi \bar{\psi}$ fluxes in either of the two planes containing $l$ (these operators become identical upon projection to the low-energy Hilbert space).  Therefore, we see that the deconfined $\psi \bar{\psi}$ particle is identified with a bound state of two $B_c = -1$ cage excitations -- this is identical to what happens with the toric code $m$ particles in the coupled-layer construction of the X-cube model.  Also as in that case, the $B_c = -1$ cage excitations are deconfined fractons, which can be created in isolation by taking a product of the the flux creation operator over a rectangular membrane, as shown in Fig.~\ref{fig:fluxmembrane}. Moreover, again as in the X-cube model, the fractons can be viewed as the ends of open $\psi \bar{\psi}$ strings.  Since these fractons are descendants of the $\psi \bar{\psi}$ flux, which is an Abelian anyon, the fractons in this theory do not carry any topological degeneracy.

However, there exist non-Abelian dim-1 particles, which arise in a similar fashion to the Abelian dim-1 particles in the coupled-layer construction of the X-Cube model.  We consider two perpendicular planes oriented normal to $\mu$ and $\nu$.  The $\sigma^{\mu}$ and $\sigma^{\nu}$ excitations are confined, because these excitations acquire a phase $\theta = \pi$ when brought around a $\psi \bar{\psi}$ string.  However, for a bound state $\sigma^{\mu} \sigma^{\nu}$, these phases cancel, and this bound state is a deconfined particle in the fracton phase.  Because the two constituents of the bound state are restricted to move in their respective planes, $\sigma^{\mu} \sigma^{\nu}$ is a dim-1 particle constrained to move along the line where the two planes intersect.  Similarly, the bound states $\sigma^{\mu}\bar{\sigma}^{\nu}$ and $\bar{\sigma}^{\mu}\bar{\sigma}^{\nu}$ are also deconfined dim-1 particles. We discuss the non-Abelian nature of these excitations in more detail in the following section.

\begin{figure}[t]
\includegraphics[width=0.4\textwidth]{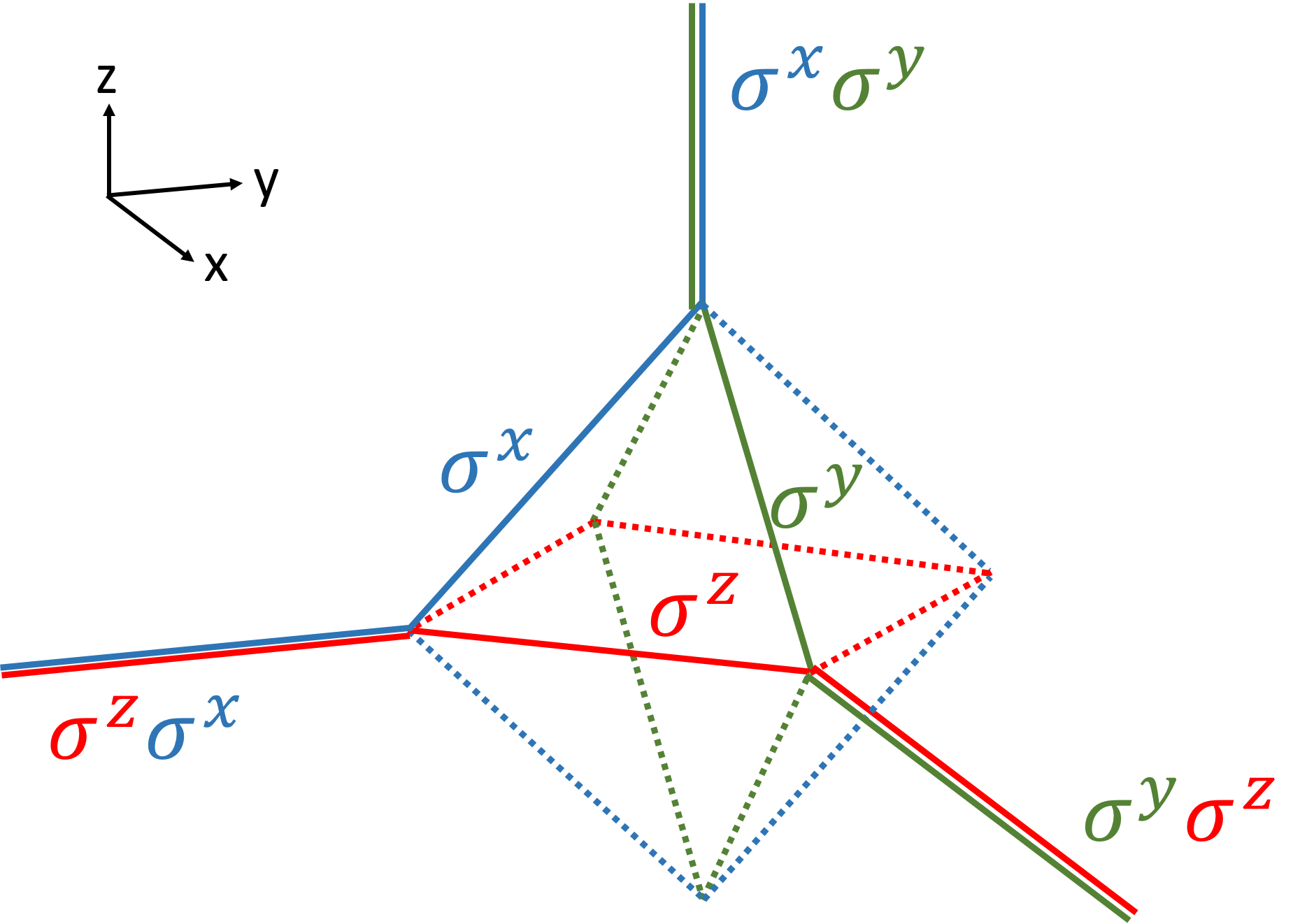}
\centering
\caption{String operators for the dim-1 particles $\sigma^x \sigma^y$, $\sigma^y \sigma^z$, and $\sigma^x \sigma^z$ meeting at a single junction. $\sigma^\mu$ strings are shown as solid lines while dotted lines represent null strings.}
\label{fig:vertex}
\end{figure}

Wilson string operators for these dim-1 particles can be constructed from the Wilson strings for $\sigma$ and $\bar{\sigma}$, whose form is specified in Appendix~\ref{app:Ising-details}. For instance, a $\sigma^{\mu} \sigma^{\nu}$ Wilson string is simply a product of $\sigma^{\mu}$ and $\sigma^{\nu}$ Wilson strings, projected to the low-energy Hilbert space. The constituent $\sigma^{\mu}$ and $\sigma^{\nu}$ strings can only ``turn corners'' in the planes perpendicular to $\mu$ and $\nu$, respectively, where the corners are turned as the string passes through a diamond plaquette.  This observation allows us to construct a junction of string operators for $\sigma^x \sigma^y$, $\sigma^y \sigma^z$, and $\sigma^x \sigma^z$ meeting at a single vertex.  We simply take a product of three $\sigma^{\mu}$ strings, each of which turns a corner at the same ``vertex'' (\emph{i.e.} octahedron formed from diamond plaquettes), so that the strings emanating away from the vertex are those of the dim-1 particles, and then project this product into the low-energy Hilbert space. This process is illustrated in Fig.~\ref{fig:vertex}.

We emphasize that the flux-string condensation picture provides a complete list of ``elementary'' deconfined excitations within the flux-string condensed phase, by which we mean that any excitation can be obtained by fusing together the elementary ones.  The elementary excitations are the dim-2 particles that survive flux-string condensation in each string-net layer, the non-Abelian dim-1 particles $\sigma^{\mu} \sigma^{\nu}$, and the $B_c = -1$ Abelian fracton excitations, that arise as open ends of flux-strings.

\subsection{Non-Abelian excitations in fracton phases}
\label{sec:excitations}

In order to understand the non-Abelian nature of the deconfined dim-1 excitations in the doubled Ising cage-net model, we first have to discuss more generally what it means for gapped excitations in a fracton phase to be non-Abelian.  In any gapped phase of matter, it is expected that point-like excitations can be assigned to superselection sectors, or particle types.  Two excitations belong to the same superselection sector if and only if there is some local process that can transform one into the other.  This expectation immediately implies a notion of fusion:  given two point-like excitations, we can consider a region containing both excitations, and ask what superselection sector the resulting composite excitation belongs to.  An excitation is Abelian when the superselection sector of any of its composites is uniquely determined by the particle types of the constituent particles.  Non-Abelian excitations are simply those excitations that are not Abelian, and it follows that non-Abelian excitations participate in multiple fusion channels.  That is, given a non-Abelian excitation of type $a$, there is always some non-Abelian excitation of type $b$, so that multiple superselection sectors are possible upon fusing $a$ and $b$.

If some non-Abelian excitations are placed at fixed positions, there are additional non-local degrees of freedom corresponding to the different fusion channels, and these degrees of freedom form a Hilbert space of degenerate states.  This is a robust topological degeneracy, as the fusion channel of any pair of well-separated particles cannot be changed by a local process.  We can define the quantum dimension $d_a$ of an excitation $a$ by fusing together $N$ copies of $a$.  The dimension $D$ of the resulting Hilbert space is expected to grow exponentially with $N$, and the quantum dimension is defined by $D \sim d_a^N$, asymptotically when $N$ is large.

\begin{figure}[t]
\includegraphics[width=8cm]{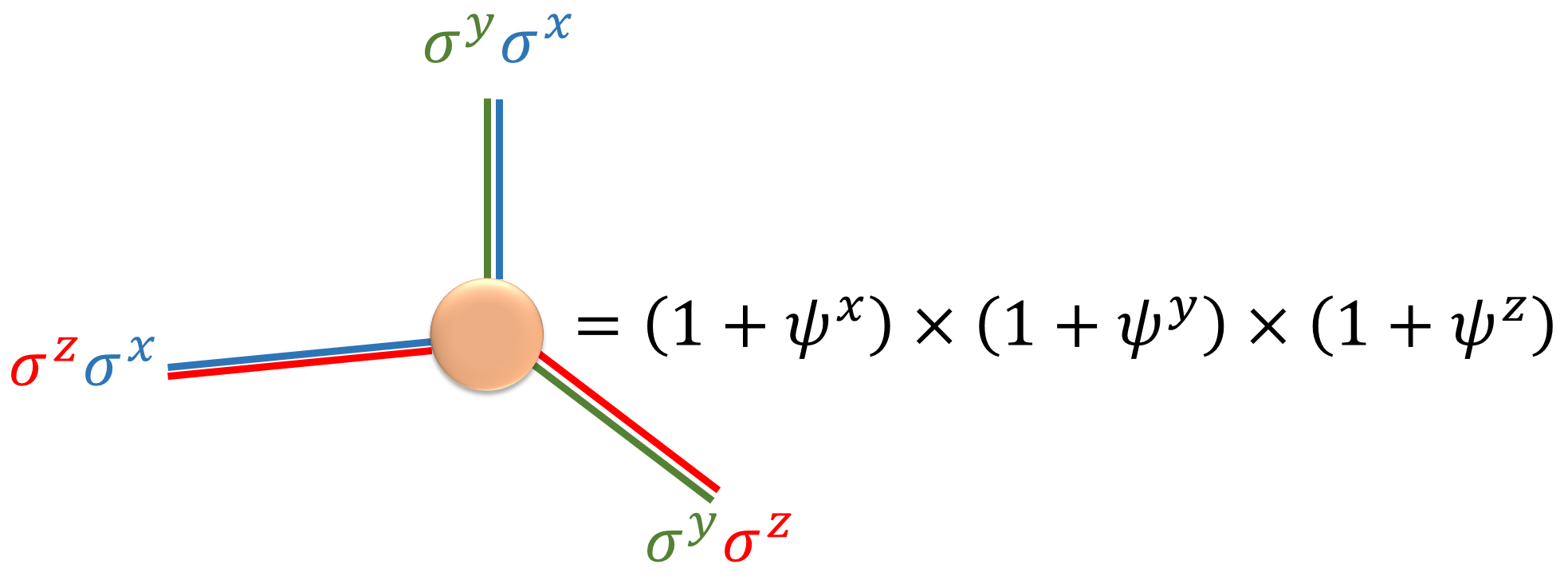}
\centering
\caption{The fusion of $\sigma^{x}\sigma^{y}$, $\sigma^{y}\sigma^{z}$, and $\sigma^{z}\sigma^{x}$ anyons results in $(0+\psi^{x}) \times (0+\psi^{y}) \times (0+\psi^{z})$, clearly reflecting the topological degeneracy, and hence, non-Abelian nature of these excitations. Here the microscopic vertex, illustrated in Fig.~\ref{fig:vertex}, has been replaced with a coarse-grained vertex for clarity.}
\label{fig:fusion_1d}
\end{figure}

In order to establish the presence of non-Abelian sub-dimensional excitations in a fracton phase, we need to understand enough about the superselection structure to show that multiple fusion outcomes are possible.  In the cage-net models, this is made possible by the flux string condensation picture, which provides an understanding of excitations in the fracton phase in terms of those of the underlying system of decoupled layers.  From this, we can see that the dim-1 particles of the doubled Ising cage-net model are non-Abelian by fusing two of them.  For instance, 
$\sigma^{\mu} \sigma^{\nu} \times \sigma^{\mu} \sigma^{\nu} = (0 + \psi^{\mu}) \times (0 + \psi^{\nu} )$, with four different fusion outcomes possible.  In addition, it is interesting to consider the fusion of $\sigma^{x}\sigma^{y}$, $\sigma^{y}\sigma^{z}$, and $\sigma^{z}\sigma^{x}$ anyons, as shown in Fig.~\ref{fig:fusion_1d}.  In this case, there are $2^{3}=8$ possible fusion outcomes.

We also expect that there should exist ``braiding-like" processes which reveal the non-Abelian nature of the dim-1 excitations. In general, if we consider two non-Abelian excitations $a$ and $b$ in a definite fusion channel, we expect that there will be some process by which this fusion channel can be changed. Such a process must necessarily be non-local,  because acting with any operator with support in a region containing $a$ and $b$ cannot change the particle type in this region, \emph{i.e.} this cannot change the fusion channel.

\begin{figure}[t]
\includegraphics[width=8cm]{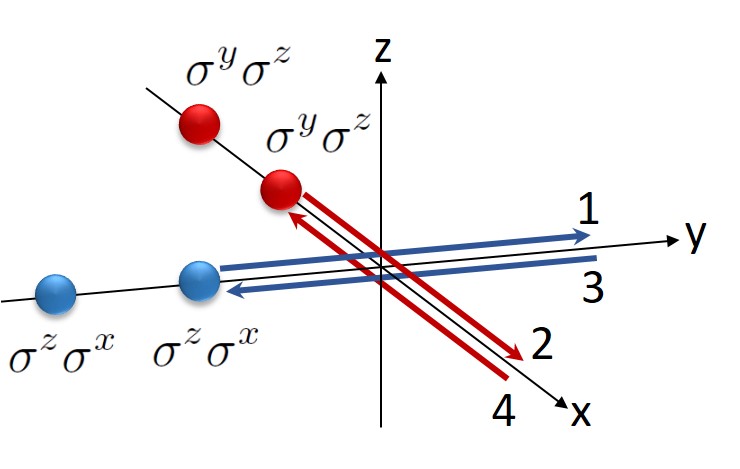}
\centering
\caption{An $S$-matrix measurement for two dim-1 particles $\sigma^z \sigma^x$ and $\sigma_y \sigma^z$ in the $xy$ plane. First, a pair of $\sigma^z \sigma^x$ ($\sigma^y \sigma^z$) particles, in blue (red), mobile along the $y$ ($x$) axis are created from vacuum. Next, we perform a full braiding of $\sigma^z \sigma^x$ and $\sigma^y \sigma^z$ which involves moving the excitations back and forth along the $y$ and $x$ axes respectively in the order (from 1-4) depicted here. Finally, all excitations are annihilated back into the vacuum.}
\label{fig:braiding1d}
\end{figure}

More specifically, we consider the topologically non-trivial process depicted in Fig.~\ref{fig:braiding1d}, which amounts to a full braid of dim-1 excitations confined to move on perpendicular lines within the same plane. Consider creating from the vacuum a pair of $\sigma^z \sigma^x$ excitations, and a pair of $\sigma^y \sigma^z$ excitations, which are mobile along the $y$ and $x$ axes, respectively. We can now consider a process where we first (step 1 in Fig.~\ref{fig:braiding1d}) move a $\sigma^z \sigma^x$ particle along $y$ and then (step 2) move a $\sigma^y \sigma^z$ particle along $x$. Next we move $\sigma^z \sigma^x$ back along its original path (step 3), after which $\sigma^y \sigma^z$ also moves back to its initial position (step 4). Finally, all excitations are annihilated back into the vacuum. The (normalized) expectation value (on the vacuum state) of this measurement can be represented diagrammatically by
\beq
\label{monodromy}
\mathcal{M}_{\sigma^z \sigma^x,\sigma^y \sigma^z} = \frac{1}{d_{\sigma^z \sigma^x} d_{\sigma^y \sigma^z}} \bmm \includegraphics[height=0.5in]{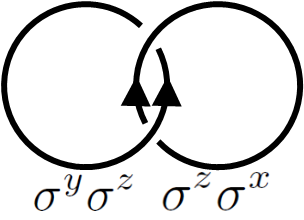} \emm \text{.}
\eeq 
For anyons in 2d, this expression is the monodromy scalar component, which is related to the $S$-matrix by rescaling~\cite{bonderson}.  Here, we interpret the diagram as representing an operator that effects the process described above, where the particle world lines braid as shown.  With this interpretation, there is an issue of normalization, but we will show that $\mathcal{M}_{\sigma^z \sigma^x,\sigma^y \sigma^z} = 0$, so this issue does not matter for our discussion.

As discussed previously, the $\sigma^\mu \sigma^\nu$ Wilson string is simply the product of $\sigma^\mu$ and $\sigma^\nu$ Wilson strings, projected to the low-energy Hilbert space. Due to this factorization, it is straightforward to show that the non-trivial braiding between $\sigma^z \sigma^x$ and $\sigma^y \sigma^z$ is actually a result of the non-trivial braiding between the two $\sigma^z$ strings involved, both of which live in the $xy$ plane. In other words, the result of braiding the dim-1 particles is equivalent to that of braiding $\sigma$ anyons in a single doubled Ising string-net layer. The quantum dimension of the dim-1 particles factorizes~\cite{twisted}
\beq
d_{\sigma^\mu \sigma^\nu} = d_{\sigma^\mu} d_{\sigma^\nu},
\eeq
and Eq.~\eqref{monodromy} simplifies to
\beq
\mathcal{M}_{\sigma^z \sigma^x,\sigma^y \sigma^z} = \frac{1}{d_{\sigma^z}^2 d_{\sigma^x} d_{\sigma^y}} \bmm \includegraphics[height=0.3in]{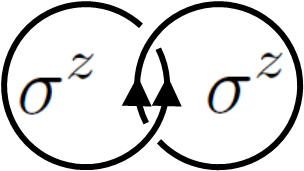}\, \includegraphics[height=0.3in]{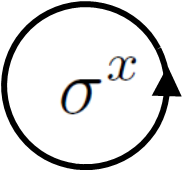}\, \includegraphics[height=0.3in]{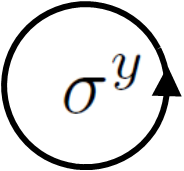} \emm ,
\eeq
since the $\sigma^x$ and $\sigma^y$ strings do not undergo any non-trivial braiding during the process depicted in Fig.~\ref{fig:braiding1d}. From the definition of the quantum dimension, it is clear that Eq.~\eqref{monodromy} reduces to
\beq
\mathcal{M}_{\sigma^z \sigma^x,\sigma^y \sigma^z} = \frac{1}{d_{\sigma^z} d_{\sigma^z}} \bmm \includegraphics[height=0.3in]{monodromy_i} \emm = \mathcal{M}_{\sigma \sigma} = 0,
\eeq
where $\mathcal{M}_{\sigma \sigma}$ is the monodromy scalar component for $\sigma$ anyons in a single doubled Ising string-net layer. Physically, $\mathcal{M}_{\sigma^z \sigma^x,\sigma^y \sigma^z}$ is the amplitude for the two pairs of dim-1 particles to be in the vacuum sector just before the particles are annihilated at the end of the process.  The particles begin the process in the vacuum sector, and if the particles are Abelian, their fusion channel cannot be changed during the process.  Therefore, $\mathcal{M}_{\sigma^z \sigma^x,\sigma^y \sigma^z} = 0$ implies the particles are non-Abelian because their fusion channel changes with unit probability.  We note that this conclusion only relies on the fact that a set of Abelian particles -- whether they are ordinary 2d anyons or subdimensional particles -- has a unique fusion channel.

We can also more directly study the change of fusion channel during the same process. We focus on the two $\sigma^z \sigma^x$ excitations, which are initially in the vacuum fusion channel, and consider the effect of braiding one of them around one of the $\sigma^y \sigma^z$ excitations. The left-hand side of the equation below encodes the initial configuration (\emph{i.e.} before braiding) of these three anyons, and this configuration can be rewritten in terms of an equal weight superposition,
\beq
\bmm \includegraphics[height=0.8in]{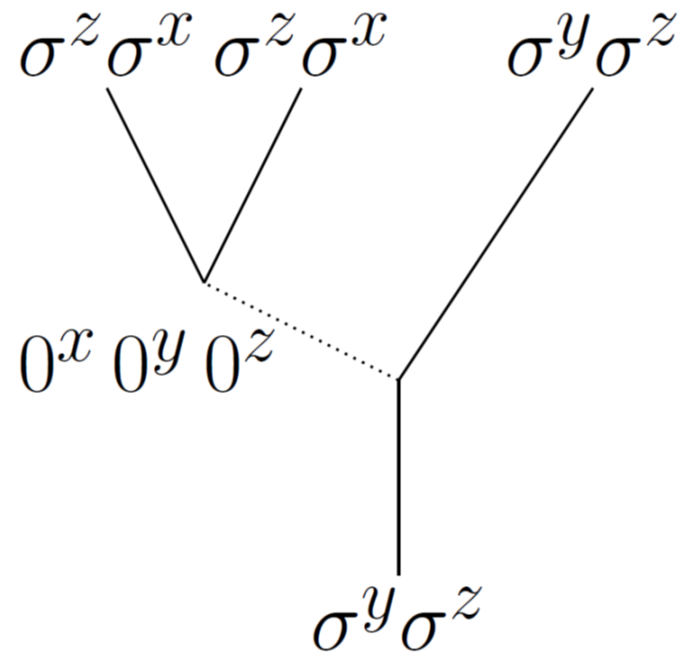} \emm = \frac{1}{\sqrt{2}}\bmm \includegraphics[height=0.8in]{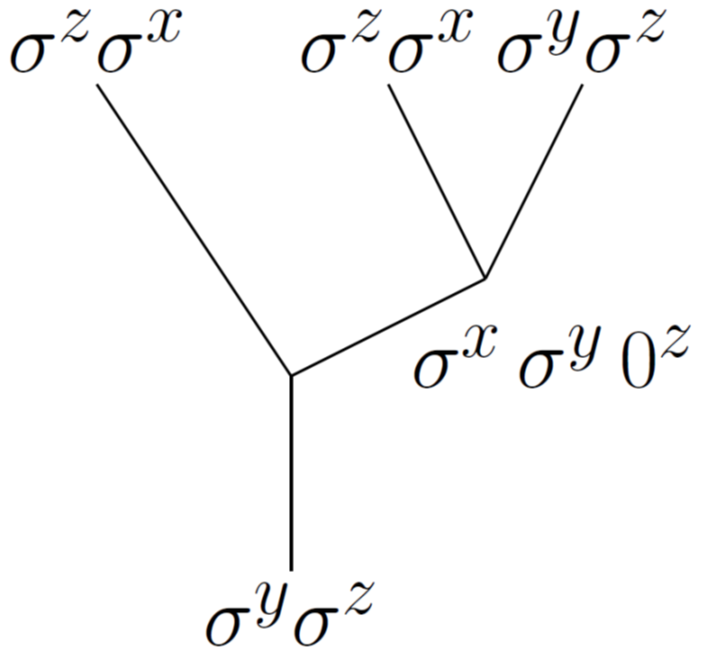} \emm +  \frac{1}{\sqrt{2}}\bmm \includegraphics[height=0.8in]{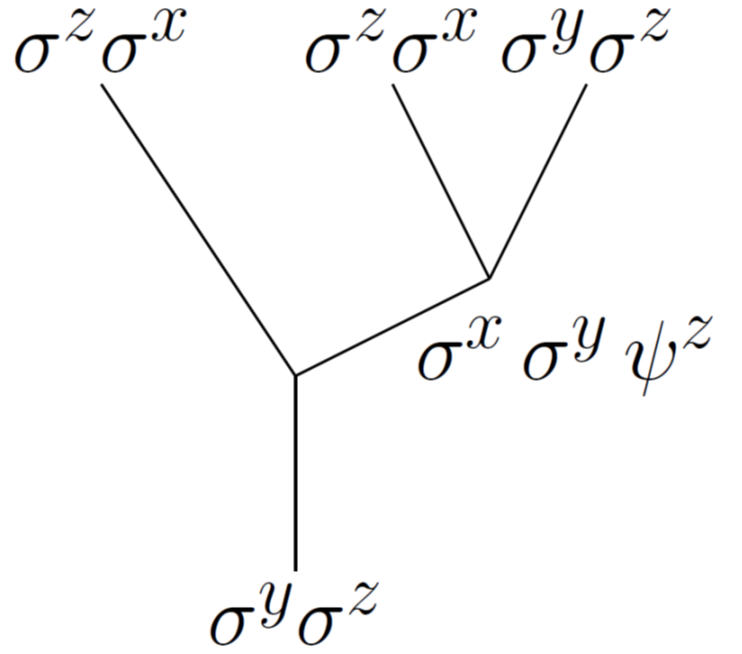} \emm.
\eeq
Here, we have made a single $F$-move on the string-configurations; similarly to our preceding discussion of $\mathcal{M}_{\sigma^z \sigma^x,\sigma^y \sigma^z}$, only the $\sigma^z$ strings belonging to the $xy$ plane contribute non-trivially to the composite $F$-move of the dim-1 bound states. In other words, this $F$-move is equivalent to one applied to an identical configuration of $\sigma$ particles in a single string-net layer. Note that for simplifying the diagrammatics, we are representing the $\sigma^\mu \sigma^\nu$ string operators as single strings and not as products of $\sigma^\mu$ and $\sigma^\nu$ strings.

The effect of braiding on the fusion channel can be seen by first making the above $F$-move on the following configuration:
\beq
\bmm \includegraphics[width=0.9in]{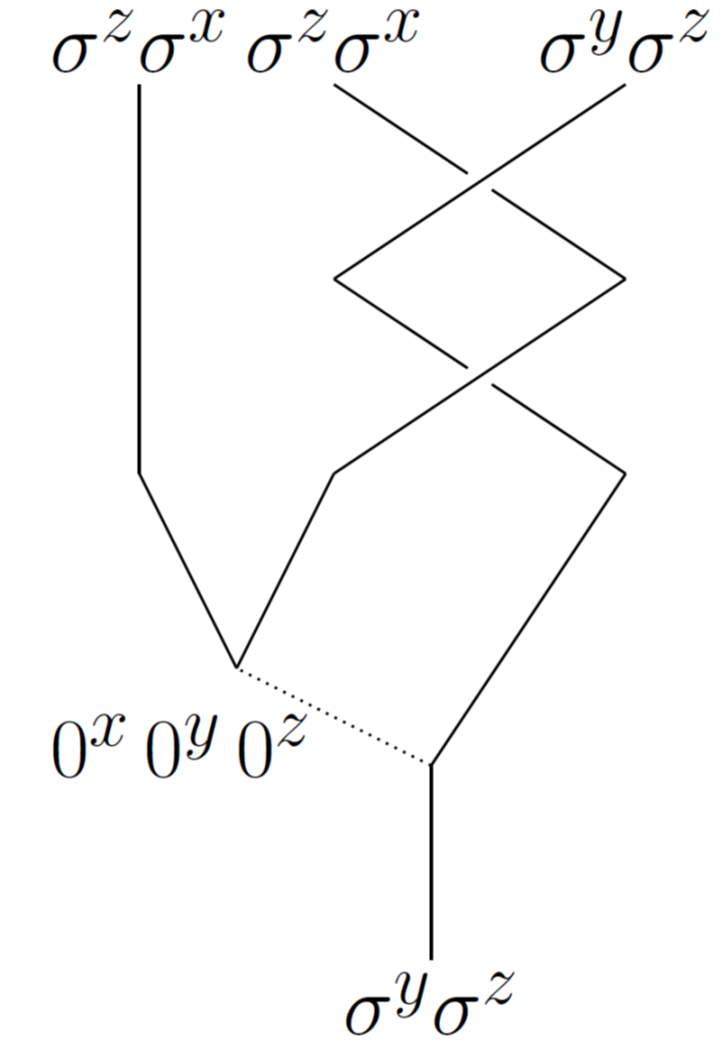} \emm = \frac{1}{\sqrt{2}}\bmm \includegraphics[width=0.9in]{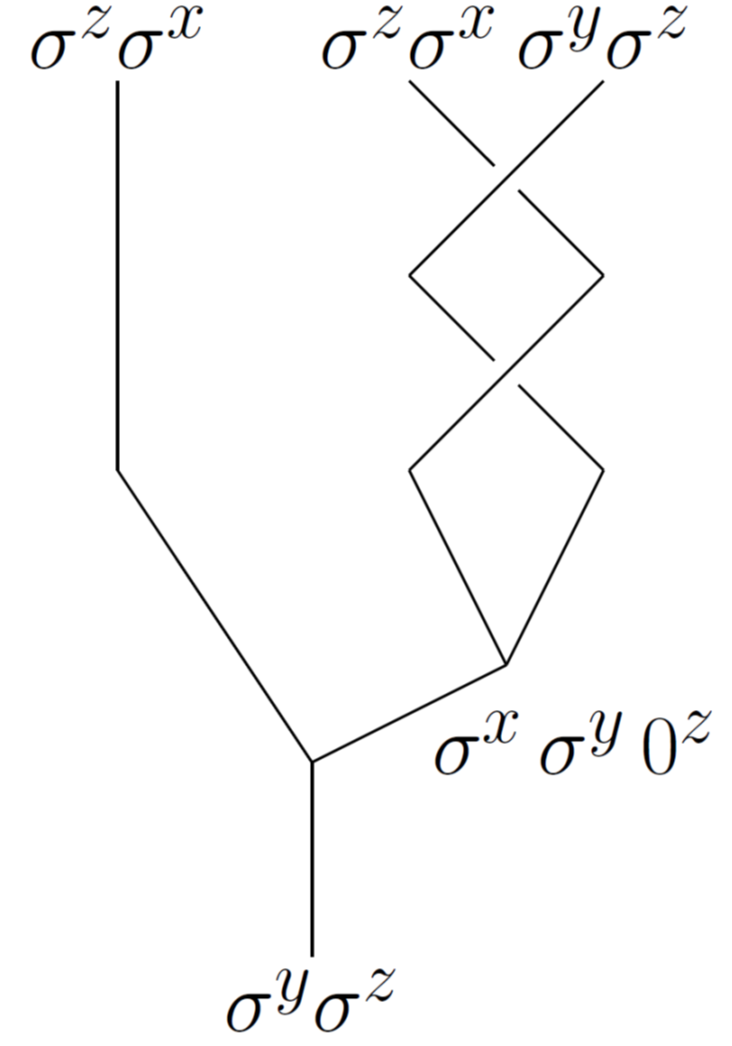} \emm +  \frac{1}{\sqrt{2}}\bmm \includegraphics[width=0.9in]{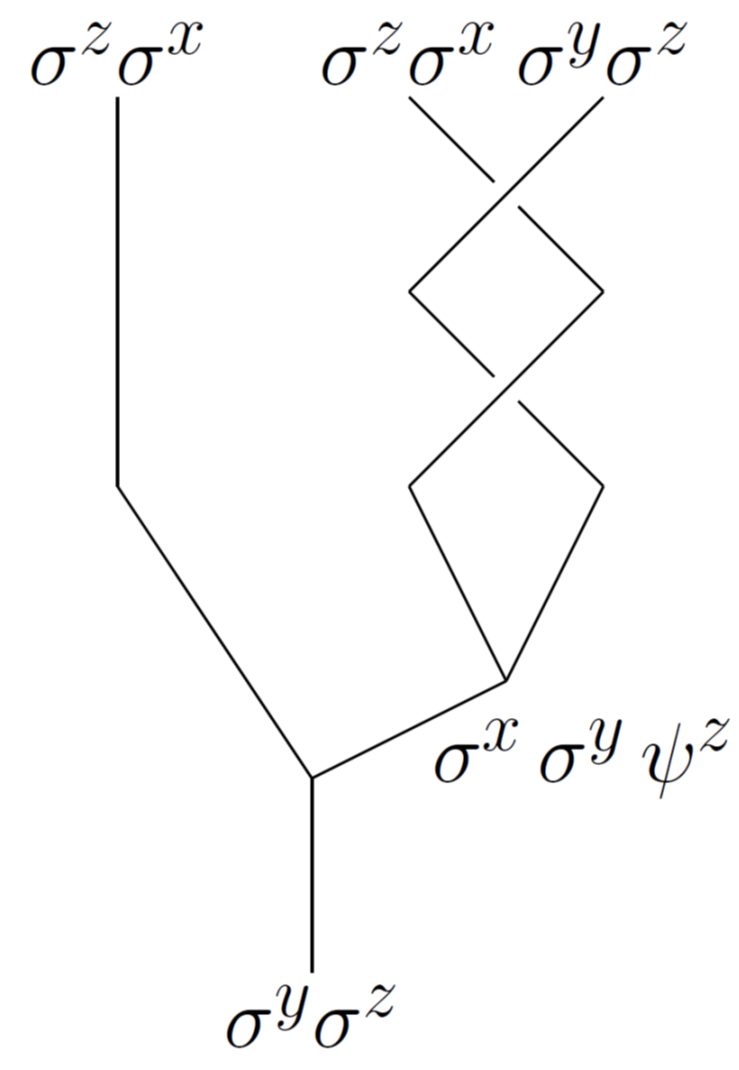} \emm.
\eeq
We can then un-braid the $\sigma^z \sigma^x$ and $\sigma^y \sigma^z$ strings, picking up two factors of the $R$-matrix in the process, and finally make another $F$-move to show that
\beq
\label{fusionchannel}
\bmm \includegraphics[width=0.9in]{fusionchannel_iv} \emm = \frac{1}{2} \left(e^{-i\pi/4} - e^{i3\pi/4} \right) \bmm \includegraphics[height=1in]{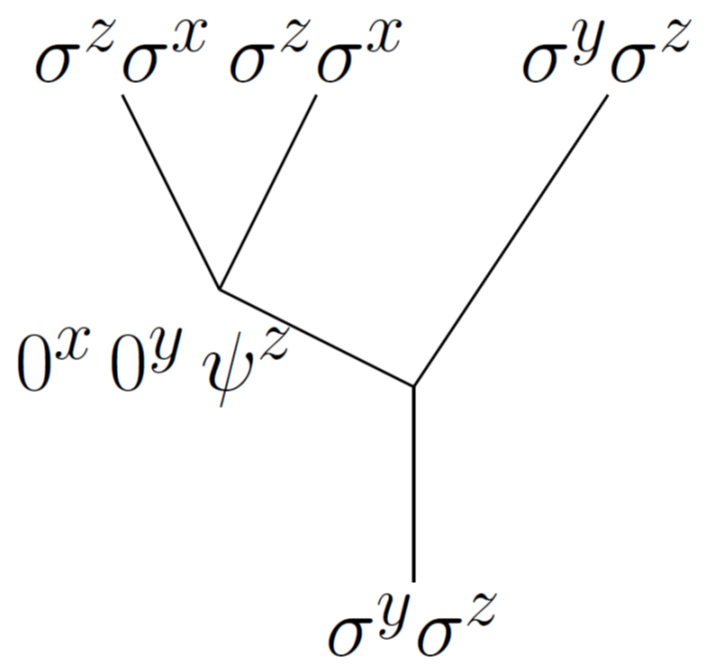} \emm.
\eeq
The Ising anyon $R$-matrix elements needed in this calculation are $R^{\sigma \sigma}_0 = e^{-i \pi / 8}$ and $R^{\sigma \sigma}_{\psi} = e^{3 i \pi / 8}$~\cite{bonderson}. We emphasize that the only non-trivial $F$- and $R$- tensors involved in this process come from the $\sigma^z$ strings living in the $xy$ planes, since the $\sigma^x$ and $\sigma^y$ strings remain unchanged throughout. Some details of the $F$ and $R$ tensors for UMTCs are discussed in Appendix~\ref{app:FandRdetails}.

The non-Abelian nature of the dim-1 excitations is evident from Eq.~\eqref{fusionchannel}, since braiding the dim-1 particles changes their fusion channel from the vacuum to the $\psi$ channel with unit probability. This implies that after braiding two dim-1 excitations with each other, they can no longer be annihilated back into the vacuum, as is also reflected in the fact that $\mathcal{M} = 0$ for these excitations. Such a process is impossible for Abelian excitations, so this unequivocally establishes the existence of non-Abelian excitations with restricted mobility in cage-net models.

The fact that the braiding between sub-dimensional excitations in the cage-net fracton phase may be reduced to that of anyons in 2d topological orders is one of the key benefits of flux-string condensation, as it allows us to simply understand the properties of the fracton phase in terms of more familiar anyon theories. In particular, it has allowed us to explicitly demonstrate that cage-net models host excitations such as $\sigma^\mu \sigma^\nu$, which obey non-Abelian braiding and fusion. 

Finally, we can also consider composites of non-Abelian dim-1 particles moving along the same direction, such as $\sigma^x(x_0) \sigma^y(y_0) \times \sigma^x(x_0 + 1) \sigma^y(y_0)$, where we have made layer indices for the $\sigma^{\mu}$ particles explicit. Specifically, $\sigma^x(x_0)$ belongs to the $yz$-plane with $x$-coordinate $x = x_0$, and similarly for $\sigma^y(y_0)$, so that $\sigma^x(x_0) \sigma^y(y_0)$ is understood as the dim-1 excitation restricted to move along the line specified by $x=x_0,y=y_0$. Thus, the composite $\sigma^x(x_0) \sigma^y(y_0) \times \sigma^x(x_0 + 1) \sigma^y(y_0)$ describes a bound state of two dim-1 particles which are separated along the $x$-axis by one lattice site. Clearly, this composite is mobile in at least one direction, $z$, since the individual dim-1 excitations forming it are mobile along $z$. In Abelian fracton models, such composites are in fact dim-2 excitations, so it is natural to ask whether this also occurs here.

We first note that this composite is only a particle of simple type if we choose a definite fusion channel for the two $\sigma^y$ particles, which belong to the same $d=2$ layer.  This is easily seen by working out the fusion of two composites without making any assumptions about the $\sigma^y$ fusion channel, and observing that the vacuum sector appears twice in the fusion outcome.  The two $\sigma^y$ particles can either be in the vacuum fusion channel, or in the $\psi^y$ channel, so there are in fact two different composites of simple type to consider.  We refer to these composites as $C_0$ and $C_{\psi}$, respectively.

To understand the mobility of these composites, we need to ask whether they can ``turn a corner'' and move along the $y$-direction.  The composite $C_0$ can indeed do this.  One way to see this is to observe that the vacuum channel appears in the fusion of $C_0$ with another composite $\sigma^x(x_0) \sigma^z(z_0) \times \sigma^x(x_0 + 1) \sigma^z(z_0)$, with the $\sigma^z$'s in the vacuum channel.  This corresponds to the existence of a local process where $C_0$, mobile along the $z$-direction, converts into the the second composite and becomes mobile along the $y$-direction.  On the other hand, there is no corresponding process for $C_{\psi}$, which is a dim-1 excitation.  The two $\sigma^y$ particles are in the $\psi^y(y_0)$ fusion channel, and this dim-2 particle is not able to move in the $y$-direction.  Indeed, upon considering the fusion of $C_\psi$ with another composite mobile along the $y$-direction, the fusion outcome always contains $\psi^y(y_0)$.  Thus, unlike Abelian fracton models, where composites of dim-1 particles have enhanced (dim-2) mobility, in the cage-net models the mobility of such composites is contingent upon the fusion channel of their constituents.

\subsection{Intrinsic and Inextricable Nature of non-Abelian excitations}
\label{sec:intrinsic}

In the preceding section, we have established the existence of non-Abelian excitations with restricted (dim-1) mobility in the cage-net model. Here, we demonstrate that these excitations are an intrinsically three dimensional feature, and not 2d anyons in disguise.  In order to do this, we must exclude the possibility that either the restricted mobility or non-Abelian nature of the dim-1 excitations descends trivially from dim-2 excitations.

There are two scenarios we need to rule out.   The first is illustrated by a system of inter-penetrating but decoupled layers of string-net models stacked along all three principal axes of the cubic lattice.  In this case, there would be deconfined $\sigma$ particles moving freely along each layer. However, we can consider two perpendicular planes oriented normal to $\mu$ and $\nu$, such that the bound state $\sigma^\mu \sigma^\nu$ can only move along the line where the planes intersect. While this bound state is a non-Abelian dim-1 particle, it is trivially so, because it is a bound state of deconfined dim-2 excitations.  To distinguish such trivial bound states from restricted-mobility excitations of a fundamentally three-dimensional phase of matter, we introduce the concept of an \textit{intrinsic} dim-1 excitation, which is one that is not the fusion result of deconfined dim-2 excitations.

In the second scenario to be ruled out, a non-Abelian dim-1 excitation is a bound state of an intrinsic Abelian dim-1 excitation $a$ and a non-Abelian dim-2 excitation $b$.  Due to the restricted mobility of $a$ and the non-Abelian nature of $b$, the bound state $ab$ is a non-Abelian dim-1 excitation, but its non-Abelian nature trivially descends from that of $b$.  We are thus led to introduce the notion of \textit{inextricably} non-Abelian dim-1 excitations as those which are not the fusion result of a deconfined Abelian dim-1 excitation and a deconfined non-Abelian dim-2 excitation~\footnote{The concepts of \textit{intrinsic} and \textit{inextricable} sub-dimensional excitations were concurrently introduced by some of us in Ref.~\cite{twisted}.}. 

We now show that the non-Abelian dim-1 excitations of the doubled Ising cage-net model are both intrinsic and inextricably non-Abelian, thus demonstrating these excitations are a fundamentally three-dimensional phenomenon.  The existence of such particles constitutes one of the central results of this paper.

To begin the argument, we consider an arbitrary point-like excitation in the cage-net model.  The $\psi^z(z_0)\, \bar{\psi}^z(z_0)$ excitation is mobile in the $z = z_0$ plane and can be braided around a cylinder that contains the excitation of interest, resulting in a statistical phase of $(-1)^{n_z(z_0)}$ with $n_z(z_0) \in \{0,1\}$.  We expect that $n_z(z_0) = 0$ if the excitation is sufficiently far away from the $z = z_0$ plane, but the statistical phase factor is nonetheless well-defined.  We then define the $\mathbb{Z}_2$ quantum number 
\beq
N_z = \sum_z n_z(z) \mod 2 \text{,}
\eeq
where the non-zero contributions only arise from those values of $z$ not too far from the $z$-coordinate of the excitation of interest.
Similarly, we define $N_x$ and $N_y$, for the $x$ and $y$-directions respectively, to get a triple of $\mathbb{Z}_2$ quantum numbers
\beq
N = (N_x, N_y, N_z) \in \mathbb{Z}_2^3.
\eeq
These quantum numbers are useful when keeping track of the braiding statistics between a point-like excitation and the condensed $\psi \bar{\psi}$-strings, as we will see below.

\begin{figure}[t]
\includegraphics[width=\columnwidth]{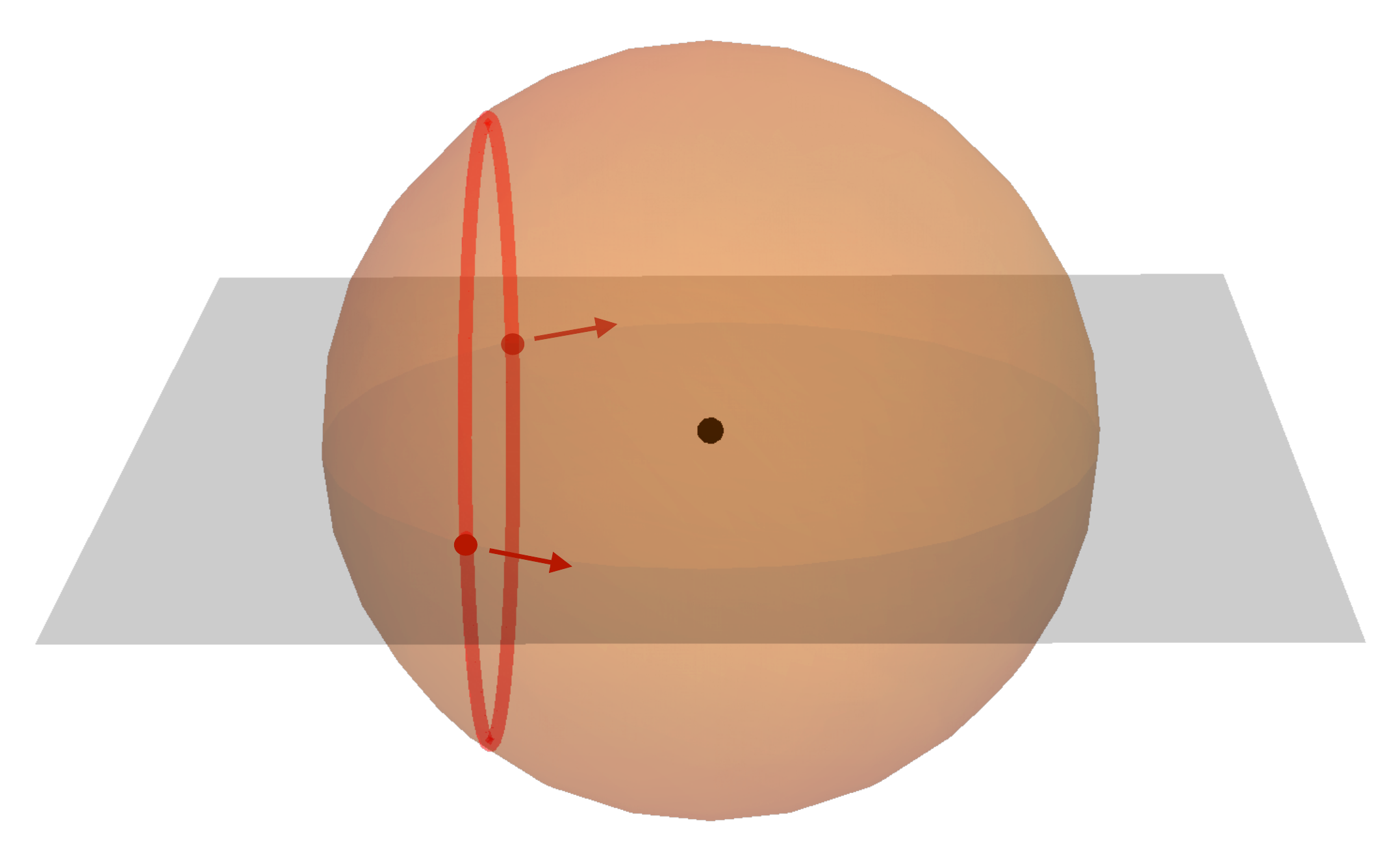}
\centering
\caption{Depiction of a process where a closed loop of flux string (red line) is created from the vacuum on the left side of the sphere, and moves along the sphere's surface until disappearing back into the vacuum on the right side of the sphere.  During this process the flux string encircles a point-like excitation with $N = (0,0,1)$ (black dot).  A non-trivial statistical phase of $-1$ results from braiding of $\psi^z \bar{\psi}^z$ excitations (red dots) around the $N = (0,0,1)$ excitation during this process, so that such excitations are confined in the flux-string condensed phase.  The $\psi^z \bar{\psi}^z$ excitations move within the gray-shaded $xy$-plane.}
\label{fig:pstring_braiding}
\end{figure}

We note that the ``elementary'' non-Abelian dim-1 particles $\sigma^x \sigma^y$, $\sigma^x \sigma^z$ and $\sigma^y \sigma^z$ have $N = (1,1,0), (1,0,1)$ and $(0,1,1)$, respectively, while the dim-2 excitations of each string-net layer that survive flux-string condensation (\emph{i.e.} $\psi, \bar{\psi}, \psi \bar{\psi}$ and  $\sigma \bar{\sigma}$), as well as the Abelian fractons, have $N = (0,0,0)$.  Because all excitations can be obtained by fusion of these, it follows that deconfined excitations realize only the $\mathbb{Z}_2 \times \mathbb{Z}_2$ subgroup of $\mathbb{Z}_2^3$ generated by $N = (1,1,0)$ and $N = (1,0,1)$.  It is instructive to reach the same conclusion by observing that any excitation with one of the other four possible values of $N$ is necessarily confined due to its statistical interactions with the flux-string condensate.  For instance, Fig.~\ref{fig:pstring_braiding} shows a process where a loop of flux-string is created from the vacuum, and encloses a $N = (0,0,1)$ excitation before disappearing into the vacuum again, resulting in a statistical phase of $-1$ due to braiding with the $\psi^z \bar{\psi}^z$ excitations in the flux-string.

\begin{figure}[t]
\includegraphics[width=\columnwidth]{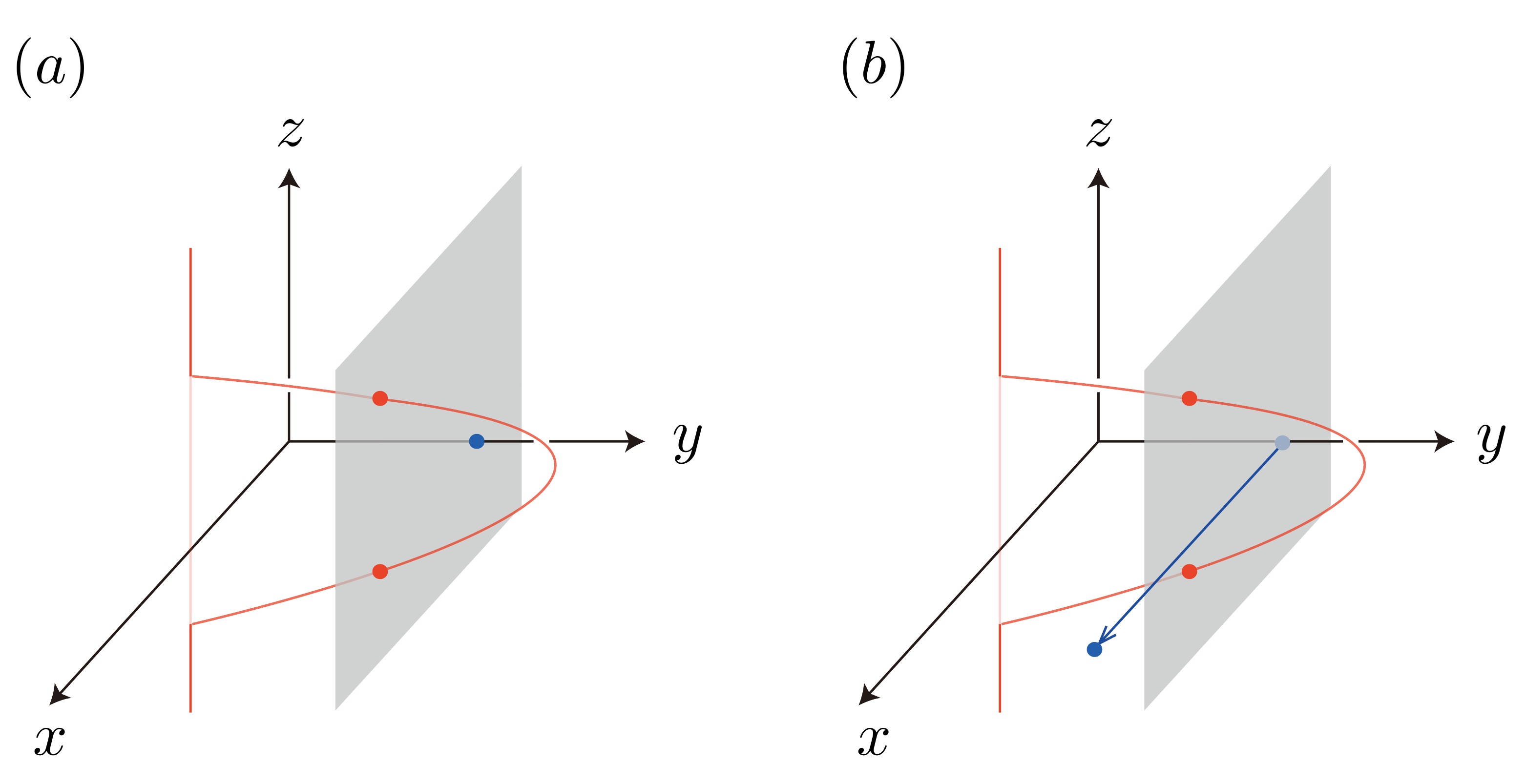}
\centering
\caption{Depiction of a process where a point particle (blue dot) with $N = (1,1,0)$, which is assumed to be mobile in the $x$-direction, braids with a flux-string (red line).  (a) shows the first step of the process, where a flux-string that initially runs along the $z$-direction is bent into the $y$-direction within a $yz$-plane.  In doing so, the flux-string pierces  the $xz$-plane containing the particle, and the red dots show the locations of $\psi^y \bar{\psi}^y$ excitations in this plane.  In the second step as shown in (b), the point particle moves in the $x$-direction, passing through the loop formed by the bent flux-string.  In the third step, the loop is bent back to its original configuration, undoing the first step.  Finally, the particle is moved back to its original position, undoing step two.  A statistical phase of $-1$ results from the braiding of the point particle with the $\psi^y \bar{\psi}^y$ excitations during this process.  It follows that the $N = (1,1,0)$ excitation is forbidden to move along the $x$-direction in the flux-string condensed phase.}
\label{fig:pstring_1d}
\end{figure}

Next, we show that the mobility of an excitation is directly tied to its value of  $N \in \mathbb{Z}_2 \times \mathbb{Z}_2$.  In particular, an excitation with $N = (1,1,0)$ can have no component of its motion along the $x$ or $y$-direction.  If it did, there would be a braiding-like process with a flux-string that gives a statistical phase of $-1$, as illustrated in Fig.~\ref{fig:pstring_1d}.   Such motion is thus  forbidden in the presence of the flux-string condensate, and the $N = (1,1,0)$ excitation can only move along the $z$-direction.  Corresponding statements hold for the other two non-trivial elements of $ \mathbb{Z}_2 \times \mathbb{Z}_2$.  Incidentally, this argument shows that the sub-dimensionality of $N \neq 0$ excitations is in fact a kind of confinement arising from statistical interactions with the flux-string condensate.

It follows immediately from this discussion that all dim-2 excitations have $N = (0,0,0)$.   Therefore, because $N$-values add under fusion, it follows that any excitation with $N \neq 0$ cannot be obtained by fusing together dim-2 particles.  In particular, the  ``elementary" non-Abelian dim-1 particles such as $\sigma^x \sigma^y$ cannot be obtained by fusing together dim-2 excitations, and are thus intrinsic dim-1 excitations.

To complete our argument, we need to show that these dim-1 excitations are inextricably non-Abelian. We proceed by contradiction, supposing that the dim-1 non-Abelian excitations are not inextricably non-Abelian. By definition, there must then exist an Abelian dim-1 excitation and a non-Abelian dim-2 excitation which fuse to one of the excitations of interest, which, as we proved above, must carry a non-trivial $N$ quantum number. Since all dim-2 excitations carry trivial $N$, this implies that the Abelian dim-1 excitation must have a non-trivial $N$ associated with it. As we now show, this is not allowed in the doubled Ising cage-net model---all excitations with non-zero $N$ are non-Abelian.

First, note that any excitation with non-zero $N$ is obtainable by fusing together ``elementary'' non-Abelian dim-1 particles, possibly together with dim-2 excitations including $\sigma \bar{\sigma}$.  It follows that $N_x$ counts the total number of $\sigma^x$ and $\bar{\sigma}^x$ particles modulo two appearing in the fusion product, with corresponding statements for $N_y$ and $N_z$.  Following the discussion of composite excitations in Sec.~\ref{sec:excitations}, in order to obtain an excitation of simple type, for instance, any two $\sigma^{x}(x)$ particles with the same layer index $x$ must be combined into a definite fusion channel.  A composite is non-Abelian as long as there remain some unpaired $\sigma^{\mu}$ or $\bar{\sigma}^{\mu}$ excitations, and this must be the case whenever $N \neq 0$.  For instance, if $N_x = 1$, then the total number of $\sigma^x$ and $\bar{\sigma}^x$ particles is odd, so they cannot all be paired into definite fusion channels with other excitations.  Thus we have shown that all $N \neq 0$ excitations are non-Abelian, and the ``elementary'' dim-1 non-Abelian particles are thus inextricably non-Abelian.

An important corollary of the intrinsic and inextricable character of the dim-1 excitations is that the cage-net Ising fracton model cannot be equivalent as a foliated fracton phase \cite{shirley} to any Abelian fracton phase, including the X-cube model.    Equivalence of foliated fracton phases is defined up to stacking with layers of 2d topologically ordered states.  That is two fracton phases $A$ and $B$ are equivalent as foliated fracton phases if $A$ stacked with layers is adiabatically connected to $B$, stacked with a possibly different set of layers.  It is immediately apparent from the above discussion that the doubled Ising cage-net model is not equivalent in this sense to any Abelian fracton phase.  Therefore, we have established the existence of non-Abelian foliated fracton phases.  

\subsection{Generalization to SU(2)$_k$ Cage-Net Fracton Models}

The construction of the doubled Ising cage-net fracton model, based on the doubled Ising string-net model, can be straightforwardly generalized to SU(2)$_k$ cage-net models, by using the doubled SU(2)$_k$ string-net models considered in Ref.~\cite{burnell2}. In the chiral SU(2)$_k$ theory, the particle types are labelled by $j = 0, 1/2, 1, \dots, k/2$, so that excitations of the doubled theory (and the string-net model) are labelled by pairs $(j_1, j_2)$. The $(k/2,k/2)$ excitation is an Abelian boson whose condensation was discussed in Ref.~\cite{burnell2}.

We consider condensation of $(k/2,k/2)$ flux strings. In order to write down an SU(2)$_k$ cage-net fracton model, one could in principle implement the $(k/2,k/2)$-string condensation explicitly, following a procedure similar to the one introduced in Sec.~\ref{sec:su2_2-cage-net}. However, rather than repeating this explicit construction, we will instead obtain the excitation content of the fracton phase from the flux string condensation picture.

First, paralleling our discussion of the doubled Ising cage-net model, the $(k/2,k/2)$ flux is equivalent to a pair of fractons and remains a well-defined deconfined dim-2 particle in the flux-string condensed fracton phase. Here also, four fractons are created at the corners of a membrane comprised of the flux creation operators for the $(k/2,k/2)$ flux, which can be found using Eq.~\eqref{fun-string} and the $S$-matrix for SU(2)$_k$ anyons. Since the $(k/2,k/2)$ flux is an Abelian anyon, the fractons in all SU(2)$_k$ cage-net models will lack any topological degeneracy and will remain Abelian. 

In order to identify the dim-2 excitations from each layer that survive flux-string condensation, we simply need to know the Abelian statistics between an arbitrary excitation $(j_1, j_2)$ and $(k/2,k/2)$.  From the ${\rm SU}(2)_k$ $S$-matrix~\cite{bonderson,burnell2}, this is found to be
\begin{equation}
\exp\Big( i \Theta_{(j_1,j_2), (k/2,k/2)} \Big) = (-1)^{2 (j_1 + j_2)} \text{.}
\end{equation}
Therefore, the dim-2 deconfined excitations in each layer are $(j_1, j_2)$ where $j_1 + j_2$ is an integer.  On the other hand, excitations where $j_1 + j_2 = 1/2 \mod 1$ have $\theta = \pi$ Abelian mutual statistics with the flux strings, and are confined.

As in the doubled Ising case, we can obtain deconfined dim-1 excitations from these confined dim-2 excitations. In a given $d=2$ layer, we consider any confined excitation, \emph{i.e.} any $(j_1, j_2)$ with $j_1 + j_2 = 1/2 \mod 1$.  A bound state of this excitation with another confined excitation $(j_3, j_4)$ in a perpendicular layer will have trivial statistics with the flux strings, and will be a deconfined dim-1 excitation. Similarly to the doubled Ising case, the spectrum of deconfined excitations will generically contain non-Abelian dim-1 particles. 

\subsection{Cage-Net Wave Functions}
\label{sec:wfn}

\begin{figure}[t]
\includegraphics[width=8cm]{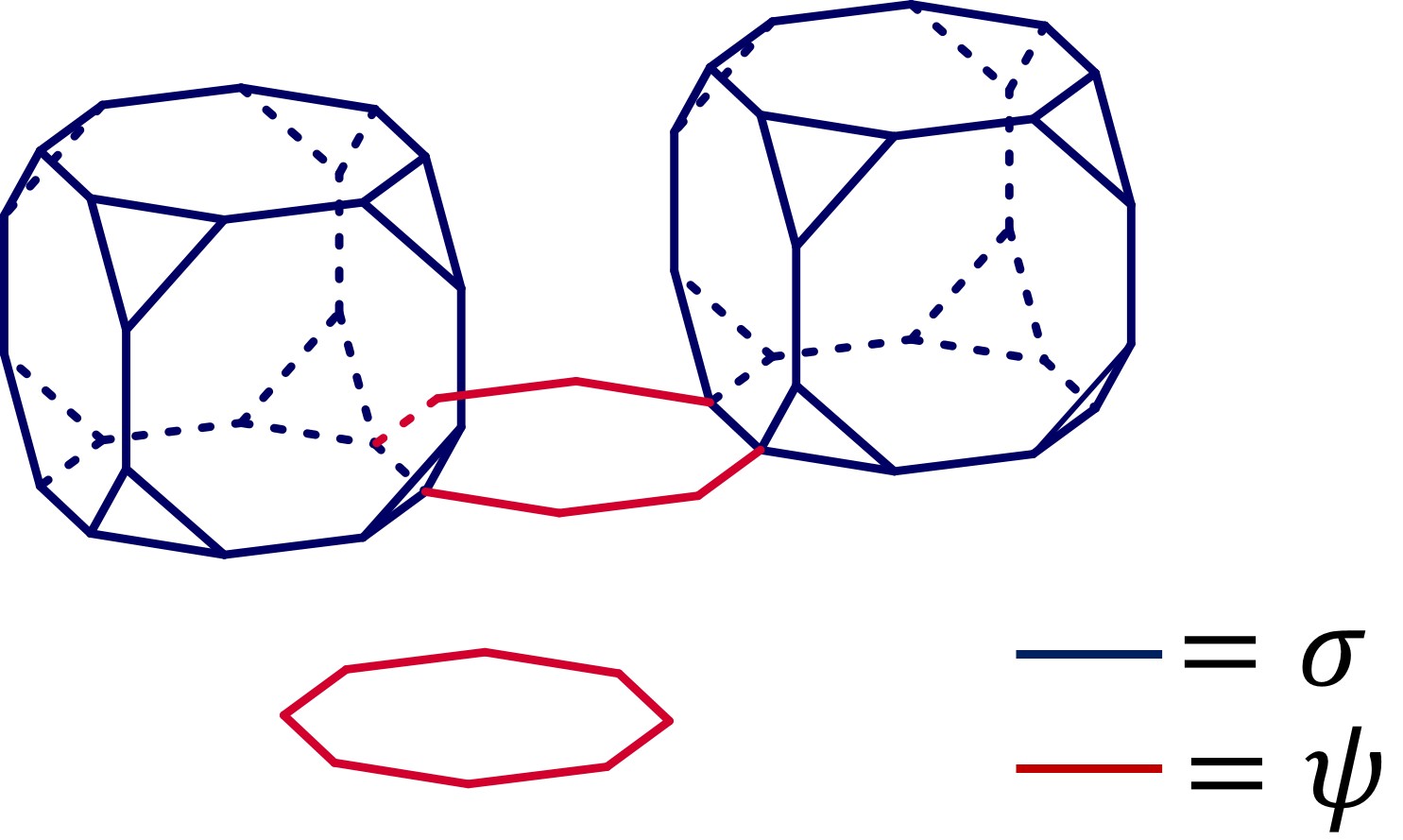}
\centering
\caption{A typical configuration in the ground state of the doubled Ising cage-net model. Cages are built out of the $\sigma$ strings, shown here by blue lines, while loops within each $d=2$ layer are built from $\psi$ strings, shown here in red.}
\label{fig:cagenet_wf}
\end{figure}

We now discuss the ground state wave function of the doubled Ising cage-net model. Recall from our discussion of the X-Cube model (see Sec.~\ref{type1}) that the string configurations which minimized the vertex term were ``cages" or skeletons of cubes. Similarly, for the doubled Ising cage-net model~\eqref{eqn:h_fraction_ising}, the configurations which minimize the vertex terms are cages built out of $\sigma$ strings. Each elementary cage of $\sigma$ strings can be thought of as six loops of $\sigma$ strings in the six octagonal plaquettes bounding the cage. In addition, there are also $\psi$ strings running within the $d=2$ layers. Due to the branching rule $\delta_{\psi\sigma\sigma}=1$, the $d=2$ loops are not decoupled from the $d=3$ cages; instead, $\psi$ strings can end on the $\sigma$ cages. A representative example of such a configuration is shown in Fig.~\ref{fig:cagenet_wf}, where the blue (red) strings correspond to the $\sigma$ ($\psi$) strings. The cage term Eq.~\eqref{eqn:H_ising_cage} gives dynamics to the cages and makes them fluctuate, while plaquette terms give dynamics to the $\psi$ strings. Thus, the ground state wave function of the non-Abelian doubled Ising fracton phase can be interpreted as a fluctuating network of $\sigma$ cages and $\psi$ strings, which we dub a ``cage-net condensate."

This picture of the fracton phase as a cage-net condensate illuminates the geometric nature of fracton order. For instance, in the X-Cube model, a generating set of string configurations associated with each vertex $v$ is shown in Fig.~\ref{vertexallowed}. By slightly coarse-graining the truncated cubic lattice on which the doubled Ising fracton model is defined, a similar generating set of string configurations, which form the fundamental building blocks for the cage configurations, can be delineated.

In contrast with conventional topologically ordered phases~\cite{levinwen,BraneNet}, the set of allowed string configurations in cage-net fracton phases have the property that certain types of strings are \textit{not} allowed to turn a corner at a vertex $v$. Instead, a string parallel to the principal axis $\mu$ can either pass straight through the vertex $v$ or it can turn in an orthogonal direction $\nu \perp \mu$ at the cost of creating another string attached to $v$ but along the axis mutually orthogonal to $\mu$ and $\nu$. Thus, there is a certain rigidity associated with the allowed string configurations in a phase with fracton order, which distinguishes these phases in a concrete way from conventional topologically ordered phases. This rigidity, or sensitivity to local geometry, explains the dependence of the ground state degeneracy in a fracton phase on the system size or on local curvature. Indeed, the sub-dimensionality of excitations is also a direct consequence of the fact that strings in a fracton phase are disallowed from simply turning corners without creating additional topological excitations. The geometric nature of fracton order has also been emphasized in Refs.~\cite{shirley,slagle3}.

\section{Conclusions and Outlook}
\label{concls}

In this work, we have introduced a class of gapped $d=3$ non-Abelian fracton models, dubbed ``cage-net" fracton models, based on coupled layers of $d=2$ string-net models. In our framework, fracton phases are obtained by condensing extended one-dimensional flux-strings made up of point-like excitations, thereby generalizing the familiar concept of anyon condensation~\cite{bais,eliens,kong,neupert,burnell,QDcondensation}. As specific examples, we have considered in detail the doubled Ising cage-net model and its straightforward extension to SU(2)$_k$ cage-net models.

A key feature of these models is the presence of non-Abelian sub-dimensional  excitations. In particular, we have demonstrated that while fractons are always Abelian in our models, there exist deconfined dim-1 non-Abelian excitations in the spectrum.  Strikingly, these results provide a route towards realizing quantum phases of matter with non-Abelian excitations in a three dimensional system with local interactions. Fracton models may thus provide an intriguing platform for future studies of both topological quantum computation and of quantum information storage.

In the doubled Ising cage-net model, we showed that the dim-1 non-Abelian excitations are both intrinsically sub-dimensional and inextricably non-Abelian, and so their existence is fundamentally a three-dimensional phenomenon. This result implies that, as a foliated fracton phase~\cite{shirley}, the doubled Ising cage-net model is not equivalent to any Abelian fracton phase.  We have thus established the existence of non-Abelian foliated fracton phases.  In the future, it will be interesting to see if characterizations of foliated fracton phases in terms of quotient superselection sectors and the corresponding interferometric operators~\cite{foliate} can be generalized to non-Abelian foliated fracton phases. In addition, suitable measures of entanglement~\cite{foliated-entanglement} may distinguish  non-Abelian foliated fracton phases from their Abelian cousins, at least to some extent.

It is worth emphasizing that to definitively show the existence of non-Abelian excitations in a particular gapped phase of matter, one needs to first derive the spectrum of deconfined excitations and to then show that some subset of these excitations can participate in multiple fusion channels. Calculating these properties in cage-net fracton models is made particularly straightforward as a result of the flux-string condensation procedure employed here. Indeed, this is what allows us to demonstrate the intrinsically dim-1 and inextricably non-Abelian nature of the dim-1 excitations in the doubled Ising cage-net model. This contrasts with the prior work of Ref.~\cite{nonabelian}, which constructed 3d models based on coupled layers of 2d quantum double models, that were claimed to support non-Abelian immobile fracton excitations. Ref.~\cite{nonabelian} did not study fusion or braiding of the excitations in these models, and the non-Abelian nature of the fractons in these models was thus not demonstrated. Therefore, it remains an open question whether non-Abelian fractons can be obtained through a construction based on coupled layers of $d=2$ topological orders or through flux-string condensation. We emphasize that nothing precludes the existence of non-Abelian immobile excitations; indeed, a different model introduced and studied in Ref.~\cite{nonabelian}, and other models very recently introduced by some of us with Martin-Delgado~\cite{twisted}, support fracton phases where such excitations have been shown to exist.

Another consequence of our construction is the identification of the ground state wave function of some fracton phases with a condensate of fluctuating cage-net configurations, providing insights into the inherent geometric nature of fracton orders. A detailed study of cage-nets may provide a route towards uncovering the general mathematical framework underlying the class of fracton models studied here. This may further provide a new route towards obtaining fracton models directly in three-dimensional space, \emph{i.e.} not from a coupled-layer construction. Alternatively, it will be interesting to understand whether there are type-I fracton orders that cannot be obtained from flux-string condensation, which remains an open question.

As a further extension of our work, it would be interesting to study the ground state wave-function structure of the more complicated ``type-II" fracton phases, of which Haah's code~\cite{haah} is the paradigmatic example and for which there appears to exist no layered construction. As an intermediate step towards this goal, finding similar string-net constructions for other ``type-I" models, such as the checkerboard model~\cite{fracton1}, is likely to provide further insights into the nature of fracton order. We also note that while we have focused on a particular $d=3$ lattice in this work, following the methods of Refs.~\cite{shirley,slagle3}, it should be possible to define these models on general three-dimensional manifolds with an appropriate foliation structure.

\section*{Acknowledgments}
We are grateful to Barry Bradlyn, Daniel Bulmash, Fiona Burnell, Xie Chen, Trithep Devakul, Lukasz Fidkowski, Liang Fu, Michael Levin, Han Ma, Miguel Angel Martin-Delgado, Michael Pretko, Rahul Nandkishore, Albert Schmitz, Wilbur Shirley, Shivaji Sondhi,  Sagar Vijay, and Dominic Williamson for stimulating conversations and correspondence. This research is supported by the U.S. Department of Energy, Office of Science, Basic Energy Sciences (BES) under Award number DE-SC0014415 (SJH and MH), by the Spanish MINECO grant FIS2015-67411 (HS), and by the CAM research consortium QUITEMAD+, Grant number S2013/ICE-2801 (HS). AP acknowledges support from the University of Colorado at Boulder.

\appendix

\section{$R$ and $F$ tensors in UMTCs of the form $\mathcal{C} \times \bar{\mathcal{C}}$}
\label{app:FandRdetails}

Here, we briefly review the definition of the $R$ tensor in UMTCs, and give the form of the $R$ and $F$ tensors in UMTCs of the form $\mC \times \bar{\mC}$. A more general and detailed discussion can be found in Ref.~\cite{bonderson}. The class of string-net models considered in the main-text are those which take as input a UMTC $\mC$, which admits a well defined braiding. Diagrammatically, the braiding is defined by 
\begin{equation}
\bmm\includegraphics[height=0.55in]{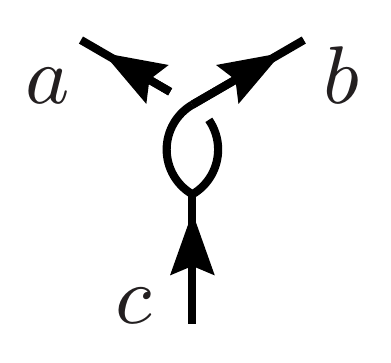}\emm = R^{ab}_{c} \bmm\includegraphics[height=0.5in]{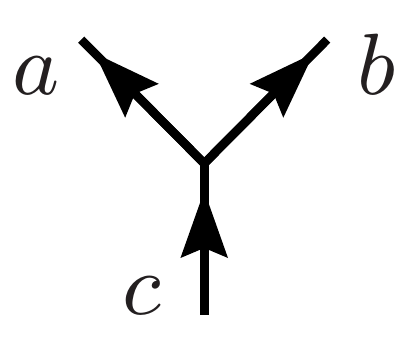}\emm \text{,}
\label{eqn:R}
\end{equation}
where the $R$ tensor encodes the information of exchanging two anyons $b$ and $a$, respectively, which fuse to an anyon $c$ (we assume no fusion multiplicities). While the $R$ tensor is not required in the construction of string-net models, we now discuss how it becomes the data describing braiding properties of anyons in string-net models with input a UMTC $\mC$. 

As discussed in Sec.~\ref{unimod}, anyons in string-net models belong to the Drinfeld center of $\mC$, which for a UMTC is $\mathcal{Z}(\mathcal{C}) = \mathcal{C} \times \bar{\mathcal{C}}$. For this sub-class of models, anyons are labelled by an ordered pair $(a,b)$, where $a \in \mathcal{C}$ and  $b \in  \bar{\mathcal{C}}$. Since $\mathcal{C} \times \bar{\mathcal{C}}$ is also a UMTC admitting a well-defined braiding, it is also equipped with an $R$ tensor $R^{(a,a')(b,b')}_{(c,c')}$, defined in terms of the same diagram as in Eq.~(\ref{eqn:R}), but now with anyons labelled by ordered pairs. It is well-known that for the output category $\mathcal{C} \times \bar{\mathcal{C}},$  the R-tensor equals $R^{ab}_{c} (R^{ab}_{c})^{*}$~\cite{gu2014}. Therefore, the $R$ tensor $R^{ab}_{c}$ in the input UMTC $\mathcal{C}$ in fact encodes all the information regarding the braiding of anyons in the corresponding string-net models. It is also useful to note that the $F$ tensor in $\mathcal{C} \times \bar{\mathcal{C}}$ is given by $F^{(i,i')(j,j')(m,m')}_{(k,k')(l,l')(n,n')} = F^{ijm}_{kln} F^{i'j'm'}_{k'l'n'}$\cite{gu2014}.


\section{Doubled Ising string-net model:  $F$ tensor and string operators}
\label{app:Ising-details}

This appendix contains further details regarding the doubled Ising string-net model. In particular, we give the $F$ tensor and the $\sigma$-type simple string operators. The non-trivial elements of the $F$ tensor are given by~\cite{bonderson} 
\begin{eqnarray}
F^{\sigma \sigma m}_{\sigma \sigma n} &=& \frac{1}{\sqrt{2}}
\begin{pmatrix} 
1 & 1 \\
1 & -1 
\end{pmatrix} \text{,}
\\
F^{\psi \sigma \sigma}_{\psi \sigma \sigma} &=& F^{\sigma \psi  \sigma}_{\sigma \psi  \sigma}=-1 \text{,}
\end{eqnarray}
where $m,n = 0,\psi$. All other elements of $F$ are $1$ as long as the branching rules are satisfied in the relevant configurations, and are $0$ otherwise. 

We now briefly review the definition of the string operators in the string-net models, following Ref.~\cite{levinwen}. A string operator $W_\alpha$ creates a pair of quasi-particles at its ends. Graphically, the string operators are defined on a ``fattened" lattice, where the links of the lattice are fattened into strips of finite width. The action of a string operator $W_\alpha(P)$ on a fixed basis state is represented by a dashed string labelled by $\alpha$, along the path $P$, on the fattened lattice. The resulting string configuration is then reduced to a string configuration in the un-fattened lattice by using the following local rules:
\begin{eqnarray}
	\Bigg| \bmm\includegraphics[height=0.5in]{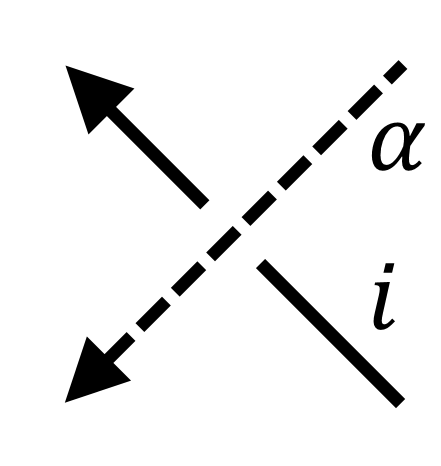}\emm \Bigg\> &=& \sum_{jst} \Omega_{\alpha,sti}^j
	\Bigg| \bmm\includegraphics[height=0.5in]{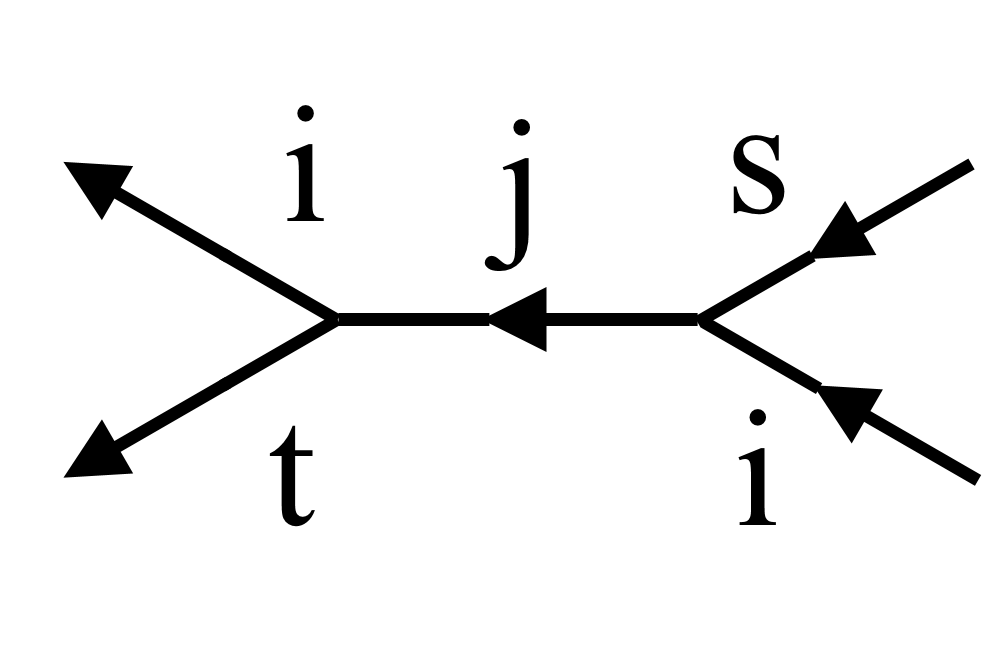}\emm \Bigg\>,
\label{eqn:string_1}
\\
	\Bigg| \bmm\includegraphics[height=0.5in]{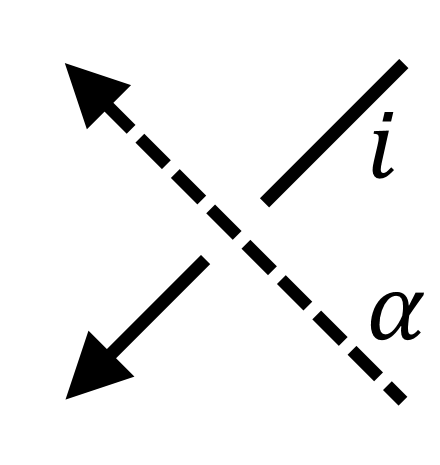}\emm \Bigg\> &=& \sum_{jst} \bar{\Omega}_{\alpha,sti}^j
	\Bigg| \bmm\includegraphics[height=0.5in]{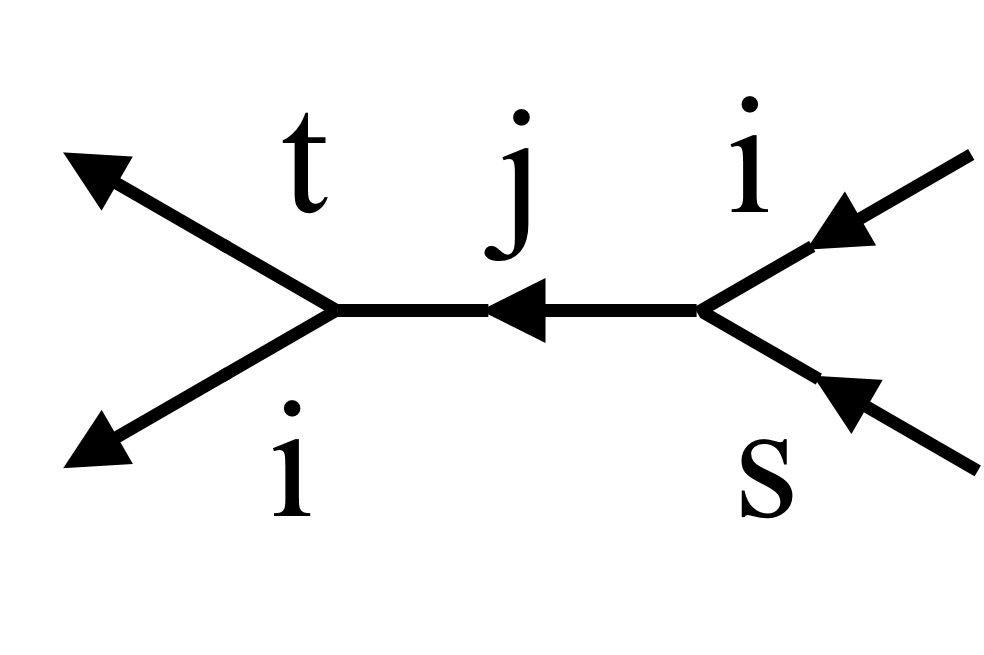}\emm \Bigg\>.
\label{eqn:string_2}
\end{eqnarray}
In order for these string operators to be path-independent, the $\Omega$ and $\bar{\Omega}$ tensors need to satisfy the following set of self-consistency conditions,
\begin{eqnarray}
\sum_{s}\bar{\Omega}^{m}_{rsj} F^{sl^{*}i}_{kjm^{*}} \Omega^{l}_{sti} \frac{v_{j}v_{s}}{v_{m}} &=& \sum_{n} F^{ji^{*}k}_{t^{*}nl^{*}} \Omega^{n}_{rtk} F^{jl^{*}n}_{krm^{*}} \text{,}
\label{eqn:string_cond1}
\\
\bar{\Omega}^{j}_{sti} &=& \sum_{k} \Omega^{k}_{sti^{*}} F^{it^{*}k}_{i^{*}sj^{*}} \text{.}
\label{eqn:string_cond2}
\end{eqnarray}
The solutions to these equations correspond to topologically distinct excitations, whose string operators are called ``simple" and are labelled by the excitations. 

We now find the $\sigma$-type string operators for the doubled Ising string-net model studied in the main text. We consider a simple ansatz for the $\Omega$ tensor \cite{levinwen},
\beq
\Omega^{i}_{stj} = \omega^{i}_{j}\delta_{s\sigma}\delta_{t\sigma}.  \label{eqn:simplestring}
\eeq
Using this ansatz, Eqs.~\eqref{eqn:string_cond1} and \eqref{eqn:string_cond2} become
\begin{eqnarray}
\bar{\omega}^{m}_{j} F^{\sigma l i}_{kjm} \omega^{l}_{i} \frac{v_{j}v_{\sigma}}{v_{m}} &=& \sum_{n} F^{jik}_{\sigma n l} \omega^{n}_{k} F^{jln}_{k \sigma m} \text{,}
\label{eqn:string_cond3}
\\
\bar{\omega}^{j}_{i} &=& \sum_{k} \omega^{k}_{i} F^{i \sigma k}_{i \sigma j} \text{.}
\label{eqn:string_cond4}
\end{eqnarray}
Solving these equations, we find that there are four distinct string solutions, listed in Table~\ref{tab:string_solution}.

Now that we have obtained the $\Omega$ tensors, we can also calculate the corresponding topological spins for the excitations by using the relation
\begin{eqnarray}
e^{i\theta_{\alpha}} &=& \frac{\sum_{s} \Omega^{0}_{\alpha,sss} d_{s}^{2}}{\sum_{s} \Omega^{s}_{\alpha,ss0} d_{s} }\text{,}
\\
&=&\sqrt{2} \omega^{0}_{\sigma}\text{.}
\label{eqn:topo_spin}
\end{eqnarray}

The resulting topological spins are listed in the $6$th column of Table~\ref{tab:string_solution}. By matching our solutions with the known topological spins of excitations in the doubled Ising topological order~\cite{gu2014}, we can identify the quasiparticle type created by each string, which is given in the first column of Table~\ref{tab:string_solution}. The $S$-matrix for the four anyons $\bar{\sigma}, \sigma,\psi \bar{\sigma}$, and $\sigma \bar{\psi}$ is 
\begin{equation}
S=
\begin{pmatrix} 
0 & 1 & 0 & -1 \\
1 & 0 & -1 & 0  \\
0 & -1 & 0 & 1 \\
-1 & 0 & 1 & 0 
\end{pmatrix} \text{,}
\end{equation}
which can be obtained by using the following formula:
\begin{eqnarray}
S_{\alpha \beta} &=& \frac{1}{D} \sum_{ijk} \Omega^{k}_{\alpha, iij}  \Omega^{k}_{\beta, jji} d_{i}d_{j}\text{,}
\\
&=&\sum_{k} \omega^{k}_{\alpha, \sigma}  \omega^{k}_{\beta, \sigma}\text{.}
\end{eqnarray}

All four string operators we have found anti-commute with the $\psi \bar{\psi}$ string
\beq
W_{\psi \bar{\psi}} = \prod_{l \perp P} (-1)^{n_{\sigma}(l)},
\eeq
where the product runs over links $l$ perpendicular to some path $P$ (see Fig.~\ref{fig:04} for an example). To prove this, we only need to show that the string operators have the property that $n_{\sigma}(l)$ is toggled between $0$ and $1$ along the path of links $l \perp P$ on which the string operator acts. Using the graphical rules Eqs.~\eqref{eqn:string_1} and \eqref{eqn:string_2} of string operators and the solutions of $\Omega$ tensors listed in Table~\ref{tab:string_solution}, it is straightforward to check that this is indeed the case.

\begin{table*}
{\renewcommand{\arraystretch}{2}%
\begin{tabular}{c|c|c|c|c|c}
Quasiparticle type & $\omega^{\sigma}_{0}$ & $\omega^{\sigma}_{\psi}$ & $\omega^{0}_{\sigma}$ & $\omega^{\psi}_{\sigma}$ & $e^{i\theta_{\alpha}}$\\
\hline
$\bar{\sigma}$ & $1$ & $e^{-i\pi/2}$ & $\frac{1}{\sqrt{2}}e^{-i\pi/8}$ &  $\frac{1}{\sqrt{2}}e^{3i\pi/8}$ & $e^{-i\pi/8}$\\
\hline
$\sigma$ & $1$ & $e^{i\pi/2}$ & $\frac{1}{\sqrt{2}}e^{i\pi/8}$ &  $\frac{1}{\sqrt{2}}e^{-3i\pi/8}$ & $e^{i\pi/8}$\\
\hline
$\psi\bar{\sigma}$ & $1$ & $e^{-i\pi/2}$ & $\frac{1}{\sqrt{2}}e^{7i\pi/8}$ &  $\frac{1}{\sqrt{2}}e^{-5i\pi/8}$ & $e^{7i\pi/8}$\\
\hline
$\sigma\bar{\psi}$ & $1$ & $e^{i\pi/2}$ & $\frac{1}{\sqrt{2}}e^{-7i\pi/8}$ &  $\frac{1}{\sqrt{2}}e^{5i\pi/8}$ & $e^{-7i\pi/8}$\\
\hline
\end{tabular}
}
\caption{Solutions for $\sigma$-type string operators, in terms of the tensor $\omega^i_j$ (see Eq.~\eqref{eqn:simplestring}). All other elements of $\omega^i_j$ are zero. Given the tensors $\omega^i_j$, the corresponding topological spins, which are listed in the sixth column, can be obtained from Eq.~\eqref{eqn:topo_spin}. The type of quasiparticle created by the string is given in the first column, and is identified by the topological spin. \label{tab:string_solution}}
\end{table*}


\newpage 

\bibliography{library}


\end{document}